\documentclass[journal,letterpaper]{IEEEtran}

\usepackage{cite}
\usepackage{graphicx}
\usepackage{subfigure}
\usepackage{amsmath} 
\usepackage{multirow}
\usepackage{amssymb}
\usepackage{url}

\begin{document}


\title{Validation Test of Geant4 Simulation of Electron Backscattering}

\author{Sung Hun Kim, Maria Grazia Pia, Tullio Basaglia, Min Cheol Han, Gabriela Hoff, Chan Hyeong Kim and Paolo Saracco 
\thanks{Manuscript received December 8, 2014. This research was partly funded by Brazilian Project 161/2012.)}
\thanks{T. Basaglia is with 
	CERN, CH1211 Gen\`eve 23, Switzerland
	(e-mail: Tullio.Basaglia@cern.ch).}
\thanks{G. Hoff is with CAPES Foundation, Ministry of Education of Brazil, Brasília - DF 70040-020, Brazil.
	(e-mail: ghoff.gesic@gmail.com).}
\thanks{ S. H. Kim,  M. C. Han and C. H. Kim are with 
	the Department of Nuclear Engineering, Hanyang University, 
        Seoul 133-791, Korea 
	(e-mail: ksh4249@hanyang.ac.kr, mchan@hanyang.ac.kr, chkim@hanyang.ac.kr).}
\thanks{M. G. Pia and P. Saracco are with 
	INFN Sezione di Genova, Via Dodecaneso 33, I-16146 Genova, Italy 
	(phone: +39 010 3536328, fax: +39 010 313358,
	e-mail: MariaGrazia.Pia@ge.infn.it, Paolo.Saracco@ge.infn.it).}

}

\maketitle

\begin{abstract}
Backscattering is a sensitive probe of the accuracy of electron scattering algorithms 
implemented in Monte Carlo codes.
The capability of the Geant4 toolkit to describe realistically the fraction of
electrons backscattered from a target volume is extensively and quantitatively
evaluated in comparison with experimental data retrieved from the literature.
The validation test covers the energy range between approximately 100 eV and 20 MeV, 
and concerns a wide set of target elements.
Multiple and single electron scattering models implemented in Geant4, as well as
preassembled selections of physics models distributed within Geant4, are analyzed
with statistical methods.
The evaluations concern Geant4 versions from 9.1 to 10.1.
Significant evolutions are observed over the range of Geant4 versions, not always 
in the direction of better compatibility with experiment.
Goodness-of-fit tests complemented by categorical
analysis tests identify a configuration based on Geant4 Urban multiple
scattering model in Geant4 9.1 version and a configuration based on
single Coulomb scattering in Geant4 10.0 as the physics options best
reproducing experimental data above a few tens of keV.
At lower energies only single scattering demonstrates some capability to 
reproduce data down to a few keV.
Recommended preassembled physics configurations appear incapable of describing
electron backscattering compatible with experiment.
With the support of statistical methods, a correlation is established between the 
validation of Geant4-based simulation of backscattering and of energy deposition.

\end{abstract}

\section{Introduction}
\label{sec_intro}

\IEEEPARstart{T}{he} simulation of backscattering is a sensitive playground to evaluate
the capability of a Monte Carlo transport code to describe electron multiple scattering 
accurately.
Multiple scattering modeling affects not only directly associated observables,
such as the fraction of electrons impinging on a target that are backscattered, 
their energy spectrum and angular distribution, but also the simulated energy
deposition in the target volume.

Quantitative evaluations \cite{tns_sandia2013, tns_sandia2009} of the capability of
Geant4 \cite{g4nim, g4tns} to reproduce high precision measurements of the
energy deposited by low energy electrons in various targets \cite{sandia79,
sandia80} hint at a significant contribution of multiple scattering
implementations to determine the accuracy of the simulated energy deposition.

The study documented in this paper analyzes quantitatively the simulation of
electron backscattering based on Geant4.
It evaluates the performance of several Geant4 physics configuration options,
in an extended series of Geant4 versions,
with respect to a large sample of experimental data collected from the literature,
which cover the energy range from 100~eV to  22~MeV approximately.
Compatibility with experiment is established by means of goodness-of-fit
statistical methods, while the different ability of Geant4 physics modeling
options to reproduce experimental data is quantified by the statistical analysis
of categorical data derived from the outcome of goodness-of-fit tests.
Finally, the results of this validation process are correlated with the outcome 
of the validation of electron energy deposition in \cite{tns_sandia2013}, and the
significance of this correlation is quantified.

The scope of the paper is limited to testing the electron backscattering 
fraction simulated by Geant4.
Apart from the considerable amount of material needed to document this
subject alone, the results reported in section \ref{sec_results} suggest that
validation tests of more complex observables, such as the spectrum and angular
distribution of backscattered electrons, would be more meaningful once the Geant4
multiple scattering domain has benefited from the opportunities for improvement
highlighted in this paper for future versions of the toolkit.

The results of this validation test provide guidance to optimize the selection
and configuration of electron scattering models in experimental application scenarios
based on several Geant4 versions examined in this paper.
They also provide a quantitative ground for future improvement of the physics
modeling and the software development process in Geant4 electromagnetic physics
domain.

\section{Electron backscattering}
\label{sec_status}

Electron backscattering has been the subject of experimental and theoretical interest
for several decades.
A review on this topic is outside the scope of this paper; the brief overview in
this section has the purpose of summarizing information relevant to the
simulation validation test documented in this paper.

\subsection{Experimental data}
\label{sec_exp}
A large number of experiments have measured various features related
to electron backscattering:
the fraction of backscattered particles from targets of various thickness and
material composition, the angular distribution of backscattered electrons and
their energy spectrum, surface effects on solid targets and other effects
associated with material properties.
These experimental observables provide information to investigate the underlying 
physical effects.

Typical experiment arrangements involve an electron beam of defined energy, high
purity targets and a detector to measure the electron current in the backward
hemisphere with respect to the target.
A hemispherical grid is often used to discriminate electrons with energy above a
preset threshold, with the intent of reducing the contamination from secondary
electrons in the detected electron sample.
The fraction of backscattered electrons is determined as the ratio between the
number of electrons reaching the detector and the total number of primary
electrons.
Experimental uncertainties are associated with the electron beam energy, 
beam current and current leakage. 
The detected electron sample can be contaminated by particles scattered
from components of the experimental apparatus other than the target,
or by secondary electrons.
Detailed descriptions of measurement setups can be found in \cite{martin2003}
and \cite{gomati2008}.

The present study collected more than 3000 experimental data
\cite{martin2003}-\cite{yadav2007} from the literature, concerning the
measurement of the electron backscattering fraction from infinite or
semi-infinite targets (i.e. of size exceeding the electron range in the target
material, or half of its value).
The experimental sample encompasses 48 target elements, which span 
the range of atomic numbers from beryllium to uranium.
Electron energies vary from 78~eV to 22.2~MeV. 
The analysis reported in the following sections concerns
measurements performed with a normally incident beam.

Apparent inconsistencies are visible in the experimental data sample, regarding
measurements performed by different experimental groups in similar nominal
configurations of target composition and primary electron energy: they hint at
the presence of systematic effects.
Several references do not report any experimental uncertainties, nor 
evaluate possible sources of systematic errors affecting the measurements.
In some cases experiments performed with similar techniques report
largely different uncertainties, which hint to possibly underestimated
errors.

The experimental data are shown in the figures of this paper with error bars 
corresponding to the uncertainties reported in the related references.
The apparent absence of error bars associated with some data points reflects 
either the omission of experimental uncertainties in their published source 
or values smaller than the size of the data point markers in the plots.

\subsection{Simulation with general purpose Monte Carlo codes}

The simulation of electron backscattering has been a subject of interest for
general purpose Monte Carlo codes, such as EGS \cite{egs5, egsnrc}, FLUKA
\cite{fluka}, Geant4 \cite{g4nim, g4tns}, ITS \cite{its}, MCNP \cite{mcnp6} and
Penelope \cite{penelope}, as it is a sensitive instrument to demonstrate the
capabilities of their electron transport models, namely the treatment of
multiple and single electron scattering.

The models implemented in these codes are based on a relatively limited set of
theoretical approaches \cite{berger_1988}, developed by Goudsmit and Saunderson
\cite{goudsmit1, goudsmit2}, Moli\`ere \cite{moliere_1947}, Lewis \cite{lewis},
and Wentzel \cite{wentzel}, complemented by algorithms specifically tailored to
the application of the theory to the Monte Carlo particle transport environment.
Some of these algorithms are documented in the literature (e.g.
\cite{fernandezvarea_1993}, \cite{ferrari_1992}), other are only described in
the software documentation of the Monte Carlo systems; nevertheless, limited
information is usually available about their theoretical grounds, methods
of approximation and implementation details, and the
evolution of the software is often the source of inconsistencies between
the description of the algorithms in the literature or the software documentation
and their actual implementation.

Comparisons of electron backscattering simulations with experimental data are
reported in the literature: some examples are \cite{kirihara_2010} and \cite{ali_2008},
concerning EGS5 \cite{egs5} and EGSnrc, respectively, \cite{jeraj_1999},
concerning MCNP4B \cite{mcnp4b} and EGS4 \cite{egs4}, \cite{fasso_2000}, concerning FLUKA.
The comparisons available in the literature are limited to a single source or a
small collection of experimental data: this limitation exposes them to the risk
of biased conclusions, if the reference data are affected by unidentified
systematic effects, or are not adequately representative of the variety of
scenarios encountered in particle transport applications regarding electron energies
and target composition.
They rest on a qualitative appraisal of
the compatibility between simulation and experiment.

To the best of our knowledge, this paper is the first quantitative report,
based on rigorous statistical methods, of the comparison of 
electron backscattering simulation based on a major Monte Carlo system 
with respect to an extensive collection of experimental data.



\section{Electron Multiple Scattering Simulation in Geant4}
\label{sec_g4models}

The electromagnetic physics package of Geant4 encompasses processes and models,
which deal with electron single and multiple scattering with the interacting
medium.
In addition, the physics\_lists package of Geant4 collects a set of
classes, which instantiate predefined selections of physics processes and models
for several particle types: they can be directly employed in simulation
applications by users not willing to develop physics selections specific to
their own experimental scenario.

\subsection{Processes and models}

\begin{figure*} 
\centerline{\includegraphics[angle=0,width=17cm]{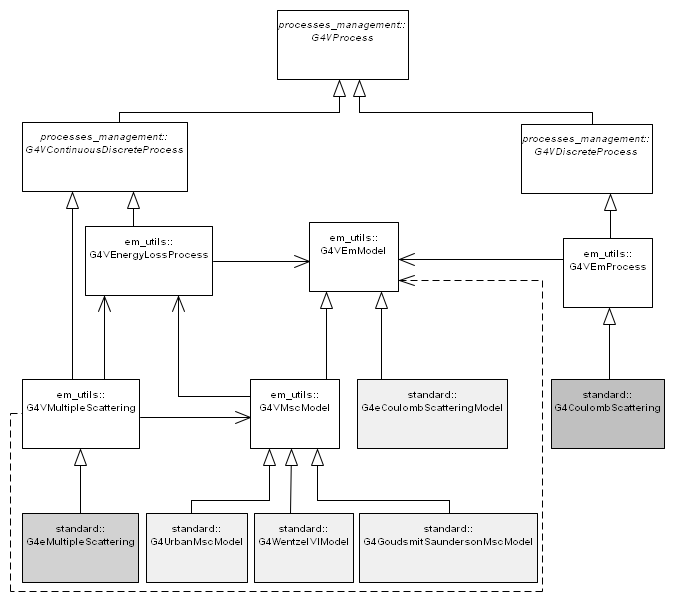}}
\caption{UML (Unified Modeling Language) class diagram illustrating Geant4
multiple and single scattering processes (dark grey) and models (light grey)
involved in this study, and their relationship with Geant4 abstract bases classes
(in italic) interacting with the tracking kernel. The diagram is pertinent to Geant4 10.0 and 10.1 versions.}
\label{fig_umlmsc}
\end{figure*}

Geant4 deals with the scattering of electrons with matter by means of two
processes, multiple and single Coulomb scattering, which are specializations of
the continuous-discrete and discrete processes handled by Geant4 tracking
algorithm, respectively.
Part of the functionality associated with these processes is delegated to multiple or
single scattering models.
A process can be configured with one or more associated models,
which implement different algorithms and may cooperate to
describe electron scattering over different energy ranges.
The UML (Unified Modeling Language) class diagram in Fig. \ref{fig_umlmsc}
illustrates the main features of  multiple and single electron scattering in 
Geant4 10.0 and 10.1.

The Urban multiple scattering model \cite{urban2002,urban,urban2006},
intended to be based on Lewis' theory \cite{lewis}, is instantiated by default in Geant4
electron multiple scattering process.
Several variants of this model have been released in the course of the evolution
of Geant4:
G4MscModel, 
G4UrbanMscModel,
G4MscModel71,  
G4UrbanMscModel90,
G4UrbanMscModel2, 
G4UrbanMscModel92,
G4Urban\-Msc\-Model93, 
G4UrbanMscModel95 and
G4UrbanMscModel96.
Table~\ref{tab_mscattdef} lists the default model associated with the Geant4 
versions considered in this paper.
Some relevant parameters characterizing the multiple scattering algorithm
are described  \cite{elles}.

\begin{table}
  \centering
  \caption{Geant4 default multiple scattering settings}
    \begin{tabular}{clc}
   \hline
    \textbf{Geant4 Version} & \textbf{Multiple Scattering Model}  \\
     \hline
  	9.1 & G4UrbanMscModel 		 \\
  	9.2 & G4UrbanMscModel2 		 \\
	9.3 & G4UrbanMscModel92 	\\
	9.4 & G4UrbanMscModel93 	 \\
	9.6 & G4UrbanMscModel95 	 \\
	10.0 & G4UrbanMscModel 		 \\
         10.1 & G4UrbanMscModel		 \\

    \hline
    \end{tabular}%
  \label{tab_mscattdef}%
\end{table}%

%

Simulation of single electron scattering is implemented in Geant4 G4Coulomb\-Scattering
process and G4Coulomb\-Scattering\-Model model \cite{em_chep2008}.
This model is based on Wentzel calculations \cite{wentzel}; it has been 
available in Geant4 since the 9.0 version.

An algorithm combining multiple and single scattering is implemented in the 
G4WentzelVIModel class \cite{msc_chep2009}.

A multiple scattering model based on Goudsmit-Saunderson calculations
\cite{goudsmit1,goudsmit2} has been available in Geant4 for the simulation of
electron multiple scattering since version 9.3 \cite{kadri_goudsmit}.
It is implemented in the G4GoudsmitSaundersonModel class.

\subsection{Multiple scattering in Geant4 prepackaged PhysicsLists}

Geant4 software documentation \cite{g4appldevguide} recommends the use of
predefined classes in simulation applications, which implement
selections of electromagnetic processes and models for several particle
types.
These classes, derived from G4VPhysicsConstructor, are supplied within
the physics\_lists package of the Geant4 toolkit; they are used in the
prepackaged PhysicsLists hosted in the same package or may be
instantiated in user defined PhysicsLists.
According to \cite{em_mc2010}, they are ``intensively validated''.

Geant4~10.0 and 10.1 version encompass six generic electromagnetic
PhysicsConstructors, which are briefly described below, as well as a
few specialized ones (e.g. for optical physics and for the interactions of
polarized particles), which are outside the scope of this paper.

The selection of electromagnetic processes and models implemented in the
G4EmStandardPhysics PhysicsConstructor is described in the
user documentation of Geant4 10.0 \cite{g4appldevguide} as ``default
electromagnetic physics''.
From an inspection of the source code it appears that for electrons the
Urban multiple scattering model is activated up to 100~MeV, while the
WentzelVI model is selected above that energy threshold.
Single Coulomb scattering is activated in association with the
WentzelVI model.
A polar angle threshold is set at 180 degrees.
Step limit and range factor settings are kept unchanged with respect to the
default implementations in the multiple scattering classes it instantiates.

According to Geant4 user documentation \cite{g4appldevguide},
G4EmStandardPhysics\_option1 provides ``fast but less accurate electron
transport'' due to the choice of the \textit{Simple} method of step limitation
by multiple scattering, reduced step limitation by the ionisation process and
enabled \textit{ApplyCuts} option.
It uses G4UrbanMscModel for multiple scattering of electrons and positrons.
From an inspection of the source code it appears that the model selections, 
energy and angle thresholds concerning multiple scattering are the same as in 
G4EmStandardPhysics.

G4EmStandardPhysics\_option2 is defined in \cite{g4appldevguide} as
``experimental electromagnetic physics with disabled \textit{ApplyCuts} option''.
From an inspection of the source code it appears that the model selections,
energy and angle thresholds concerning electron scattering scattering are the
same as in G4EmStandardPhysics.

Geant4 user documentation \cite{g4appldevguide} states that
G4EmStandardPhysics\_option3 selects electromagnetic physics for
simulation with high accuracy due to \textit{UseDistanceToBoundary} multiple
scattering step limitation, reduced \textit{finalRange} parameter of stepping
function optimized per particle type, alternative model
G4KleinNishinaModel for Compton scattering, Rayleigh scattering, and
G4IonParameterisedLossModel for ion ionisation.
From an inspection of the source code it appears that for electrons the default
configuration of the G4eMultipleScattering process is selected, which in 
turn instantiates the Urban multiple scattering model.

According to Geant4 user documentation \cite{g4appldevguide}, the combination of
``best electromagnetic models for simulation with high accuracy'' includes
\textit{UseDistanceToBoundary} multiple scattering step limitation, reduced
\textit{finalRange} parameter of stepping function optimized per particle type,
low-energy sub-library models G4\-Livermore\-Photo\-Electric\-Model,
G4\-Low\-EP\-Compton\-Model below 20 MeV,
G4\-Penelope\-Gamma\-Conversion\-Model below 1 GeV,
G4\-Penelope\-Ionisation\-Model below 100 keV, and
G4\-Ion\-Parameterised\-Loss\-Model for ion ionisation.
This combination is implemented in G4\-Em\-Standard\-Physics\_option4.
From an inspection of the source code it appears that the model selections, 
energy and angle thresholds concerning multiple scattering are the same as in 
G4EmStandardPhysics.
Some documentation of the performance of G4EmStandardPhysics\_option4
with respect to experimental data is available in \cite{em_sna2013}.
Despite our best efforts, we could not retrieve references in the literature
supporting quantitatively the statement that 
G4EmStandardPhysics\_option4 corresponds to ``best
electromagnetic models for simulation with high accuracy''.

G4EmLivermorePhysics selects models
for electrons and photons based on the EEDL \cite{eedl} and EPDL \cite{epdl97}
data libraries.
From an inspection of the source code it appears that model selections,
energy and angle thresholds, step limitation and \textit{RangeFactor} settings
concerning multiple scattering are the same as in
G4EmStandardPhysics\_option4.

Two PhysicsConstructors specific to simulations using single scattering and 
G4WentzelVIModel, named G4\-Em\-Standard\-Physics\_SS and
G4\-Em\-Standard\-Physics\_WVI respectively, are released for the first
time in Geant4 10.1.

\section{Simulation}



\subsection{Simulation application}
\label{sec_app}

A Geant4-based application was developed to simulate the electron backscattering
experiments involved in the validation test.
The application models the experimental scenarios, encodes a selection of
physics configurations corresponding to the options documented in section
\ref{sec_g4models}, drives the simulation execution and assembles a set of
significant observables produced by the simulation into for further analysis.

\begin{figure}
\centerline{\includegraphics[angle=0,width=9cm]{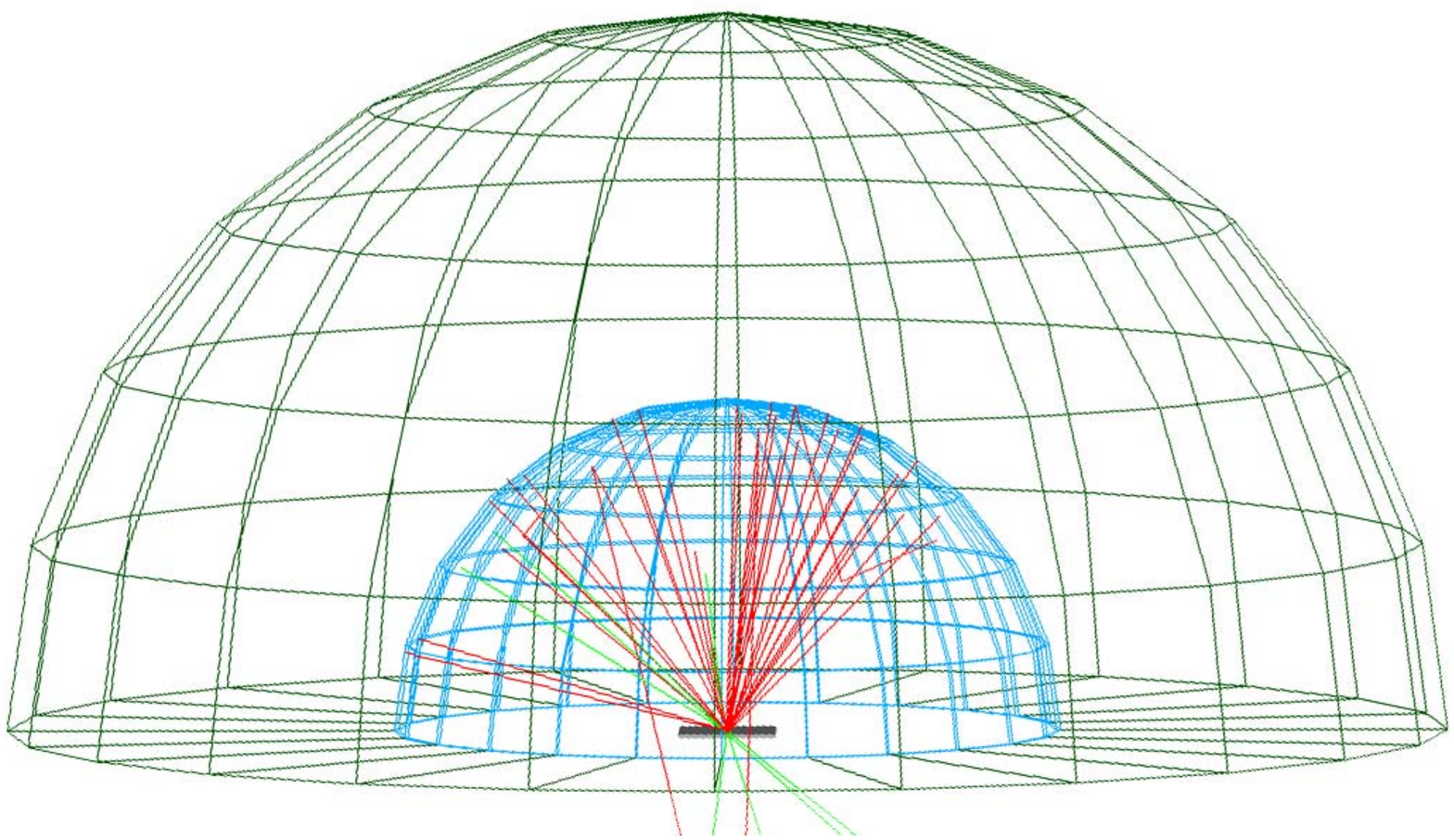}}
\caption{A sketch of the geometrical setup implemented in the simulation, produced by 
Geant4 visualization package. Red lines represent backscattered electrons; green lines 
correspond to photons escaping from the target.}
\label{fig_geom}
\end{figure}

The geometry configuration reproduces the relevant features of the experimental
setups described in \cite{martin2003}-\cite{yadav2007}: a semi-infinite or
infinite target of pure elemental composition and a hemispherical sensitive
volume representing the electron detector.
The target and the detector are surrounded by a very low density material
(equivalent to galactic vacuum) to avoid contamination of the test results
from spurious interactions.
An electron beam impinges on the target.
A sketch of the geometry configuration is illustrated in Fig. \ref{fig_geom}.
Relevant parameters, such as the target shape and thickness, the angular
acceptance of the detector, the energy and angular spread of the beam
 are retrieved from the experimental references
\cite{martin2003}-\cite{yadav2007} whenever available.
Electrons entering the sensitive volume are considered detected, when their
energy exceeds a preset threshold.

As most references do not report the experimental configuration in detail, some
assumptions are made in the simulation, when the corresponding parameters are
not explicitly documented: the target is assumed to be a cylinder of thickness
and diameter larger than the incident electron range retrieved from the ESTAR
\cite{estar} database, the sensitive detector volume is assumed
to cover the whole backward hemisphere, a detection threshold of 50~eV is
applied, consistent with common experimental practice to exclude secondary
electrons, the detection efficiency above this threshold is assumed to be 100\%,
the electron beam is assumed to be monochromatic and incident at 90$^{\circ}$.
These assumptions correspond to modeling an infinite target, whose geometrical
characteristics would not affect the resulting backscattering measurement, and
to presuming that the backscattering values reported in the experimental
references were previously corrected for geometrical acceptance and detection 
efficiency as appropriate.

The backscattering fraction is calculated as the ratio between the number of
events with at least one detected electron and the number of primary electrons.
Detected electrons may be backscattered primary electrons or secondary electrons
with energy above the preset threshold.
Events with more than one detected electron contributed to the backscattering
fraction as a single count.



\begin{figure*} 
\centerline{\includegraphics[angle=0,width=17cm]{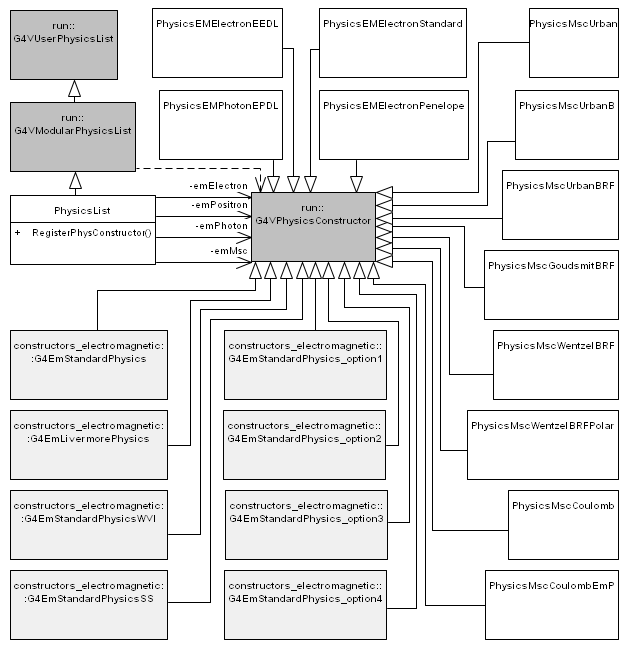}}
\caption{UML (Unified Modeling Language) class diagram illustrating the physics configuration in the simulation:
simulation application classes (white), Geant4 kernel classes (dark grey) and PhysicsConstructor classes
released in the Geant4 physics\_lists package (light grey).}
\label{fig_uml}
\end{figure*}

The physics configurations concerning electron scattering evaluated in this
paper are summarized in Table \ref{tab_msconf}.
The class names reported in this table appear as in Geant4 10.0;
different settings in Geant4 10.0 and 10.1 are identified by the
respective version number in parentheses.
The UML (Unified Modeling Language) class diagram in Fig. \ref{fig_uml} illustrates the
PhysicsConstructors involved in the application and their relationship with
classes in Geant4 kernel responsible for the configuration of physics
selections.

\begin{table*}[htbp]
  \centering
  \caption{Multiple and single scattering configurations evaluated in the electron backscattering validation test }
    \begin{tabular}{lcllcc}
    \hline
    {\bf Configuration} & {\bf Description} 		& {\bf Process class} 			& {\bf Model class} 		& {\bf StepLimitType} & {\bf RangeFactor} \\
    \hline
    \textbf{Urban} 		& Urban model, user step limit  		& G4eMultipleScattering 	& G4UrbanMscModel		& default 				& default \\
    \textbf{UrbanB} 		& Urban model, user step limit       		& G4eMultipleScattering	& G4UrbanMscModel		& DistanceToBoundary 	& default \\
    \textbf{UrbanBRF} 	& Urban model      		& G4eMultipleScattering 	& G4UrbanMscModel 		& DistanceToBoundary 	& 0.01 \\
    \textbf{GSBRF} 		& Goudsmit-Saunderson  	& G4eMultipleScattering	& G4GoudsmitSaundersonModel & DistanceToBoundary & 0.01 \\
    \textbf{WentzelBRF} 	& WentzelVI model 	& G4eMultipleScattering	& G4WentzelVIModel 			& DistanceToBoundary & 0.01 \\
    \textbf{WentzelBRFP} & WentzelVI, angle limit       & G4eMultipleScattering & G4WentzelVIModel 	& DistanceToBoundary & 0.01 \\
    \textbf{Coulomb} 	& Single scattering, default 		& G4CoulombScattering 	& G4eCoulombScatteringModel	& default 			& default \\
   \textbf{CoulombP} 	& Single scattering, 		& G4CoulombScattering 	& G4eCoulombScatteringModel	& default 			& default \\
				& with angle limit \textit{(10.1)}	& 					& 						& 				& \\

    {\bf } & {\bf } & \multicolumn{2}{c}{\bf PhysicsConstructor class} & {\bf} & {\bf } \\
    \textbf{EmLivermore} 		&       & \multicolumn{2}{c}{G4EmLivermorePhysics} & DistanceToBoundary & 0.01 \\
    \textbf{EmStd} 		& Predefined 		& \multicolumn{2}{c}{G4EmStandardPhysics} 			& default & default \\
    \textbf{EmOpt1} 		& electromagnetic 	& \multicolumn{2}{c}{G4EmStandardPhysics\_option1}        	& default & default \\
    \textbf{EmOpt2} 		& physics 			&\multicolumn{2}{c}{G4EmStandardPhysics\_option2}        	& default & default \\
    \textbf{EmOpt3} 		& selections 		& \multicolumn{2}{c}{G4EmStandardPhysics\_option3}        	& DistanceToBoundary & default \\
    \textbf{EmOpt4} 		&       			& \multicolumn{2}{c}{G4EmStandardPhysics\_option4}        	& DistanceToBoundary \textit{(10.0)} & 0.01 \textit{(10.0)} \\
					&				& \multicolumn{2}{c}{ }   	& SafetyPlus \textit{(10.1)} & 0.02 \textit{(10.1)} \\
 \textbf{EmSS}			&				& \multicolumn{2}{c}{G4EmStandardPhysics\_SS \textit{(10.1)} }   	& default  & default \\
 \textbf{EmWVI}			&				& \multicolumn{2}{c}{G4EmStandardPhysics\_WVI \textit{(10.1)}}   	& default & default \\   
\hline
    \end{tabular}%
  \label{tab_msconf}%
\end{table*}

\subsection{Simulation Production}
\label{sec_prod}

Simulations are produced with six Geant4 versions released between December 2007 and
December 2014.
The selection  reflects the study of Geant4 energy deposited by electrons
documented in \cite{tns_sandia2013}, with the addition of Geant4 version 10.0
and 10.1,
which were first released after the publication of that paper, and the exclusion
of Geant4 version 9.5, which exhibited a very similar
behaviour to version 9.6 in \cite{tns_sandia2013}.
Correction patches to these versions released by October 2014 are applied on 
top of the original versions.
For convenience, the Geant4 versions evaluated in this study are identified
through their original version number; the corresponding patched versions used
in the validation test are listed in Table \ref{tab_versions}.

\begin{table} 
\begin{center}
\caption{Geant4 versions subject to test}
\begin{tabular}{cc}
{\bf Version Identifier}	& {\bf Patched Geant4 Version}	\\
\hline
9.1 		& 9.1p03	\\
9.2		& 9.2p04	\\
9.3 		& 9.3p02	\\
9.4 		& 9.4p04	\\
9.6 		& 9.6p03	\\
10.0		& 10.0p03 \\
10.1		& not pertinent\\
\hline
\end{tabular}
\label{tab_versions}
\end{center}
\end{table}

The simulations were executed on workstations running the Scientific Linux~6
operating system 
with the gcc version 4.4.7 compiler and on computers running
MacOS 10.8.5 (Mountain Lion) operating system with clang 3.3.
The same Geant4 versions and application code were installed on all machines.

IThe simulation configurations corresponding to the experimental test cases were
encoded in ``macro'' files handled by Geant4 user interface.
Configurations corresponding to different physics options for the same
experimental scenario were automatically generated based on a master
configuration file to minimize the possibility of accidental encoding errors,
which could be the source of systematic effects in the validation results.
All simulation configuration files were kept under version control to ensure the
reproducibility of results.

The physics configurations produced in the Geant4 versions subject to evaluation
are listed in Table \ref{tab_prodconf}.
Simulations using the family of Urban configurations and the default single Coulomb scattering
configuration were produced in all versions, while those using the WentzelVI and
Goudsmit-Saunderson models were produced with Geant4 version 9.3 and later,
consistent with their first release.
Simulations with Coulomb scattering settings different from the default
ones (namely with the ``$\theta$~limit'' parameter set to zero) were produced with
Geant4 version 10.1.

Simulations using predefined electromagnetic Phy\-sics\-Con\-struc\-tors were
produced with Geant4 versions 9.6, 10.0 and 10.1, which reflect the
recommendations to experimental users issued in the most recent versions of
Geant4 at the time of writing this paper.
Simulations using the G4EmStandardPhysics\_WVI and G4EmStandardPhysics\_SS
predefined electromagnetic Phy\-sics\-Con\-struc\-tors were limited to 
Geant4 version 10.1, where these classes have been first released.

The secondary production threshold was set to 1~$\mu m$ (expressed in terms 
of particle range) in all simulations, with the exception of test cases with primary electron beam 
energy above 1~MeV, for which it was raised to 10~$\mu m$ to reduce the 
execution time compatible with the available computational 
resources.

A user-defined step limitation was applied in the simulations corresponding to the Urban and UrbanB configurations;
the step limit was set to the same value as the secondary production threshold setting.

For each simulation run, the number of generated events was determined to ensure that
the statistical errors on the simulated backscattering fraction was smaller or
at most comparable to the experimental uncertainty  reported in the corresponding 
reference paper or in similar experimental configurations, when the associated
reference did not document uncertainties.
Based on these considerations, the number of events in a single run varied between 
10000 and 500000, depending on the corresponding experimental configuration.

\begin{table}[htbp]
  \centering
  \caption{Physics configurations in the simulation production}
    \begin{tabular}{c|l|c|c}
    \hline
    Versions & Scattering & Electrons & Photons \\
    \hline
    \multirow{4}[0]{*}{9.1-10.1} & Urban 		& EEDL, Standard, Penelope & \multirow{4}[0]{*}{EPDL} \\
          					& UrbanB 		& EEDL 	&  \\
         					& UrbanBRF 	& EEDL       	&  \\
          					& Coulomb 	& EEDL       	&  \\
\hline
    \multirow{3}[0]{*}{9.3-10.1} & GSBRF & \multirow{3}[0]{*}{EEDL} & \multirow{3}[0]{*}{EPDL} \\
          & WentzelBRF &       &  \\
          & WentzelBRFP &       &  \\
\hline
    \multirow{6}[0]{*}{9.6-10.1} & EmLivermore & \multirow{6}[0]{*}{Embedded } & \multirow{6}[0]{*}{Embedded} \\
          & EmStd &       &  \\
          & EmOpt1 &       &  \\
          & EmOpt2 &       &  \\
          & EmOpt3 &       &  \\
          & EmOpt4 &       &  \\
\hline
     \multirow{3}[0]{*}{10.1} & EmSS & \multirow{2}[0]{*}{Embedded } & \multirow{2}[0]{*}{Embedded} \\
          & EmWVI &       &  \\
\cline{2-4}
       & CoulombP & EEDL       	& EPDL \\

    \hline
    \end{tabular}%
  \label{tab_prodconf}%
\end{table}%

\section{Data Analysis}

The data analysis addresses various issues related to the validation of
Geant4-based simulation of electron backscattering: the evaluation of the
capability of Geant4 to produce results consistent with experiment, the
comparison of the simulation accuracy achievable with different Geant4 multiple
and single scattering configurations in the user application, the correlation between
the validity of simulated electron backscattering and energy deposition,
and the evolution of the results over a series of Geant4 versions.
Appropriate statistical methods are applied to analyze the data pertinent to each 
problem.

\subsection{Analysis method}
\label{sec_analysis_gen}

The experimental data sample retrieved from the literature is dominated by low
energy measurements (mostly below a few tens keV): this intrinsic characteristic
of the data set would affect the ability to discern the energy dependence of the
capabilities of Geant4 models from an analysis of their compatibility with
experiment over the whole data sample, as the the outcome of the statistical
tests would be mainly determined by the weight of the low energy component.
Therefore, the experimental data sample was partitioned into three energy
ranges, identified in the following as ``low'', ``intermediate'' and ``high''
energy, respectively, to evaluate the capabilities of the
examined models as a function of the electron energy.

The grouping of data in three energy intervals is empirically based on the
behaviour observed in the plots reporting simulation results and experimental
data.
To mitigate the risk of biased conclusions due to the empirical nature of
its definition, it was verified that the general conclusions of the validation
test reported in the following sections are stable with respect to small
variations of the definition of low, intermediate and high energies (e.g. 10~keV
instead of 20~keV, or 75~keV instead of 100~keV).
This verification was done by performing the whole analysis procedure over four
different definitions of the three energy groups: although the numerical values
of the test statistics were slightly different for different partitioning of the
data sample, the global conclusions of the validation tests were insensitive to
how the three energy ranges were defined.
The results reported in this paper data correspond to grouping the data between
1 and 20 keV, between 20 and 100 keV and above 100 keV.
Data below 1~keV are excluded from the analysis, since all the examined models
appear incapable of reproducing the experimental measurements below this
threshold.
The apparent absence of any backscattering occurrence below this energy,
concerning some models, hints at an intrinsic limitation of their functionality.

%

In the context of the data analysis a test case is
defined by a set of experimental data, associated with a given target
composition, within a given energy range, performed by a given research group.
The previously mentioned energy intervals encompass 137, 112 and 57 test cases, respectively.

The statistical analysis of compatibility between simulation and experiment
is articulated over two stages.

The first stage of the statistical analysis consists of evaluating the compatibility 
between simulation and experiment in single test cases.
The compatibility between simulated and experimental data in a single test case
is determined by two-sample goodness-of-fit tests.
The details of this component of the analysis are described in section \ref{sec_gof}.

In the second stage of the analysis categorical statistical tests determine
whether the differences in compatibility with experiment observed across the
various categories of Geant4 models can be explained only by chance, or should
be interpreted as deriving from intrinsic behavioural characteristics.

The approach adopted in the categorical analysis is connected to the physics
configuration that characterizes how the data samples were produced,
namely whether the data subject to test are unrelated or to some extent related.
Simulations using different multiple scattering models produce unrelated
samples, as the different modeling options available in Geant4 intend to implement
distinct conceptual alternatives in the treatment of multiple scattering. 
Data samples deriving from simulations that differ only for a secondary option
(e.g. a parameter setting), but are based on the same multiple scattering model,
are to some extent related: this is the case, for instance, of the three
configurations based on the Urban model and the two configurations
based on the WentzelVI model listed in Table~\ref{tab_msconf}.
It is also the case of simulations based on single Coulomb scattering and the
WentzelVI multiple scattering model, since the latter incorporates the
former.
Statistical tests pertinent to independent or related data samples
are applied as appropriate in this stage of the analysis.

Finally, a statistical analysis is performed to evaluate whether a correlation
can be established between the compatibility with experiment of electron
backscattering and energy deposition simulated by Geant4.


The significance level for the rejection of the null hypothesis is set at 0.01
for all tests, unless otherwise specified.

The statistical data analysis reported in the following sections exploits the
Statistical Toolkit \cite{gof1,gof2} for goodness-of-fit testing and R \cite{R}
for the analysis of contingency tables and of correlation.

\subsection{Evaluation of individual test cases}
\label{sec_gof}

This stage of the analysis determines whether the experimental and simulated
data distributions associated with each test case are consistent with deriving
from the same parent distribution.

The generally poor documentation of experimental errors in the experimental
references, discussed in section \ref{sec_exp}, prevents the use of the $\chi^2$
test \cite{bock}, which involves measurement uncertainties explicitly.
Therefore goodness-of-fit tests based on the empirical distribution function,
where uncertainties do not appear directly in the calculation of the test statistic,
were used in the analysis.
Four independent goodness-of-fit tests were performed for each test case to
mitigate the risk of introducing systematic effects in the validation process
due to peculiarities of the mathematical formulation of the test statistic: the
Anderson-Darling \cite{anderson1952,anderson1954}, Cramer-von Mises
\cite{cramer,vonmises}, Kolmogorov-Smirnov \cite{kolmogorov1933,smirnov} and
Watson \cite{watson} tests.

The null hypothesis for each goodness-of-fit test is that  the simulated and experimental 
distributions derive from the same parent distribution.
The outcome of the test is classified as ``fail'', if the null hypothesis is rejected, as
``pass'' otherwise.

For convenience, the ``efficiency'' of a Geant4 simulation configuration is
defined as the fraction of test cases in which a goodness-of-fit test does not
reject the null hypothesis at 0.01 level of significance.
This variable quantifies the capability of that simulation configuration to
produce results statistically consistent with experiment over the whole set of
test cases pertinent to one of the energy ranges defined in section
\ref{sec_analysis_gen}.

Despite some qualitatively visible inconsistencies in the experimental data, 
no attempt was made to exclude the concerned data sets from the analysis, nor to remove
single outliers, since the poor documentation of experimental errors prevents a
proper evaluation of the consistency of the various measurements.
The possible leftover presence of experimental data affected by systematic
errors may artificially lower the efficiency associated with a given Geant4
model: therefore, caution should be applied to interpreting the efficiency
values reported in this papers as absolute estimates of Geant4 simulation
capabilities.
Nevertheless, since all Geant4 physics configurations are exposed to the same
test cases, the possible inclusion of experimental data samples exhibiting
systematic effects would marginally affect the conclusions regarding the
comparison of their compatibility with experiment.


\subsection{Evaluation of unrelated data samples}
\label{sec_conting}

This component of the analysis evaluates the difference in compatibility with
experiment across independent categories of data, e.g. associated with two
different multiple scattering models.
It receives its input from the results of goodness-of-fit tests described in
section \ref{sec_gof}.

The differences in the behavior of the two categories are quantified by means of
contingency tables, which are built by counting the number of test cases
identified as ``fail'' or ``pass'' in section \ref{sec_gof}.
Their configuration is illustrated in Table~\ref{tab_exconting}.

\begin{table}
  \centering
  \caption{Configuration of contingency tables for the evaluation of unrelated simulation configurations}
    \begin{tabular}{l|ll}
    \hline
          & \textbf{Category A} & \textbf{Category B} \\
    \hline
    \textbf{Pass} & N$_{pass A}$ & N$_{pass B}$ \\
    \textbf{Fail} & N$_{fail A}$ & N$_{fail B}$ \\
    \hline
    \end{tabular}%
  \label{tab_exconting}%
\end{table}%


In the analysis of contingency tables the null hypothesis is that there is no
relationship between the two categories; in physical terms it means that the two
simulation configurations under examination 
are equivalent regarding the compatibility with experiment of their respective
outcome.

A variety of tests is applied to determine the statistical significance of the
difference between the two categories of data subject to evaluation: Pearson's
$\chi^2$ test \cite{pearson} (when the number of entries in each cell of the
table is sufficiently large to justify its applicability), Fisher's exact test
\cite{fisher} and a variety of unconditional exact tests.
The use of different tests mitigates the risk of introducing systematic effects
in the validation results due to peculiarities in the mathematical formulation
of the test statistic.

Fisher's exact test is widely used in the analysis of contingency tables;
although it is based on a model in which both the row and column sums are fixed
in advance, which seldom occurs in experimental practice, it remains valid when
the row or column totals, or both, are not fixed, but in these cases it tends to
be conservative, yielding a larger p-value than the true significance of the
test \cite{agresti}.

Unconditional tests, such as Barnard's test \cite{barnard}, Boschloo's test
\cite{boschloo} and Suissa and Shuster's \cite{suissa} calculation of a
Z-pooled statistic, are deemed more powerful than Fisher's exact test in some
configurations of 2$\times$2 contingency tables \cite{andres_1994,andres_2004},
but they are computationally more intensive.
They yield consistent results in the context of the validation process described
in this paper; the p-values of two of such tests are reported among
the results of the validation analysis in section~\ref{sec_results}.


%

\subsection{Evaluation of related data samples}
\label{sec_conting}

This component of the analysis evaluates the difference in compatibility with
experiment across related categories of data.
It receives its input from the results of goodness-of-fit tests.

This analysis method is applied to two situations: when each subject (e.g. a
physics configuration in the simulation) serves in both categories being
evaluated (e.g. two Geant4 versions), and when one examines two closely related
subjects (e.g. simulations differing for a secondary feature of a multiple
scattering model, while all the other physics configuration settings are
identical).

\begin{table}
  \centering
  \caption{Configuration of contingency tables for the evaluation of related simulation configurations}
    \begin{tabular}{l|ll}
    \hline
          & \textbf{Category A: Pass} & \textbf{Category A: Fail} \\
    \hline
    \textbf{Category B: Pass} & N$_{pass\,A, \, pass\,B}$ & N$_{fail\,A, \, pass\,B}$ \\
    \textbf{Category B: Fail} & N$_{pass\,A, \, fail\,B}$ & N$_{fail\,A, \, fail\,B}$ \\
    \hline
    \end{tabular}%
  \label{tab_exmcnemar}%
\end{table}%


\tabcolsep=2pt
\begin{table*}[htbp]
  \centering
  \caption{Efficiency of physics configurations for Geant4 versions 9.1 to 10.1}
\resizebox{\textwidth}{!}{%
   \begin{tabular}{lc|cccc|cccc|cccc}
    \hline
    Configuration & Geant4 & \multicolumn{4}{|c|}{1-20 keV}       & \multicolumn{4}{c}{20-100 keV}        & \multicolumn{4}{|c}{$>100 keV$}  \\
          & version & AD    & CvM   & KS    & Watson & AD    & CvM   & KS    & Watson & AD    & CvM   & KS    & Watson \\
    \hline
    Urban & 9.1   & $<0.01$ & $<0.01$ & $<0.01$ & $<0.01$ & 0.10$\pm 0.03$  & 0.10$\pm 0.03$  & 0.10$\pm 0.03$  & 0.10$\pm 0.03$  & 0.79$\pm 0.05$  & 0.79$\pm 0.05$   & 0.77$\pm 0.05$   & 0.79$\pm 0.05$  \\
    Urban & 9.2   & $<0.01$ & $<0.01$ & $<0.01$ & $<0.01$ & 0.03$\pm 0.02$  & 0.03$\pm 0.02$  & 0.03$\pm 0.02$  & 0.03$\pm 0.02$  & 0.79$\pm 0.05$   & 0.77$\pm 0.05$   & 0.82$\pm 0.05$   & 0.88$\pm 0.04$  \\
    Urban & 9.3   & $<0.01$ & $<0.01$ & $<0.01$ & $<0.01$ & 0.09$\pm 0.03$  & 0.09$\pm 0.03$  & 0.09$\pm 0.03$  & 0.09$\pm 0.03$  & 0.74$\pm 0.06$   & 0.74$\pm 0.06$  & 0.72$\pm 0.06$  & 0.74$\pm 0.06$ \\
    Urban & 9.4   & $<0.01$ & $<0.01$ & $<0.01$ & $<0.01$ & 0.10$\pm 0.03$  & 0.10$\pm 0.03$  & 0.10$\pm 0.03$  & 0.10$\pm 0.03$  & 0.56$\pm 0.06$  & 0.56$\pm 0.06$  & 0.56$\pm 0.06$  & 0.61$\pm 0.06$ \\
    Urban & 9.6   & $<0.01$ & $<0.01$ & $<0.01$ & $<0.01$ & 0.17$\pm 0.04$  & 0.17$\pm 0.04$  & 0.17$\pm 0.04$  & 0.17$\pm 0.04$  & 0.68$\pm 0.06$  & 0.68$\pm 0.06$  & 0.68$\pm 0.06$  & 0.82$\pm 0.05$  \\
    Urban & 10.0    & $<0.01$ & $<0.01$ & $<0.01$ & $<0.01$ & $<0.01$  & $<0.01$  & $<0.01$  & $<0.01$  & 0.11$\pm 0.04$  & 0.11$\pm 0.04$  & 0.11$\pm 0.04$  & 0.09$\pm 0.04$ \\
    Urban & 10.1    & $<0.01$ & $<0.01$ & $<0.01$ & $<0.01$ & $<0.01$  & $<0.01$  & $<0.01$  & $<0.01$  & 0.07$\pm 0.04$  & 0.07$\pm 0.04$  & 0.07$\pm 0.04$  & 0.09$\pm 0.04$ \\
\hline
    UrbanB & 9.1   & $<0.01$ & $<0.01$ & $<0.01$ & $<0.01$ & 0.10$\pm 0.02$  & 0.10$\pm 0.02$  & 0.10$\pm 0.02$  & 0.10$\pm 0.02$  & 0.79$\pm 0.05$   & 0.79$\pm 0.05$   & 0.77$\pm 0.05$   & 0.79$\pm 0.05$  \\
    UrbanB & 9.2   & $<0.01$ & $<0.01$ & $<0.01$ & $<0.01$ & 0.03$\pm 0.02$  & 0.03$\pm 0.02$  & 0.03$\pm 0.02$  & 0.03$\pm 0.02$  & 0.79$\pm 0.05$   & 0.77$\pm 0.05$   & 0.82$\pm 0.05$   & 0.88$\pm 0.04$ \\
    UrbanB & 9.3   & $<0.01$ & $<0.01$ & $<0.01$ & $<0.01$ & $<0.01$  & $<0.01$  & $<0.01$  & $<0.01$  & 0.11$\pm 0.04$  & 0.11$\pm 0.04$  & 0.11$\pm 0.04$  & 0.11$\pm 0.04$ \\
    UrbanB & 9.4   & $<0.01$ & $<0.01$ & $<0.01$ & $<0.01$ & $<0.01$  & $<0.01$  & $<0.01$  & $<0.01$  & 0.07$\pm 0.04$  & 0.07$\pm 0.04$  & 0.07$\pm 0.04$  & 0.07$\pm 0.04$ \\
    UrbanB & 9.6   & $<0.01$ & $<0.01$ & $<0.01$ & $<0.01$ & $<0.01$  & $<0.01$  & $<0.01$  & $<0.01$  & 0.05$\pm 0.03$  & 0.05$\pm 0.03$  & 0.05$\pm 0.03$  & 0.07$\pm 0.04$ \\
    UrbanB & 10.0    & $<0.01$ & $<0.01$ & $<0.01$ & $<0.01$ & $<0.01$  & $<0.01$  & $<0.01$  & $<0.01$  & 0.07$\pm 0.04$  & 0.07$\pm 0.04$  & 0.07$\pm 0.04$  & 0.07$\pm 0.04$ \\
   UrbanB & 10.1    & $<0.01$ & $<0.01$ & $<0.01$ & $<0.01$ & $<0.01$  & $<0.01$  & $<0.01$  & $<0.01$  & 0.09$\pm 0.04$  & 0.07$\pm 0.04$  & 0.09$\pm 0.04$  & 0.11$\pm 0.04$ \\
\hline
    UrbanBRF & 9.1   & $<0.01$ & $<0.01$ & $<0.01$ & $<0.01$ & $<0.01$  & $<0.01$  & $<0.01$  & $<0.01$  & 0.05$\pm 0.03$  & 0.05$\pm 0.03$  & 0.05$\pm 0.03$  & 0.07$\pm 0.04$ \\
    UrbanBRF & 9.2   & $<0.01$ & $<0.01$ & $<0.01$ & $<0.01$ & $<0.01$  & $<0.01$  & $<0.01$  & $<0.01$  & 0.02$\pm 0.02$  & 0.02$\pm 0.02$  & 0.02$\pm 0.02$  & 0.04$\pm 0.03$ \\
    UrbanBRF & 9.3   & $<0.01$ & $<0.01$ & $<0.01$ & $<0.01$ & $<0.01$  & $<0.01$  & $<0.01$  & $<0.01$  & 0.07$\pm 0.04$  & 0.07$\pm 0.04$  & 0.07$\pm 0.04$  & 0.07$\pm 0.04$ \\
    UrbanBRF & 9.4   & $<0.01$ & $<0.01$ & $<0.01$ & $<0.01$ & $<0.01$  & $<0.01$  & $<0.01$  & $<0.01$  & 0.07$\pm 0.04$  & 0.07$\pm 0.04$  & 0.07$\pm 0.04$  & 0.07$\pm 0.04$ \\
    UrbanBRF & 9.6   & $<0.01$ & $<0.01$ & $<0.01$ & $<0.01$ & $<0.01$  & $<0.01$  & $<0.01$  & $<0.01$  & 0.07$\pm 0.04$  & 0.07$\pm 0.04$  & 0.07$\pm 0.04$  & 0.07$\pm 0.04$ \\
    UrbanBRF & 10.0    & $<0.01$ & $<0.01$ & $<0.01$ & $<0.01$ & $<0.01$  & $<0.01$  & $<0.01$  & $<0.01$  & 0.07$\pm 0.04$  & 0.07$\pm 0.04$  & 0.07$\pm 0.04$  & 0.07$\pm 0.04$ \\
    UrbanBRF & 10.1    & $<0.01$ & $<0.01$ & $<0.01$ & $<0.01$ & $<0.01$  & $<0.01$  & $<0.01$  & $<0.01$  & 0.09$\pm 0.04$  & 0.09$\pm 0.04$  & 0.09$\pm 0.04$  & 0.09$\pm 0.04$ \\
\hline
    Coulomb & 9.1   & 0.01$\pm0.01$  & 0.01$\pm0.01$   & 0.01$\pm0.01$   & 0.01$\pm0.01$   & 0.16$\pm0.03$  & 0.16$\pm0.03$   & 0.17$\pm0.04$   & 0.18$\pm0.04$   & 0.60$\pm0.06$   & 0.60$\pm0.06$ & 0.60$\pm0.06$   & 0.68$\pm0.06$  \\
    Coulomb & 9.2   & 0.01$\pm0.01$   & 0.01$\pm0.01$   & 0.01$\pm0.01$   & 0.01$\pm0.01$   & 0.04$\pm 0.02$  & 0.04$\pm 0.02$  & 0.04$\pm 0.02$  & 0.04$\pm 0.02$  & 0.61$\pm 0.06$  & 0.61$\pm 0.06$  & 0.65$\pm 0.06$  & 0.65$\pm 0.06$ \\
    Coulomb & 9.3   & 0.19$\pm 0.03$  & 0.18$\pm 0.03$  & 0.18$\pm 0.03$  & 0.20$\pm 0.03$  & 0.08$\pm 0.03$  & 0.08$\pm 0.03$  & 0.08$\pm 0.03$  & 0.09$\pm 0.03$  & 0.63$\pm 0.06$  & 0.63$\pm 0.06$  & 0.63$\pm 0.06$  & 0.67$\pm 0.06$ \\
    Coulomb & 9.4   & 0.03$\pm 0.02$  & 0.03$\pm 0.02$  & 0.04$\pm 0.02$  & 0.04$\pm 0.02$  & 0.22$\pm0.04$  & 0.21$\pm0.04$  & 0.21$\pm0.04$  & 0.21$\pm0.04$  & 0.68$\pm 0.06$  & 0.68$\pm 0.06$  & 0.72$\pm 0.06$  & 0.74$\pm 0.06$ \\
    Coulomb & 9.6   & 0.48$\pm0.04$  & 0.48$\pm0.04$  & 0.48$\pm0.04$  & 0.46$\pm0.04$  & 0.42$\pm 0.05$  & 0.42$\pm 0.05$  & 0.42$\pm 0.05$  & 0.41$\pm 0.05$  & 0.79$\pm 0.05$   & 0.77$\pm 0.05$   & 0.72$\pm 0.06$   & 0.81$\pm 0.05$  \\
    Coulomb & 10.0 & 0.49$\pm0.04$  & 0.49$\pm0.04$  & 0.48$\pm0.04$  & 0.48$\pm0.04$  & 0.40$\pm 0.05$  & 0.39$\pm 0.05$  & 0.39$\pm 0.05$  & 0.38$\pm 0.05$  & 0.79$\pm 0.05$   & 0.77$\pm 0.05$   & 0.79$\pm 0.05$   & 0.89$\pm 0.04$  \\
   Coulomb & 10.1 & $<0.01$ & $<0.01$ & $<0.01$ & $<0.01$ &  $<0.01$ & $<0.01$ & $<0.01$ & $<0.01$ & $<0.02$ & $<0.02$ & $<0.02$ & $<0.02$   \\
\hline
   CoulombP & 10.1 & 0.47$\pm0.04$  & 0.47$\pm0.04$  & 0.47$\pm0.04$  & 0.46$\pm0.04$  & 0.45$\pm 0.05$  & 0.44$\pm 0.05$  & 0.44$\pm 0.05$  & 0.43$\pm 0.05$  & 0.81$\pm 0.05$   & 0.81$\pm 0.05$   & 0.77$\pm 0.05$   & 0.88$\pm 0.04$  \\
\hline
    GSBRF & 9.3   & $<0.01$ & $<0.01$ & $<0.01$ & $<0.01$ & 0.01$\pm 0.01$  & 0.01$\pm 0.01$  & 0.01$\pm 0.01$  & 0.01$\pm 0.01$  & 0.07$\pm 0.03$  & 0.07$\pm 0.03$  & 0.07$\pm 0.03$  & 0.07$\pm 0.03$ \\
    GSBRF & 9.4   & $<0.01$ & $<0.01$ & $<0.01$ & $<0.01$ & 0.01$\pm 0.01$  & 0.01$\pm 0.01$  & 0.01$\pm 0.01$  & 0.01$\pm 0.01$  & 0.07$\pm 0.03$  & 0.07$\pm 0.03$  & 0.07$\pm 0.03$  & 0.07$\pm 0.03$ \\
    GSBRF & 9.6   & $<0.01$ & $<0.01$ & $<0.01$ & $<0.01$ & 0.01$\pm 0.01$  & 0.01$\pm 0.01$  & 0.01$\pm 0.01$  & 0.01$\pm 0.01$  & 0.58$\pm 0.06$  & 0.58$\pm 0.06$  & 0.60$\pm 0.06$  & 0.58$\pm 0.06$ \\
    GSBRF & 10.0    & $<0.01$ & $<0.01$ & $<0.01$ & $<0.01$ & 0.01$\pm 0.01$  & 0.01$\pm 0.01$  & 0.01$\pm 0.01$  & 0.01$\pm 0.01$  & 0.58$\pm 0.06$  & 0.56$\pm 0.06$  & 0.58$\pm 0.06$  & 0.54$\pm 0.06$ \\
   GSBRF & 10.1    & $<0.01$ & $<0.01$ & $<0.01$ & $<0.01$ & 0.01$\pm 0.01$  & 0.01$\pm 0.01$  & 0.01$\pm 0.01$  & 0.01$\pm 0.01$  & 0.39$\pm 0.06$  & 0.40$\pm 0.06$  & 0.39$\pm 0.06$  & 0.39$\pm 0.06$ \\

\hline
    WentzelBRF & 9.3   & 0.18$\pm 0.03$  & 0.18$\pm 0.03$  & 0.19$\pm 0.03$  & 0.20$\pm 0.03$  & 0.09$\pm0.03$  & 0.09$\pm0.03$  & 0.09$\pm0.03$  & 0.10$\pm0.03$  & 0.60$\pm 0.06$  & 0.60$\pm 0.06$  & 0.60$\pm 0.06$  & 0.65$\pm 0.06$ \\
    WentzelBRF & 9.4   & 0.02$\pm 0.01$  & 0.02$\pm 0.01$  & 0.02$\pm 0.01$  & 0.04$\pm 0.02$  & 0.21$\pm0.04$  & 0.20$\pm0.04$  & 0.20$\pm0.04$  & 0.20$\pm0.04$  & 0.61$\pm 0.06$  & 0.60$\pm 0.06$ & 0.60$\pm 0.06$  & 0.68$\pm 0.06$ \\
    WentzelBRF & 9.6   & 0.46$\pm 0.04$  & 0.46$\pm 0.04$  & 0.46$\pm 0.04$  & 0.46$\pm 0.04$  & 0.44$\pm0.05$  & 0.43$\pm0.05$  & 0.43$\pm0.05$  & 0.42$\pm0.05$  & 0.79$\pm 0.05$  & 0.77$\pm 0.05$  & 0.75$\pm 0.06$  & 0.81$\pm 0.05$ \\
    WentzelBRF & 10.0 & 0.49$\pm 0.04$  & 0.48$\pm 0.04$  & 0.48$\pm 0.04$  & 0.49$\pm 0.04$  & 0.44$\pm0.05$  & 0.43$\pm0.05$  & 0.43$\pm0.05$  & 0.42$\pm0.05$  & 0.81$\pm 0.05$  & 0.79$\pm 0.05$  & 0.82$\pm 0.05$  & 0.88$\pm 0.04$ \\
    WentzelBRF & 10.1& $<0.01$ & $<0.01$ & $<0.01$ & $<0.01$ & 0.01$\pm 0.01$  & 0.01$\pm 0.01$  & 0.01$\pm 0.01$  & 0.01$\pm 0.01$   & 0.42$\pm 0.06$  & 0.42$\pm 0.06$  & 0.42$\pm 0.06$  & 0.46$\pm 0.06$ \\
\hline
    WentzelBRFP & 9.3   & 0.02$\pm 0.01$  & 0.02$\pm 0.01$  & 0.02$\pm 0.01$  & 0.03$\pm 0.02$  & 0.04$\pm 0.02$  & 0.04$\pm 0.02$  & 0.04$\pm 0.02$  & 0.04$\pm 0.02$  & 0.33$\pm 0.06$  & 0.33$\pm 0.06$  & 0.33$\pm 0.06$  & 0.37$\pm 0.06$ \\
    WentzelBRFP & 9.4   & 0.02$\pm 0.01$  & 0.02$\pm 0.01$  & 0.02$\pm 0.01$  & 0.02$\pm 0.01$  & 0.03$\pm 0.02$  & 0.03$\pm 0.02$  & 0.03$\pm 0.02$  & 0.03$\pm 0.02$  & 0.21$\pm 0.05$   & 0.21$\pm 0.05$   & 0.21$\pm 0.05$   & 0.21$\pm 0.05$  \\
    WentzelBRFP & 9.6   & $<0.01$ & $<0.01$ & $<0.01$ & $<0.01$ & 0.01$\pm 0.01$  & 0.01$\pm 0.01$  & 0.01$\pm 0.01$  & 0.01$\pm 0.01$  & 0.30$\pm 0.05$  & 0.28$\pm 0.05$  & 0.30$\pm 0.05$  & 0.33$\pm 0.05$ \\
    WentzelBRFP & 10.0    & $<0.01$ & $<0.01$ & $<0.01$ & $<0.01$ & 0.01$\pm 0.01$  & 0.01$\pm 0.01$  & 0.01$\pm 0.01$  & 0.01$\pm 0.01$  & 0.44$\pm 0.06$  & 0.42$\pm 0.06$  & 0.44$\pm 0.06$  & 0.46$\pm 0.06$ \\
    WentzelBRFP & 10.1    & $<0.01$ & $<0.01$ & $<0.01$ & $<0.01$ & 0.01$\pm 0.01$  & 0.01$\pm 0.01$  & 0.01$\pm 0.01$  & 0.01$\pm 0.01$  & 0.42$\pm 0.06$  & 0.42$\pm 0.06$  & 0.42$\pm 0.06$  & 0.46$\pm 0.06$ \\
\hline
    EmLivermore & 9.6		& $<0.01$ & $<0.01$ & $<0.01$ & $<0.01$ & $<0.01$ & $<0.01$ & $<0.01$ & $<0.01$ 	& 0.07$\pm 0.04$  & 0.05$\pm 0.03$  & 0.07$\pm 0.04$  & 0.07$\pm 0.04$ \\
    EmLivermore & 10.0    	& $<0.01$ & $<0.01$ & $<0.01$ & $<0.01$ & $<0.01$ & $<0.01$ & $<0.01$ & $<0.01$ 	& 0.05$\pm 0.03$  & 0.05$\pm 0.03$  & 0.05$\pm 0.03$  & 0.05$\pm 0.03$ \\
    EmLivermore & 10.1    	& $<0.01$ & $<0.01$ & $<0.01$ & $<0.01$ & $<0.01$ & $<0.01$ & $<0.01$ & $<0.01$  	& 0.07$\pm 0.04$  & 0.07$\pm 0.04$  & 0.07$\pm 0.04$  & 0.09$\pm 0.04$ \\
\hline
    EmStd & 9.6   	& $<0.01$ & $<0.01$ & $<0.01$ & $<0.01$ & $<0.01$ & $<0.01$ & $<0.01$ & $<0.01$  	& 0.40$\pm 0.06$  & 0.40$\pm 0.06$  & 0.42$\pm 0.06$  & 0.44$\pm 0.06$ \\
    EmStd & 10.0    	& $<0.01$ & $<0.01$ & $<0.01$ & $<0.01$ & $<0.01$ & $<0.01$ & $<0.01$ & $<0.01$  & 0.07$\pm 0.04$  & 0.07$\pm 0.04$  & 0.07$\pm 0.04$  & 0.09$\pm 0.04$ \\
    EmStd & 10.1    	& $<0.01$ & $<0.01$ & $<0.01$ & $<0.01$ & $<0.01$ & $<0.01$ & $<0.01$ & $<0.01$  & 0.05$\pm 0.03$  & 0.05$\pm 0.03$  & 0.05$\pm 0.03$  & 0.05$\pm 0.03$ \\
\hline
    EmOpt1 & 9.6   	& $<0.01$ & $<0.01$ & $<0.01$ & $<0.01$ & $<0.01$ & $<0.01$ & $<0.01$ & $<0.01$  & 0.33$\pm 0.06$  & 0.33$\pm 0.06$  & 0.33$\pm 0.06$  & 0.35$\pm 0.06$ \\
    EmOpt1 & 10.0    	& $<0.01$ & $<0.01$ & $<0.01$ & $<0.01$ & $<0.01$ & $<0.01$ & $<0.01$ & $<0.01$  & 0.39$\pm 0.06$  & 0.39$\pm 0.06$  & 0.39$\pm 0.06$  & 0.40$\pm 0.06$ \\
    EmOpt1 & 10.1    	& $<0.01$ & $<0.01$ & $<0.01$ & $<0.01$ & $<0.01$ & $<0.01$ & $<0.01$ & $<0.01$  & 0.14$\pm 0.05$  & 0.12$\pm 0.05$  & 0.14$\pm 0.05$  & 0.18$\pm 0.05$ \\
\hline
    EmOpt2 & 9.6   	& $<0.01$ & $<0.01$ & $<0.01$ & $<0.01$ & $<0.01$ & $<0.01$ & $<0.01$ & $<0.01$  & 0.32$\pm 0.06$  & 0.32$\pm 0.06$  & 0.32$\pm 0.06$  & 0.35$\pm 0.06$ \\
    EmOpt2 & 10.0   	& $<0.01$ & $<0.01$ & $<0.01$ & $<0.01$ & $<0.01$ & $<0.01$ & $<0.01$ & $<0.01$  & 0.37$\pm 0.06$  & 0.37$\pm 0.06$  & 0.37$\pm 0.06$  & 0.40$\pm 0.06$ \\
    EmOpt2 & 10.1    	& $<0.01$ & $<0.01$ & $<0.01$ & $<0.01$ & $<0.01$ & $<0.01$ & $<0.01$ & $<0.01$  & 0.16$\pm 0.05$  & 0.16$\pm 0.05$  & 0.16$\pm 0.05$  & 0.19$\pm 0.05$ \\
\hline
    EmOpt3 & 9.6   	& $<0.01$ & $<0.01$ & $<0.01$ & $<0.01$ & $<0.01$ & $<0.01$ & $<0.01$ & $<0.01$  & 0.07$\pm 0.04$  & 0.07$\pm 0.04$  & 0.07$\pm 0.04$  & 0.07$\pm 0.04$ \\
    EmOpt3 & 10.0    	& $<0.01$ & $<0.01$ & $<0.01$ & $<0.01$ & $<0.01$ & $<0.01$ & $<0.01$ & $<0.01$  & 0.07$\pm 0.04$  & 0.07$\pm 0.04$  & 0.07$\pm 0.04$  & 0.07$\pm 0.04$ \\
    EmOpt3 & 10.1    	& $<0.01$ & $<0.01$ & $<0.01$ & $<0.01$ & $<0.01$ & $<0.01$ & $<0.01$ & $<0.01$  & 0.07$\pm 0.04$  & 0.07$\pm 0.04$  & 0.07$\pm 0.04$  & 0.07$\pm 0.04$ \\
\hline
    EmOpt4 & 9.6   	& $<0.01$ & $<0.01$ & $<0.01$ & $<0.01$ & $<0.01$ & $<0.01$ & $<0.01$ & $<0.01$  & 0.07$\pm 0.04$  & 0.07$\pm 0.04$  & 0.07$\pm 0.04$  & 0.07$\pm 0.04$ \\
    EmOpt4 & 10.0    	& $<0.01$ & $<0.01$ & $<0.01$ & $<0.01$ & $<0.01$ & $<0.01$ & $<0.01$ & $<0.01$  & 0.09$\pm 0.04$  & 0.09$\pm 0.04$  & 0.09$\pm 0.04$  & 0.09$\pm 0.04$ \\
    EmOpt4 & 10.1    	& $<0.01$ & $<0.01$ & $<0.01$ & $<0.01$ & $<0.01$ & $<0.01$ & $<0.01$ & $<0.01$	& $<0.02$ & $<0.02$ & $<0.02$ & $<0.02$  \\
    EmWVI 	& 10.1    & $<0.01$ & $<0.01$ & $<0.01$ & $<0.01$ & 0.01$\pm 0.01$  & 0.01$\pm 0.01$  & 0.01$\pm 0.01$  & 0.01$\pm 0.01$  & 0.40$\pm 0.06$  & 0.40$\pm 0.06$  & 0.44$\pm 0.06$  & 0.46$\pm 0.06$ \\

    \hline
    \end{tabular}}%
  \label{tab_eff}%
\end{table*}%
\tabcolsep=6pt

Appropriate 2$\times$2 contingency tables are built for this purpose, based on
the results of the goodness-of-fit tests documented in section \ref{sec_gof}:
for each goodness-of-fit test, they report on one diagonal the number of test
cases where both categories
subject to evaluation ``pass'' or ``fail'' the test, and on the other diagonal
the number of test cases where one category ``passes'' the $\chi^2$ test, while
the other one``fails''.
An example of their configuration is shown in Table~\ref{tab_exmcnemar}.

McNemar's test \cite{mcnemar} is applied to the analysis of related data
samples.
This test focuses on the significance of the discordant results, i.e. the number
of test cases where one category ``passes'' a goodness-of-fit test and the other
one ``fails''.
The null hypothesis for McNemar's test is that the proportion of 
discordant results is the same in the two cells corresponding to 
``pass-fail'' or ``fail-pass'' associated with the two categories subject to test.

The calculation of McNemar's test is performed using either the $\chi^2$
asymptotic distribution or the binomial distribution \cite{bennett}:
the latter is also known as ``McNemar exact test''.
Yates' \cite{yates} continuity correction may be applied to the calculation of
the $\chi^2$ statistic to account for cells with a small number of entries.
According to \cite{lui_2001}, McNemar's test uncorrected for continuity is more
powerful than the exact test, and performs well even when the number of
discordant pairs is as low as 6, while both the exact test and the corrected
McNemar's test are conservative.
The results reported in this paper concern McNemar exact test, as most of
the analyzed tables involved a small number of entries in some cells, which
prevented the calculation of the $\chi^2$ statistic.


\section{Results}
\label{sec_results}

The analysis of electron backscattering simulation is focused on the quantification
of a few salient features, based on the considerations discussed in the previous
sections:
\begin{itemize}
\item the quantification of the capability of each physics configuration to 
simulate electron backscattering compatible with experiment;
\item its evolution with Geant4 versions, with emphasis on quantifying the
capabilities of the latest versions at the time of writing this paper, to which
the interest of experimental users is more generally directed;
\item the identification of the physics configurations and Geant4 versions
achieving the highest compatibility with experiment;
\item the role played by single and multiple scattering in contributing to
the accuracy and the computational speed of the simulation.
\end{itemize}
The following sections document first an overview of the main features of the
statistical results, followed by detailed discussion of each Geant4 model
configuration subject to evaluation.

The large number of experimental test cases, Geant4 physics modeling options and
versions evaluated in the validation tests makes a full graphical illustration
of the results impractical within the scope of a journal publication.
A sample of plots, which span a wide range of electron beam energies, target
atomic numbers, Geant4 models and versions, complement the outcome of
the statistical analysis with a qualitative illustration of the results.

\subsection{Model Compatibility with Experiment}

The efficiencies resulting from goodness-of-fit tests are summarized in Table
\ref{tab_eff} for all the physics configurations and Geant4 versions evaluated
in this paper.
One can observe that the four tests produce consistent outcomes regarding the 
rejection of the null hypothesis of equivalence between simulated and experimental
distributions.
Therefore, for convenience, the categorical analyses reported in the following sections
are performed only based on the outcome of the Anderson-Darling test, unless otherwise specified.

The results listed in Table \ref{tab_eff} show that Geant4 multiple scattering
models fail to reproduce experimental electron backscattering data in the lower
energy end, while the single scattering approach adopted in the Coulomb
and WentzelBRF configurations retains the capability to describe lower
energy data, although with reduced efficiency with respect to the higher energy
range.
At energies above 100 keV the highest efficiency is achieved with the Urban configuration in
Geant4 9.1 in the context of a condensed path scheme, and with the
Coulomb scattering model in Geant4 10.0 in the context of simulating individual scattering
occurrences.

Evaluations specific to each Geant4 model, or family of models, are detailed in 
the following sections.

\subsection{Effect of different Geant4 electron physics models}

The fraction of backscattered electrons produced in the simulation is mainly
determined by multiple scattering; a small contribution to the number of electrons
reaching the detector consists of secondary particles produced in the ionization
of target atoms, which escape the target volume.

As Geant4 encompasses different models of electron ionization, their effect on
the validation of the simulation of electron backscattering has been evaluated.
For this purpose simulations where executed with the Urban multiple scattering
configuration associated with different Geant4 electron physics models: those
based on the EEDL \cite{eedl} evaluated data library \cite{lowe_e, lowe_chep,
lowe_nss} , those encompassed in the Geant4 Standard electromagnetic package
\cite{emstandard} and those reengineered in Geant4 from the Penelope
\cite{penelope} Monte Carlo code.

The efficiency obtained with the three different configurations is listed in
Table \ref{tab_stdpen} for the Geant4 versions subject to evaluation: it
derives from the results of the Anderson-Darling goodness-of-fit test
comparing simulated and experimental backscattered electron distributions.
One can observe that the results associated with different electron physics
models are very similar; a categorical analysis comparing the results
of the Standard and Penelope models with those based on EEDL 
fails to reject the hypothesis of equivalent behaviour with 0.05 significance.
Both Fisher's and Barnard's exact tests lead to the same conclusion.

From this analysis one can infer that the electron backscattering test
studied in this paper is insensitive to how electron interactions, apart from
multiple scattering, are modeled in Geant4.

The results reported in the following sections are obtained using Geant4 
electron interaction models based on EEDL. 

The interactions of secondary photons were modeled based on the EPDL
\cite{epdl97} evaluated data library in all simulations, based on validation
studies \cite{tns_rayleigh, nss_photoel, nss_compton, mc2013_photons}:
they play a negligible role in the validation test, as they could only affect
the estimated electron backscattering fraction as a contamination originating
from secondary electrons they produce.
No such contamination was observed in the simulated data sample subject 
to analysis.

\begin{table}[htbp]
  \centering
  \caption{Efficiency for the Urban multiple scattering configuration with different electron physics models}
    \begin{tabular}{ccccc}
    \hline
    Energy (keV) & Version & EEDL  & Standard & Penelope \\
    \hline
    \multirow{7}[0]{*}{1-20} 		& 9.1   & $<0.01$ & $<0.01$ & $<0.01$ \\
          					& 9.2   & $<0.01$ & $<0.01$ & $<0.01$ \\
         					& 9.3   & $<0.01$ & $<0.01$ & $<0.01$ \\
          					& 9.4   & $<0.01$ & $<0.01$ & $<0.01$ \\
          					& 9.6   & $<0.01$ & $<0.01$ & $<0.01$ \\
          					& 10.0  & $<0.01$ & $<0.01$ & $<0.01$ \\
         					& 10.1  & $<0.01$ & $<0.01$ & $<0.01$ \\
\hline
 {\multirow{7}[0]{*}{20-100 }} 	& 9.1   	& 0.10$\pm0.03$  		& 0.11$\pm0.03$  	& 0.07$\pm0.02$ \\
 						& 9.2   	& 0.03$\pm0.02$  		& 0.06$\pm0.02$ 	& 0.03$\pm0.02$ \\
						& 9.3   	& 0.09$\pm0.02$  		& 0.04$\pm0.02$  	& 0.07$\pm0.02$ \\
						& 9.4  	& 0.10$\pm0.03$  		& 0.08$\pm0.03$  	& 0.09$\pm0.03$ \\
						& 9.6   	& 0.17$\pm0.04$  		& 0.15$\pm0.03$  	& 0.12$\pm0.03$ \\
						& 10.0  	&$<0.01$				& $<0.01$ 		& $<0.01$ \\
						& 10.1  	&$<0.01$				& $<0.01$ 		& $<0.01$ \\
\hline
{\multirow{7}[0]{*}{$>$100}} 	& 9.1   	& 0.79$\pm0.05$		& 0.75$\pm0.06$  	& 0.77$\pm0.05$ \\
						& 9.2   	& 0.79$\pm0.05$  		& 0.81$\pm0.05$  	& 0.74$\pm0.06$ \\
						& 9.3   	& 0.74$\pm0.06$  		& 0.65$\pm0.06$  	& 0.67$\pm0.06$ \\
						& 9.4   	& 0.58$\pm0.06$  		& 0.61$\pm0.06$  	& 0.61$\pm0.06$ \\
						& 9.6   	& 0.68$\pm0.06$  		& 0.72$\pm0.06$  	& 0.68$\pm0.06$ \\
						& 10.0  	& 0.11$\pm0.04$  		& 0.09$\pm0.04$  	& 0.09$\pm0.04$ \\
 						& 10.1  	& 0.07$\pm0.04$  		& 0.07$\pm0.04$  	& 0.09$\pm0.04$ \\
   \hline
    \end{tabular}%
  \label{tab_stdpen}%
\end{table}%


\subsection{Geant4 Urban models}

\begin{figure} 
\centerline{\includegraphics[angle=0,width=8.5cm]{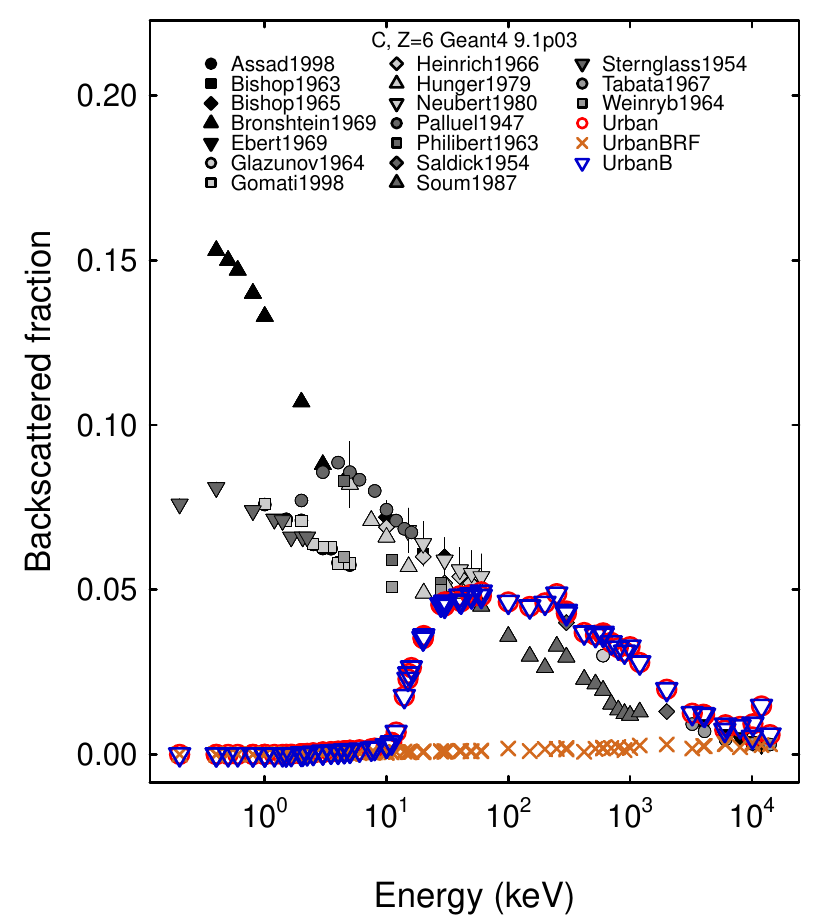}}
\caption{Fraction of electrons backscattered from a carbon target as a function
of the electron beam energy: experimental data (black and grey filled symbols)
and Geant4 9.1 simulation results with Urban (red empty circles), UrbanB (blue
empty triangles) and UrbanBRF (brown crosses) multiple scattering
configurations.}
\label{fig_urban3_v913_6}
\end{figure}

\begin{figure} 
\centerline{\includegraphics[angle=0,width=8.5cm]{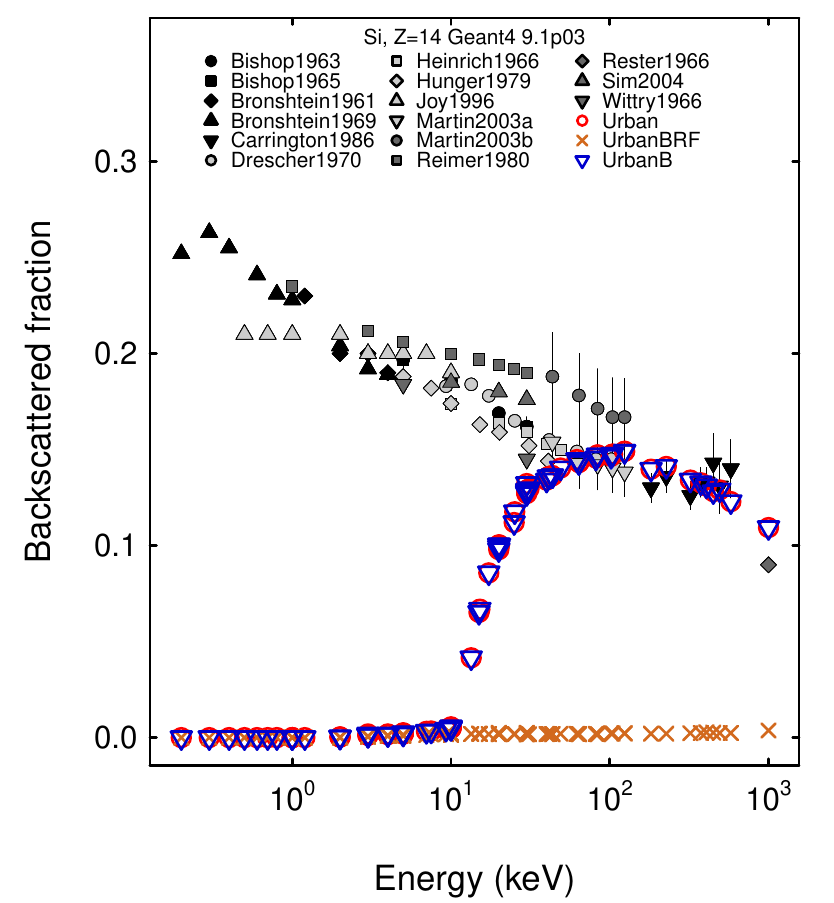}}
\caption{Fraction of electrons backscattered from a silicon target as a function
of the electron beam energy: experimental data (black and grey filled symbols)
and Geant4 9.1 simulation results with Urban (red empty circles), UrbanB (blue
empty triangles) and UrbanBRF (brown crosses) multiple scattering
configurations.}
\label{fig_urban3_v913_14}
\end{figure}

\begin{figure} 
\centerline{\includegraphics[angle=0,width=8.5cm]{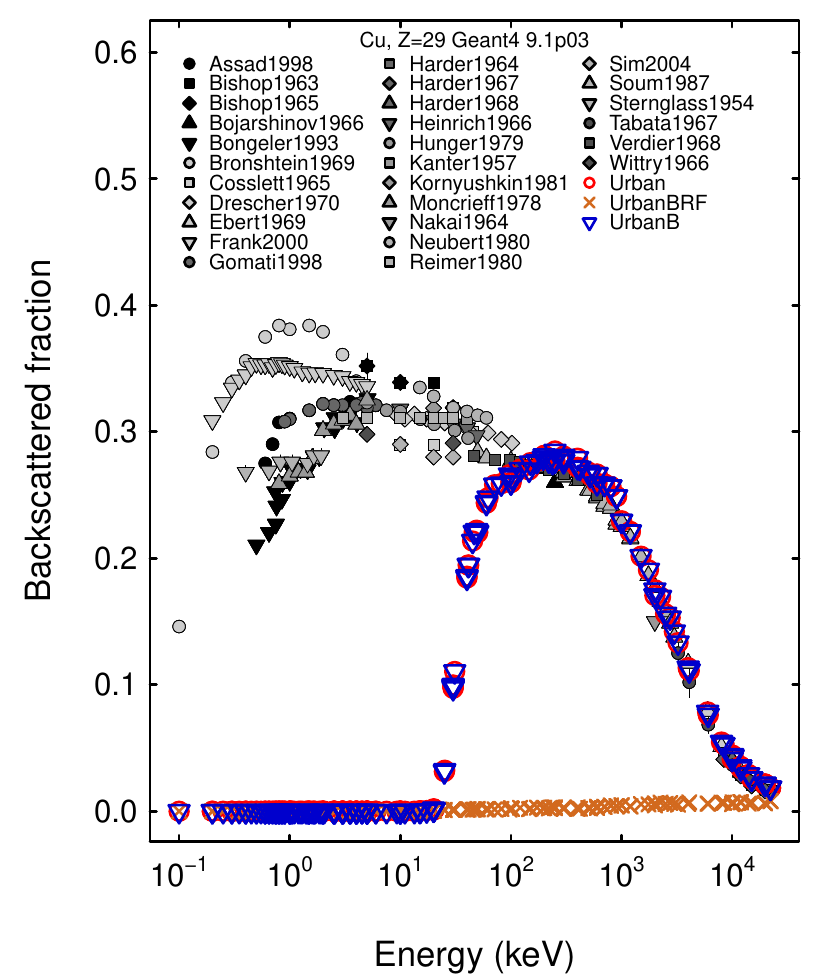}}
\caption{Fraction of electrons backscattered from a copper target as a function
of the electron beam energy: experimental data (black and grey filled symbols)
and Geant4 9.1 simulation results with Urban (red empty circles), UrbanB (blue
empty triangles) and UrbanBRF (brown crosses) multiple scattering
configurations.}
\label{fig_urban3_v913_29}
\end{figure}

\begin{figure} 
\centerline{\includegraphics[angle=0,width=8.5cm]{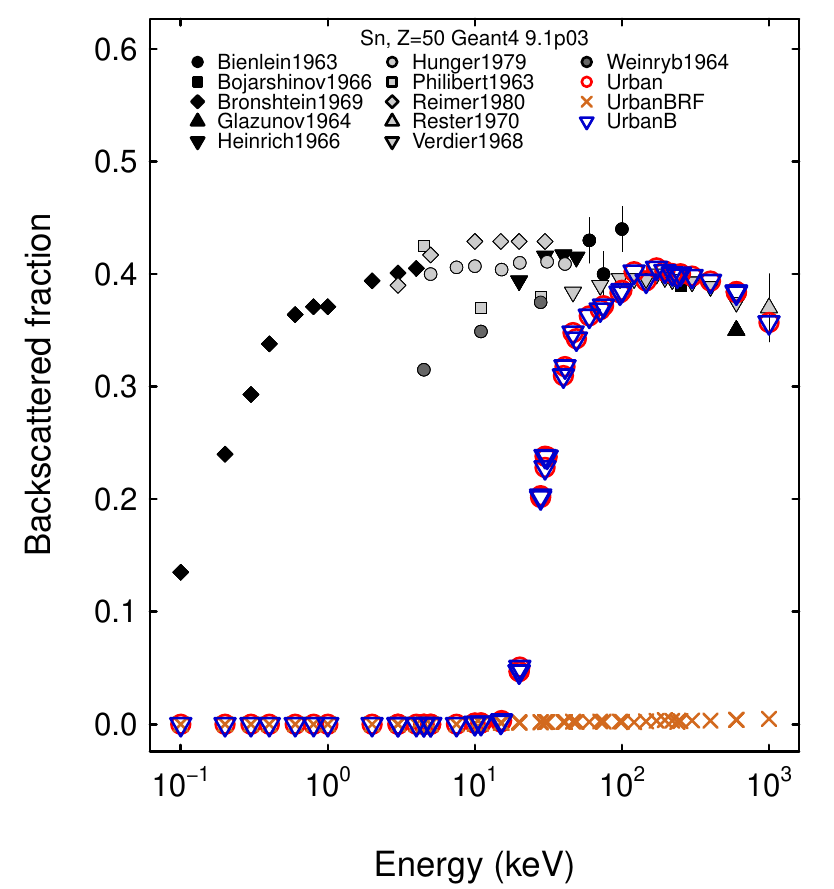}}
\caption{Fraction of electrons backscattered from a tin target as a function
of the electron beam energy: experimental data (black and grey filled symbols)
and Geant4 9.1 simulation results with Urban (red empty circles), UrbanB (blue
empty triangles) and UrbanBRF (brown crosses) multiple scattering
configurations.}
\label{fig_urban3_v913_50}
\end{figure}

\begin{figure} 
\centerline{\includegraphics[angle=0,width=8.5cm]{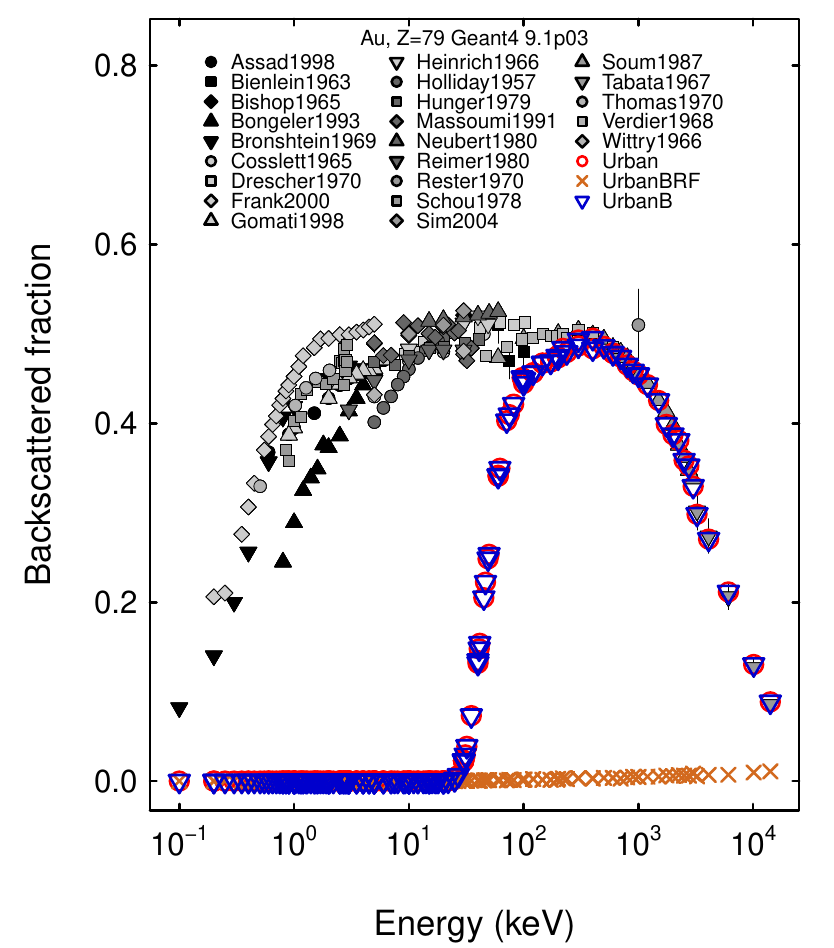}}
\caption{Fraction of electrons backscattered from a gold target as a function
of the electron beam energy: experimental data (black and grey filled symbols)
and Geant4 9.1 simulation results with Urban (red empty circles), UrbanB (blue
empty triangles) and UrbanBRF (brown crosses) multiple scattering
configurations.}
\label{fig_urban3_v913_79}
\end{figure}

\begin{figure} 
\centerline{\includegraphics[angle=0,width=8.5cm]{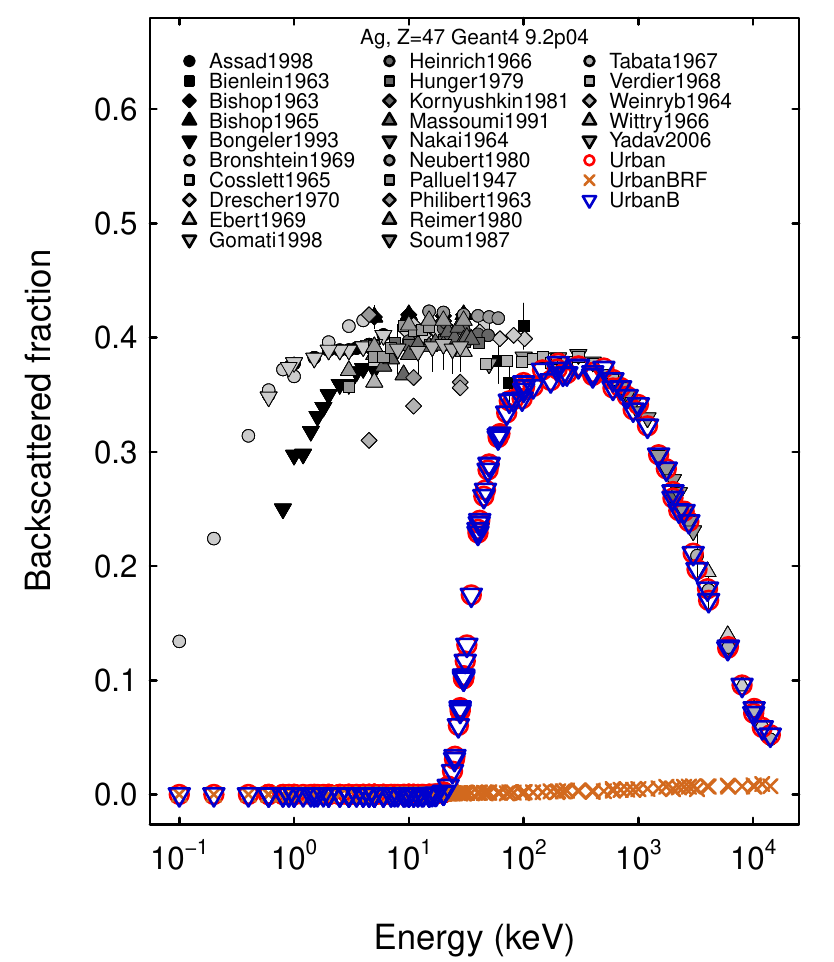}}
\caption{Fraction of electrons backscattered from a silver target as a function
of the electron beam energy: experimental data (black and grey filled symbols)
and Geant4 9.2 simulation results with Urban (red empty circles), UrbanB (blue
empty triangles) and UrbanBRF (brown crosses) multiple scattering
configurations.}
\label{fig_urban3_v924_47}
\end{figure}

\begin{figure} 
\centerline{\includegraphics[angle=0,width=8.5cm]{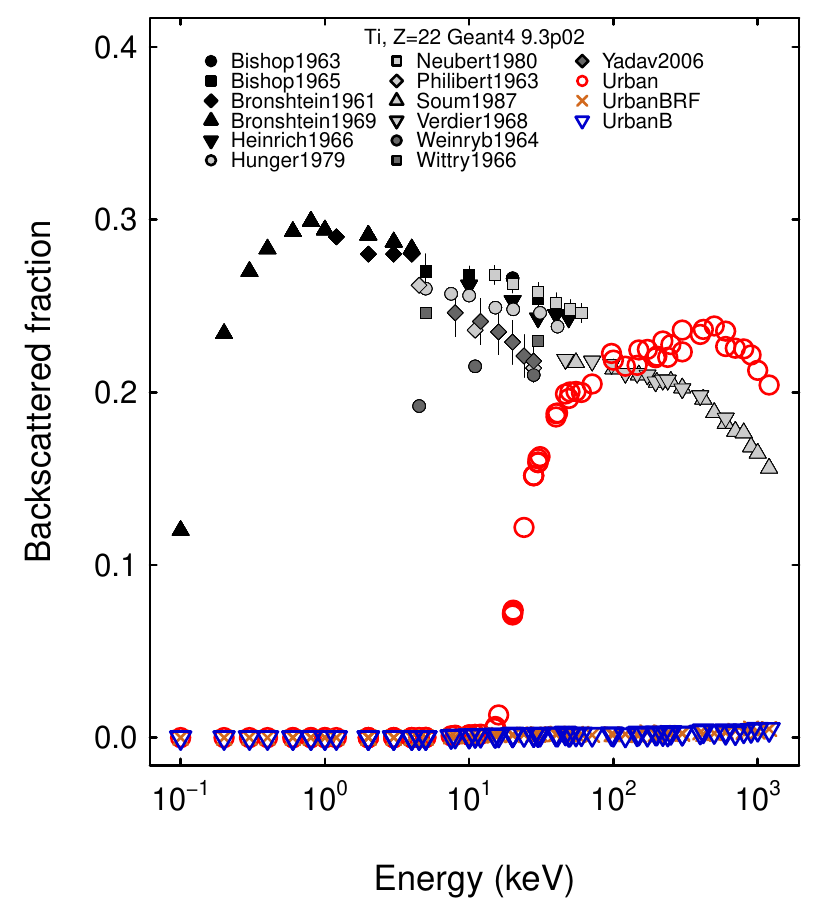}}
\caption{Fraction of electrons backscattered from a titanium target as a function
of the electron beam energy: experimental data (black and grey filled symbols)
and Geant4 9.3 simulation results with Urban (red empty circles), UrbanB (blue
empty triangles) and UrbanBRF (brown crosses) multiple scattering
configurations.}
\label{fig_urban3_v932_22}
\end{figure}

\begin{figure} 
\centerline{\includegraphics[angle=0,width=8.5cm]{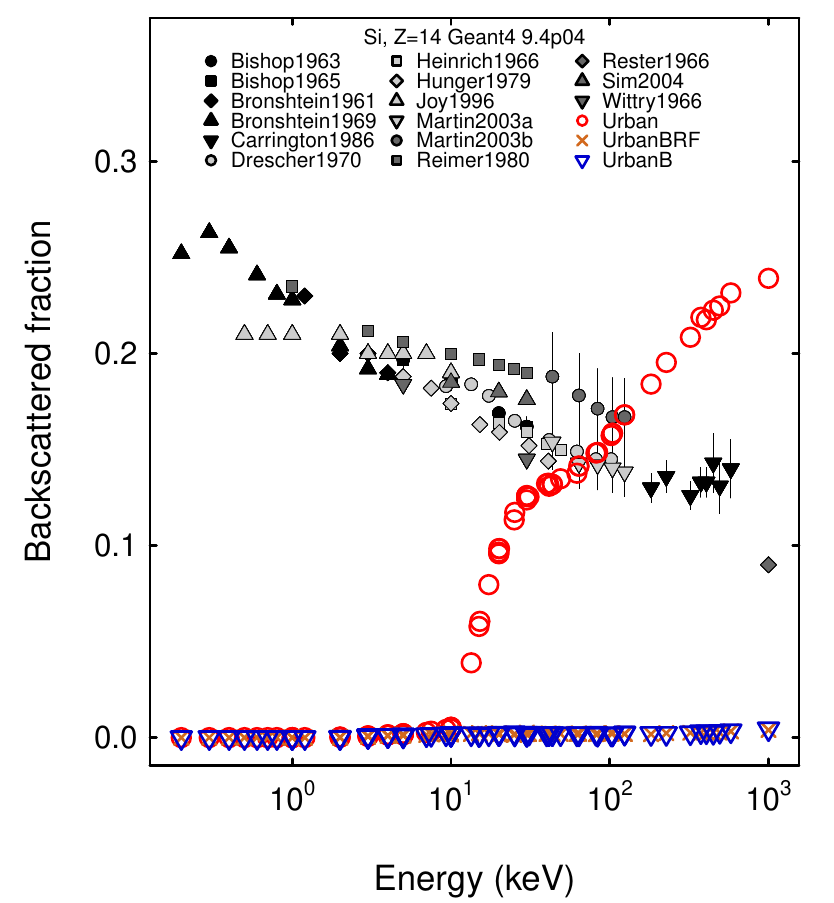}}
\caption{Fraction of electrons backscattered from a silicon target as a function
of the electron beam energy: experimental data (black and grey filled symbols)
and Geant4 9.4 simulation results with Urban (red empty circles), UrbanB (blue
empty triangles) and UrbanBRF (brown crosses) multiple scattering
configurations.}
\label{fig_urban3_v944_14}
\end{figure}

\begin{figure} 
\centerline{\includegraphics[angle=0,width=8.5cm]{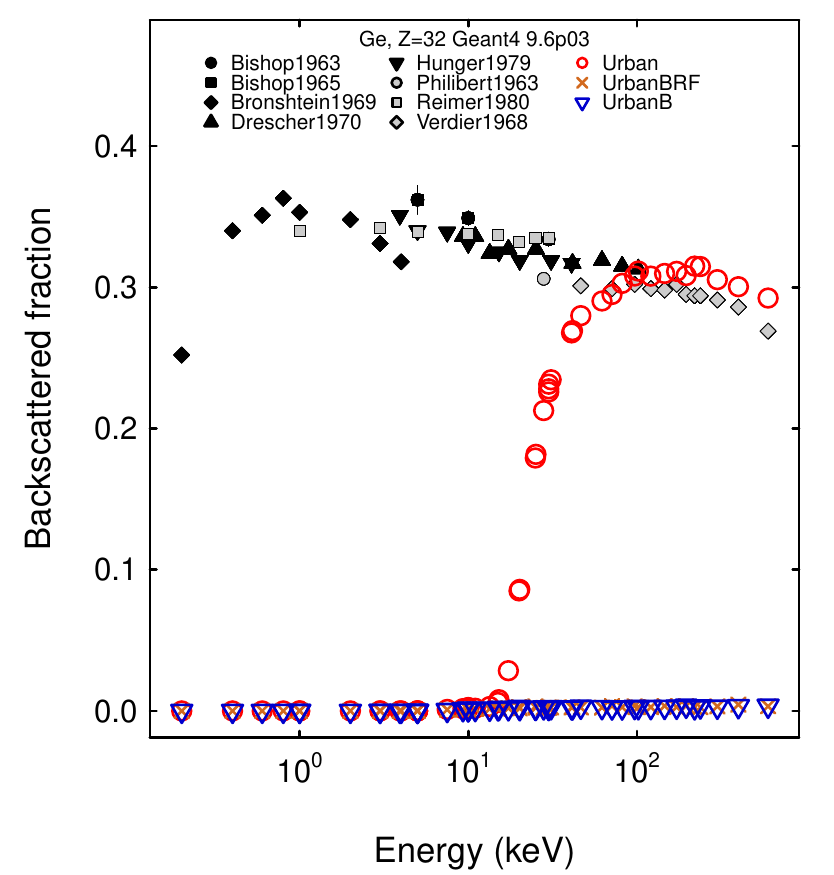}}
\caption{Fraction of electrons backscattered from a germanium target as a function
of the electron beam energy: experimental data (black and grey filled symbols)
and Geant4 9.6 simulation results with Urban (red empty circles), UrbanB (blue
empty triangles) and UrbanBRF (brown crosses) multiple scattering
configurations.}
\label{fig_urban3_v963_32}
\end{figure}

\begin{figure} 
\centerline{\includegraphics[angle=0,width=8.5cm]{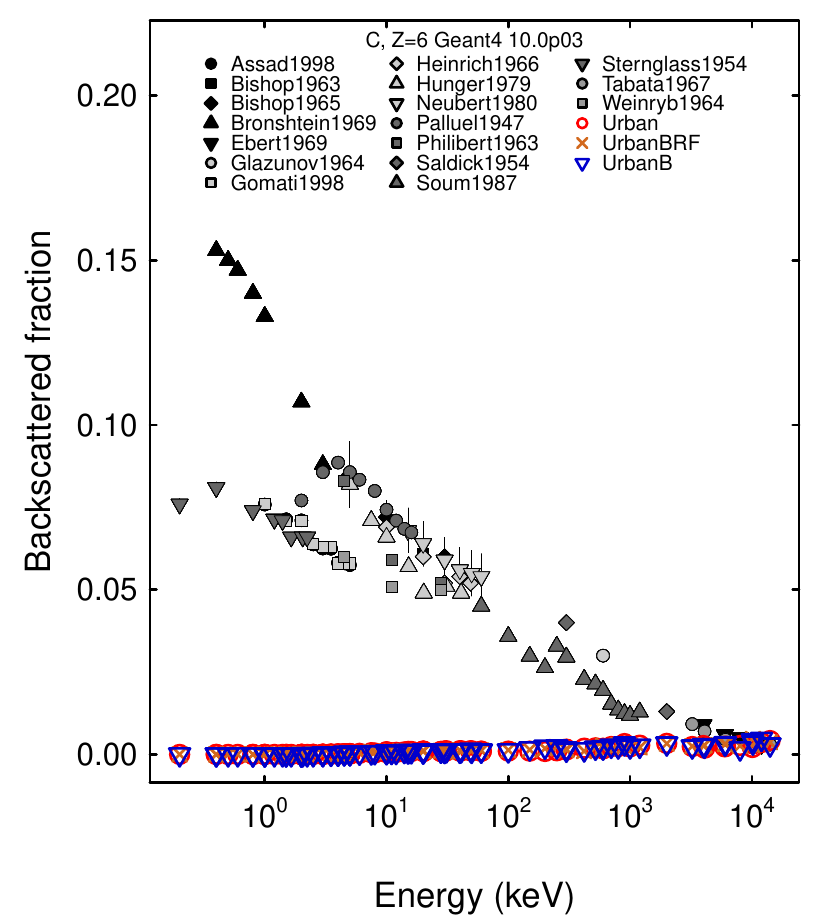}}
\caption{Fraction of electrons backscattered from a carbon target as a function
of the electron beam energy: experimental data (black and grey filled symbols)
and Geant4 10.0 simulation results with Urban (red empty circles), UrbanB (blue
empty triangles) and UrbanBRF (brown crosses) multiple scattering
configurations.}
\label{fig_urban3_v1003_6}
\end{figure}

\begin{figure} 
\centerline{\includegraphics[angle=0,width=8.5cm]{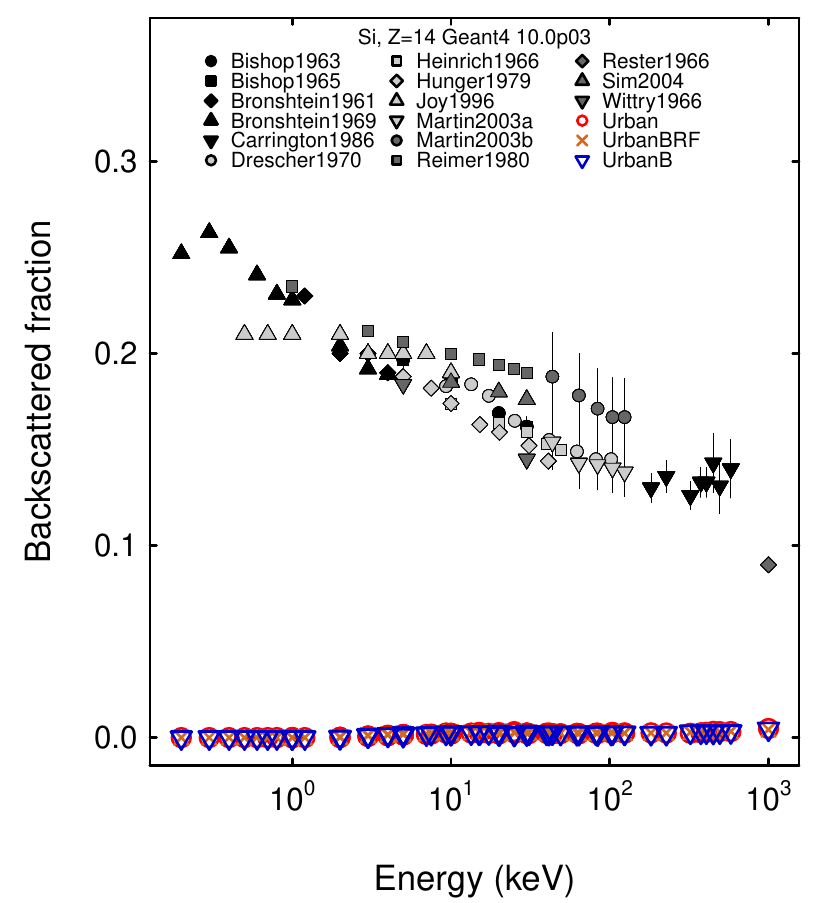}}
\caption{Fraction of electrons backscattered from a silicon target as a function
of the electron beam energy: experimental data (black and grey filled symbols)
and Geant4 10.0 simulation results with Urban (red empty circles), UrbanB (blue
empty triangles) and UrbanBRF (brown crosses) multiple scattering
configurations.}
\label{fig_urban3_v1003_14}
\end{figure}

\begin{figure} 
\centerline{\includegraphics[angle=0,width=8.5cm]{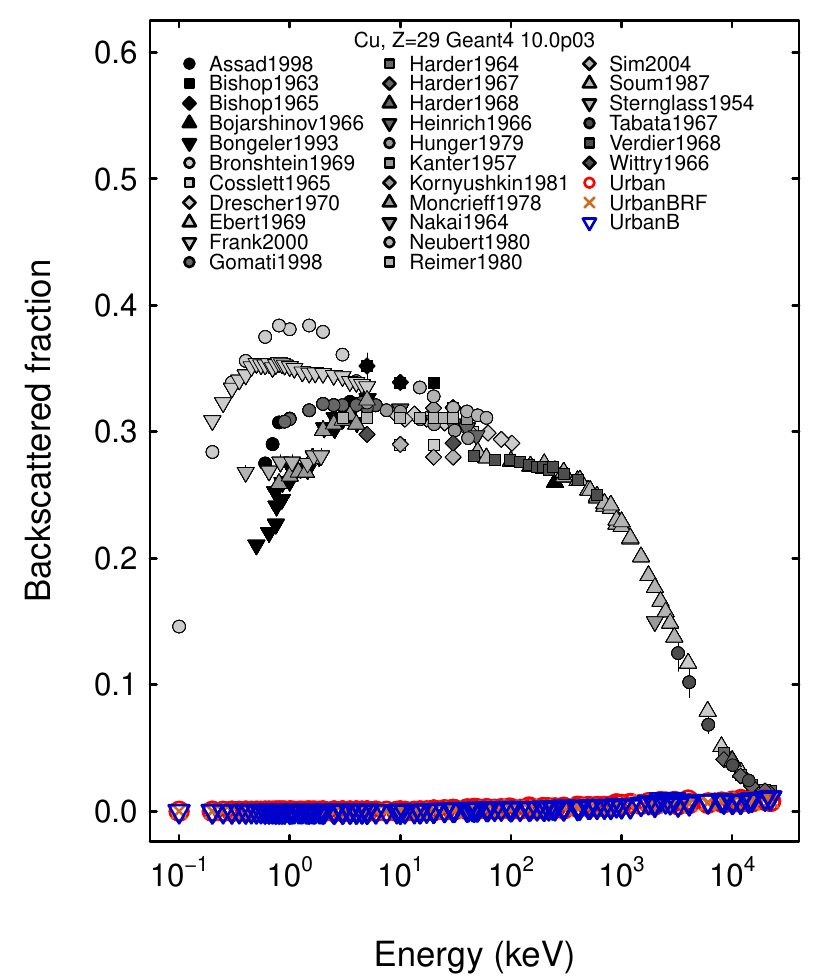}}
\caption{Fraction of electrons backscattered from a copper target as a function
of the electron beam energy: experimental data (black and grey filled symbols)
and Geant4 10.0 simulation results with Urban (red empty circles), UrbanB (blue
empty triangles) and UrbanBRF (brown crosses) multiple scattering
configurations.}
\label{fig_urban3_v1003_29}
\end{figure}

\begin{figure} 
\centerline{\includegraphics[angle=0,width=8.5cm]{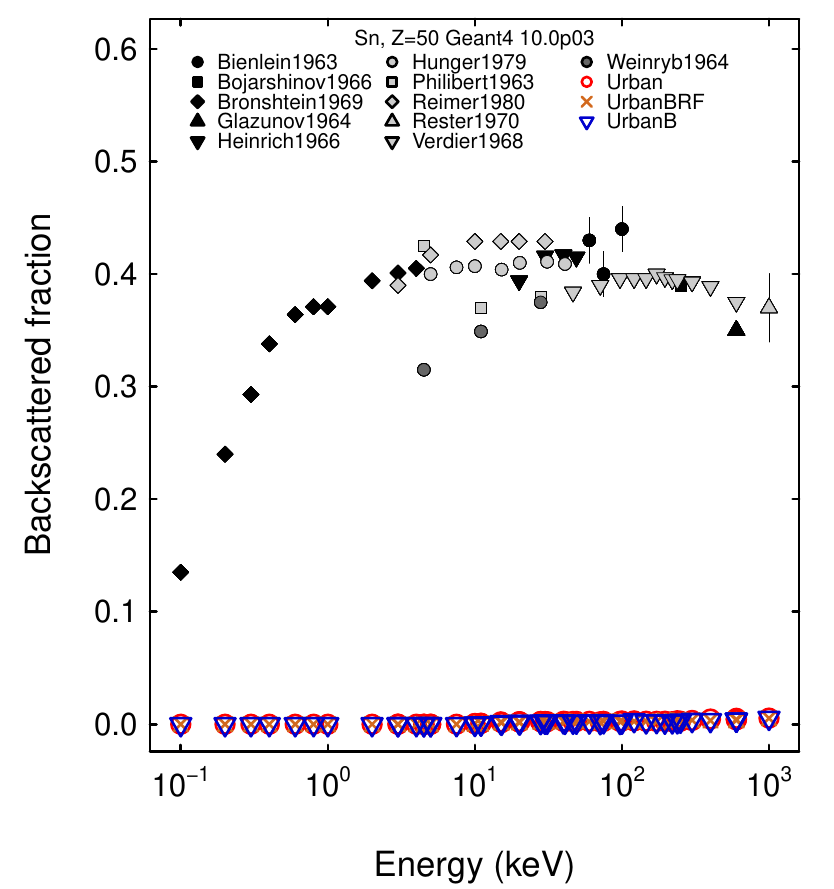}}
\caption{Fraction of electrons backscattered from a tin target as a function
of the electron beam energy: experimental data (black and grey filled symbols)
and Geant4 10.0 simulation results with Urban (red empty circles), UrbanB (blue
empty triangles) and UrbanBRF (brown crosses) multiple scattering
configurations.}
\label{fig_urban3_v1003_50}
\end{figure}

\begin{figure} 
\centerline{\includegraphics[angle=0,width=8.5cm]{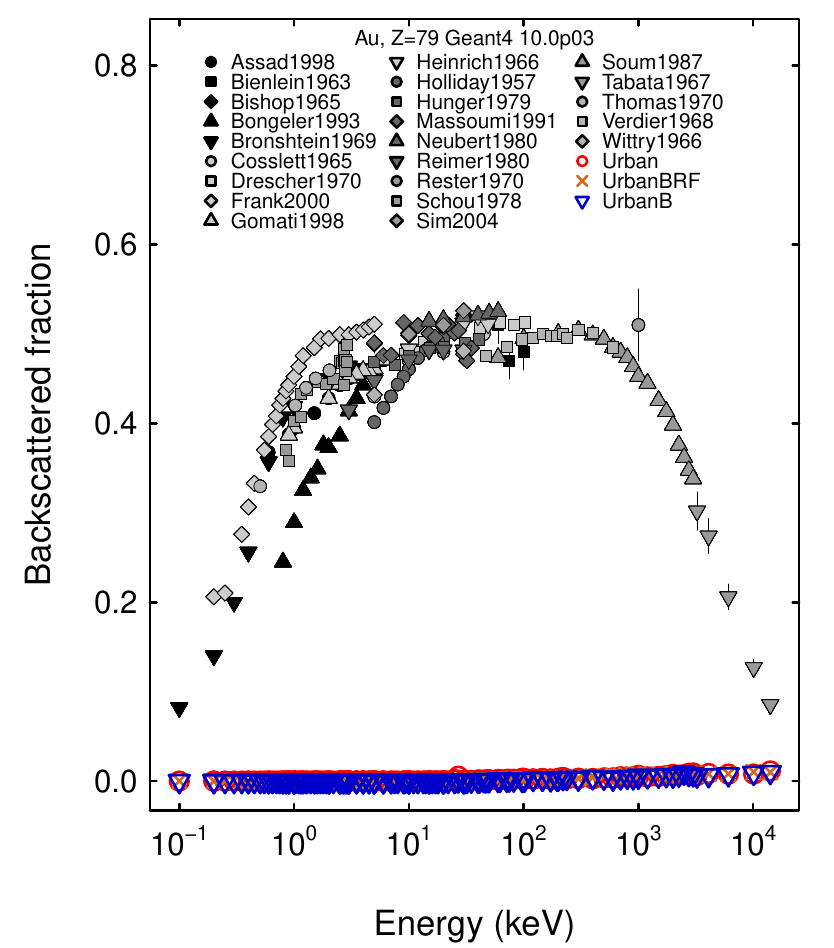}}
\caption{Fraction of electrons backscattered from a gold target as a function
of the electron beam energy: experimental data (black and grey filled symbols)
and Geant4 10.0 simulation results with Urban (red empty circles), UrbanB (blue
empty triangles) and UrbanBRF (brown crosses) multiple scattering
configurations.}
\label{fig_urban3_v1003_79}
\end{figure}

The fraction of backscattered electrons produced with the Urban,
UrbanB and UrbanBRF configurations is illustrated in Figs.
\ref{fig_urban3_v913_6}-\ref{fig_urban3_v1003_79} for Geant4
versions 9.1 to 10.0.
Plots concerning Geant4 10.1 are omitted, as they look very similar to
those produced with version 10.0.
In these figures and the following ones the labels associated with experimental data
encode the name of the first author and the publication year of the reference
they derive from, thus allowing the tracebility of the experimental data
points appearing in the plots.
In all figures, experimental errors are plotted whenever they are documented in
the associated papers; in some cases they may be smaller than the marker size. 
The statistical uncertainties of simulated data are not drawn for better clarity of the 
plots; in most cases they are smaller than the marker size.

As a qualitative appraisal, one can observe that none of the three configuration
variants is capable of describing electron backscattering accurately below a few
tens of keV.
The simulations adopting the Urban and UrbanB multiple
scattering configurations exhibit different characteristics of compatibility
with experimental data along the evolution of Geant4 from version 9.1 to 10.1,
while simulations adopting the UrbanBRF options appear to behave
consistently over all the examined versions.

\begin{figure} 
\centerline{\includegraphics[angle=0,width=8.5cm]{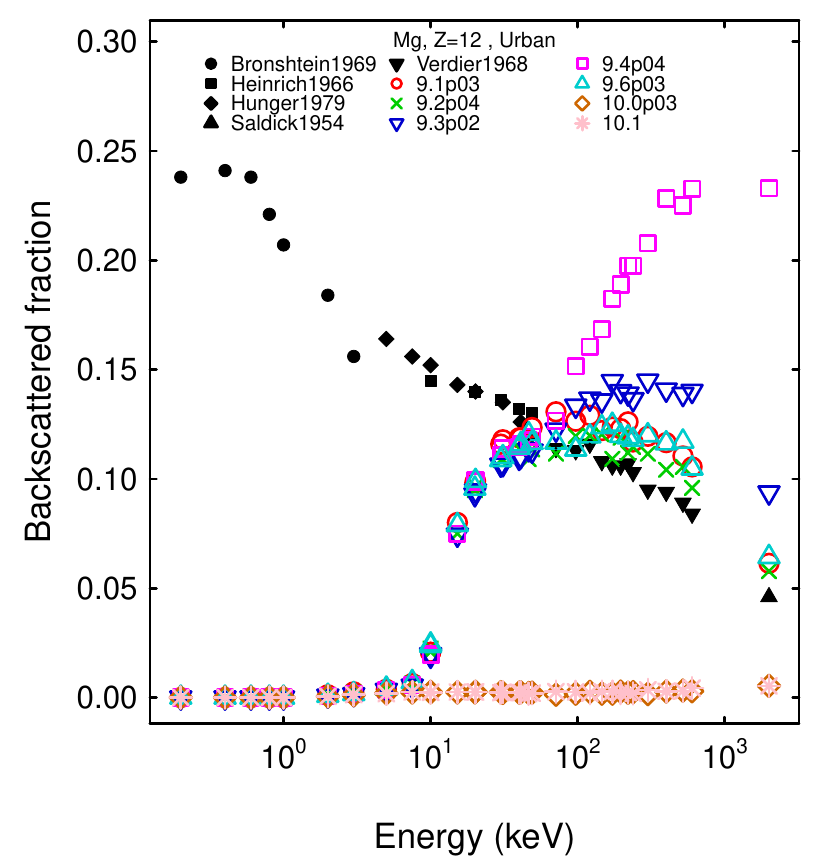}}
\caption{Fraction of electrons backscattered from a magnesium target as a function
of the electron beam energy: experimental data (black and grey filled symbols)
and simulation results (empty symbols) with the Urban multiple scattering model, 
complemented by user defined step limitation, in
Geant4 version 9.1 (red circles), 9.2 (green crosses), 9.3 (blue upside down triangles), 
9.4 (magenta squares), 9.6 (turquoise triangles), 10.0 (brown diamonds) and 10.1 (pink asterisks).}
\label{fig_Urban_12}
\end{figure}

\begin{figure} 
\centerline{\includegraphics[angle=0,width=8.5cm]{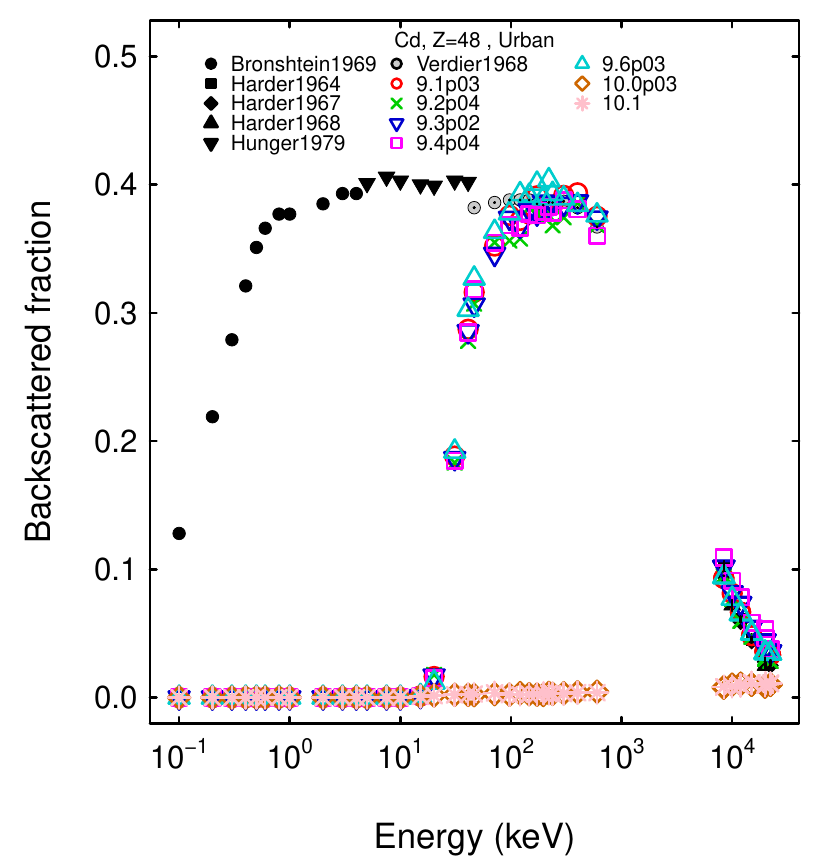}}
\caption{Fraction of electrons backscattered from a cadmium target as a function
of the electron beam energy: experimental data (black and grey filled symbols)
and simulation results (empty symbols) with the Urban multiple scattering model, 
complemented by user defined step limitation, in
Geant4 version 9.1 (red circles), 9.2 (green crosses), 9.3 (blue upside down triangles), 
9.4 (magenta squares), 9.6 (turquoise triangles), 10.0 (brown diamonds) and 10.1 (pink asterisks).}
\label{fig_Urban_48}
\end{figure}

For energies above a few tens of keV the evolution of the
compatibility with experiment of simulations using the Urban configuration seems
to follow a similar pattern to that observed in the validation of Geant4
simulation of deposited energy in \cite{tns_sandia2013} over Geant4 versions 9.1 to 9.6.
Visibly different results are obtained with Geant4 version 10.0, which manifestly
worsens the capability of the simulation to reproduce experimental data.
This behaviour persists in Geant4 10.1.
A qualitative illustration of the evolution of simulations using the Urban
configuration is summarized in Figs. \ref{fig_Urban_12} and
\ref{fig_Urban_48}.
These empirical observations are quantitatively supported by the results of
goodness-of-fit tests in Table \ref{tab_eff}.
The highest compatibility with experiment is achieved with the Urban 
multiple scattering configuration in the earlier Geant4 versions among
those evaluated in this paper.

The evolution of the capability of the Urban multiple scattering
configuration to produce backscattering simulations compatible with experiment
is quantified in Table \ref{tab_mcnemar_urban}: the second column reports the
p-values of McNemar's test comparing the compatibility with experiment of the
Urban configuration in Geant4 9.1, which according to Table
\ref{tab_eff} produces the highest efficiency, with later versions.
The hypothesis of equivalent compatibility with experiment is rejected with 0.01
significance for the Urban configuration in Geant4 versions 9.4, 10.0 and 10.1.
These results concern the energy range above 100~keV; the statistical comparison
of lower energy results is less relevant due to the reduced
ability of the Urban multiple scattering configuration to reproduce experimental
measurements below a few tens of keV.

\begin{table}[htbp]
  \centering
\caption{P-values of McNemar exact test comparing the compatibility with
experiment of the Urban configuration in Geant4 9.1 with later Geant4 versions
and with variants of the Urban configuration}
    \begin{tabular}{rccc}
    \hline
    Geant4 & \multicolumn{3}{c}{Configuration} \\
  
    Version & Urban & UrbanB & UrbanBRF \\
   \hline   
    9.1 	&       		& 1.000 		& $<$0.001 \\
    9.2 	& 1.000 		& 1.000 		& $<$0.001 \\
    9.3 	& 0.453 		& $<$0.001 	& $<$0.001 \\
    9.4 	& 0.002 		& $<$0.001 	& $<$0.001 \\
    9.6 	& 0.070 		& $<$0.001 	& $<$0.001 \\
    10.0 	& $<$0.001 	& $<$0.001 	& $<$0.001 \\
    10.1 	& $<$0.001 	& $<$0.001 	& $<$0.001 \\
    \hline
    \end{tabular}%
  \label{tab_mcnemar_urban}%
\end{table}%

The backscattering fraction resulting from the UrbanBRF options appears
to be very small in all test cases and differs significantly from 
experimental data.
In depth inspection of the results shows that the fraction of detected
electrons mainly consists of secondary particles produced by ionization of 
target atoms.
These qualitative observations are confirmed by the results of goodness-of-fit
tests listed in Table \ref{tab_eff}: the number of test cases that fail to
reject the hypothesis of compatibility between experimental and simulated
backscattering fractions is very small for the UrbanBRF multiple
scattering configuration at all energy intervals.

The UrbanB configuration produces identical results to the Urban one in
Geant4 versions 9.1 and 9.2, while it approaches the behaviour of the UrbanBRF
configuration in later Geant4 versions.
Both configurations apply the \textit{DistanceToBoundary} step limit type in the
multiple scattering algorithm, while they differ in the textit{RangeFactor}
setting, which is not modified with respect to its default value in the UrbanB
configuration and is assigned the value of 0.01 in the UrbanBRF configuration. 
The observed evolution in their compatibility with experiment hints at an
effect of the multiple scattering implementation related to this parameter.

Table \ref{tab_mcnemar_urban} also shows the p-values resulting from the
comparison of the compatibility with experiment achieved with the Urban
configuration in Geant4 9.1 and the UrbanB and UrbanBRF variants in the same
Geant4 version and later.
The hypothesis of equivalent compatibility with experiment is rejected in all
cases, with the exception of the UrbanB configuration in Geant4 9.1 and 9.2.

From this analysis one concludes that the multiple scattering settings
characterizing the UrbanBRF configuration produce significantly different
compatibility with experiment with respect to the Urban one, which achieves the
best compatibility with backscattering measurements in the earlier Geant4
versions evaluated in this paper.
It is worthwhile to note that the multiple scattering settings of the UrbanBRF
configuration reproduce those implemented in the recommended 
G4EmStandardPhysics\_option3 PhysicsConstructor, which is claimed in
\cite{g4appldevguide} to enable high accuracy simulation.

A similar categorical analysis estimates the statistical significance of the
difference in compatibility with experiment between simulations based on the
Urban configuration in Geant4 9.1 and on multiple or single scattering
models other than the Urban model in the latest Geant4 versions.
The results are summarized in Table \ref{tab_urban91_range8} regarding energies
above 100~keV.
In the energy range between 20 and 100~keV the hypothesis of equivalence is
rejected in all test cases, while below 20~keV it is rejected only in test cases
involving the Coulomb and WentzelBRF configurations.
Given the low efficiency of the Urban model below 20 keV observed in Table
\ref{tab_eff}, the test cases where the hypothesis of equivalence is not
rejected reflect the equivalent incapability of Geant4 multiple scattering
models to describe backscattering accurately at the lowest energies.

\tabcolsep=4pt
\begin{table}[htbp]
  \centering
  \caption{P-values resulting from the comparison of compatibility with experiment
   between simulations using Urban multiple scattering in Geant4 9.1 and simulations using 
   other unrelated configurations in
   versions 9.6, 10.0 and 10.1, concerning electron energy above 100 keV}
    \begin{tabular}{clcccc}
    \hline
    Geant4 		& Physics  			& Fisher & Pearson $\chi^2$ & Barnard 	& Boschloo\\
    Version       	& Configuration   	& Test   & Test                   & Test 		& Test\\
    \hline
    {\multirow{10}[0]{*}{9.6}} & Coulomb &  1     & 1     & 1 & 1\\
    & GSBRF  & 0.026 & 0.016 & 0.019 & 0.19 \\
    & WentzelBRF  & 1     & 1     & 1  & 1\\
    & WentzelBRFP  & $<$0.0001 & $<$0.0001 & $<$0.0001 & $<$0.0001 \\
    & EmLivermore  & $<$0.0001 &  & $<$0.0001 & $<$0.0001 \\
    & EmStd  & $<$0.0001 & $<$0.0001 & $<$0.0001 & $<$0.0001 \\
    & EmOpt1  & $<$0.0001 & $<$0.0001 & $<$0.0001 & $<$0.0001 \\
    & EmOpt2  & $<$0.0001 & $<$0.0001 & $<$0.0001 & $<$0.0001 \\
    & EmOpt3  & $<$0.0001 &  & $<$0.0001 & $<$0.0001 \\
    & EmOpt4  & $<$0.0001 & & $<$0.0001 & $<$0.0001 \\
\hline
     {\multirow{10}[0]{*}{10.0}} & Coulomb  & 1     & 0.815 & 0.889 & 1\\
    & GSBRF  & 0.026 & 0.016 & 0.019 & 0.19\\
    & WentzelBRF  & 1     & 0.815 & 0.889 & 1\\
    & WentzelBRFP  & $<$0.0001 &  & $<$0.0001 & $<$0.0001 \\
    & EmLivermore  & $<$0.0001 &  & $<$0.0001 & $<$0.0001 \\
    & EmStd  & $<$0.0001 & $<$0.0001 & $<$0.0001 & $<$0.0001 \\
    & EmOpt1  & $<$0.0001 & $<$0.0001 & $<$0.0001 & $<$0.0001 \\
    & EmOpt2  & $<$0.0001 & $<$0.0001 & $<$0.0001 & $<$0.0001 \\
    & EmOpt3  & $<$0.0001 &  & $<$0.0001 & $<$0.0001 \\
    & EmOpt4  & $<$0.0001 & $<$0.0001 & $<$0.0001 & $<$0.0001 \\
    \hline
    {\multirow{12}[0]{*}{10.1}} 	&Coulomb &  $<$0.0001	&  			& $<$0.0001 		& $<$0.0001\\
   & CoulombP  & 1     & 1     & 1  & 1\\
    & GSBRF  						& $<$0.0001 	& $<$0.0001 	& $<$0.0001 		& $<$0.0001\\
    & WentzelBRF  					& $<$0.0001	& $<$0.0001 	& $<$0.0001 		& $<$0.0001\\
    & WentzelBRFP  				& 0.0001 		&  			& 0.0001 	& 0.0001 \\
    & EmLivermore  				& 0.0001 		&  			& 0.0001 	& 0.0001 \\
    & EmStd  						& $<$0.0001 	& 		 	& $<$0.0001 	& $<$0.0001 \\
    & EmOpt1  					& $<$0.0001 	& $<$0.0001 	& $<$0.0001 	& $<$0.0001 \\
    & EmOpt2  					& $<$0.0001 	& $<$0.0001 	& $<$0.0001 	& $<$0.0001 \\
    & EmOpt3  					& $<$0.0001 	&  			& $<$0.0001 	& $<$0.0001 \\
    & EmOpt4  					& $<$0.0001 	& 		 	& $<$0.0001 	& $<$0.0001 \\
    & EmWVI  					& $<$0.0001 	& 		 	& $<$0.0001 	& $<$0.0001 \\
    \hline
  \end{tabular}%
  \label{tab_urban91_range8}%
\end{table}%
\tabcolsep=6pt

\subsection{Geant4 Coulomb scattering model}

The simulation application instantiates Geant4 Coulomb scattering process and
its associated model for electron scattering through the respective class
constructors and uses them in their default configuration.

\begin{figure} 
\centerline{\includegraphics[angle=0,width=8.5cm]{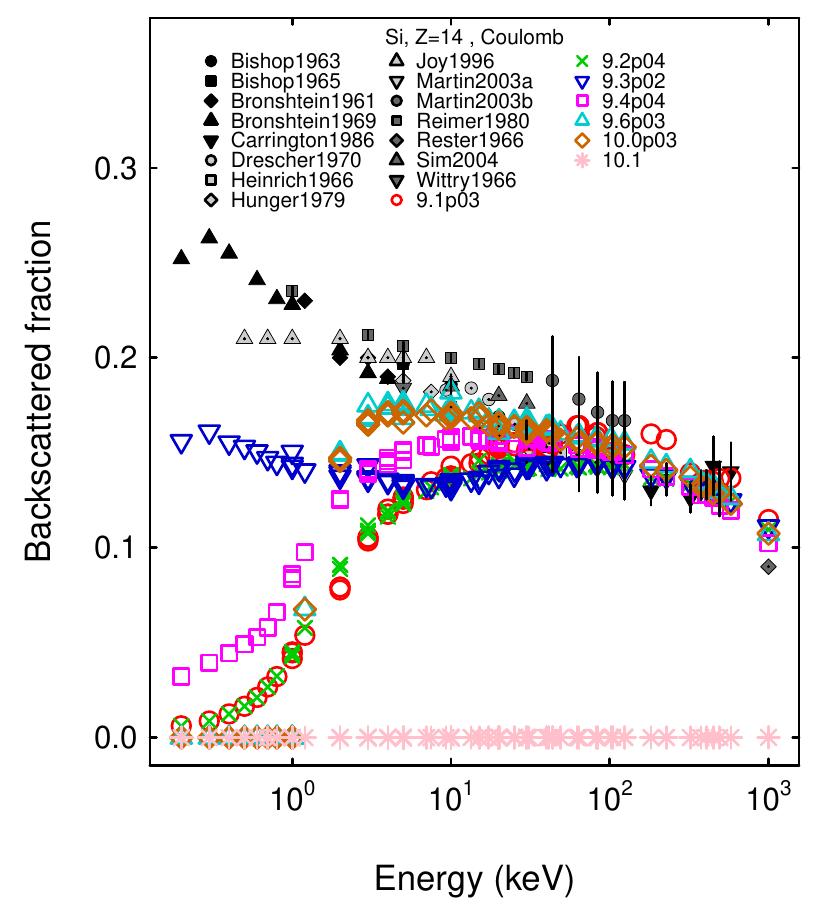}}
\caption{Fraction of electrons backscattered from a silicon target as a function
of the electron beam energy: experimental data (black and grey filled symbols)
and simulation results (empty symbols) with the Coulomb scattering model in
Geant4 version 9.1 (red circles), 9.2 (green crosses), 9.3 (blue upside down triangles), 
9.4 (magenta squares), 9.6 (turquoise triangles), 10.0 (brown diamonds) and 10.1 (pink asterisks). }
\label{fig_Coulomb_14}
\end{figure}

\begin{figure} 
\centerline{\includegraphics[angle=0,width=8.5cm]{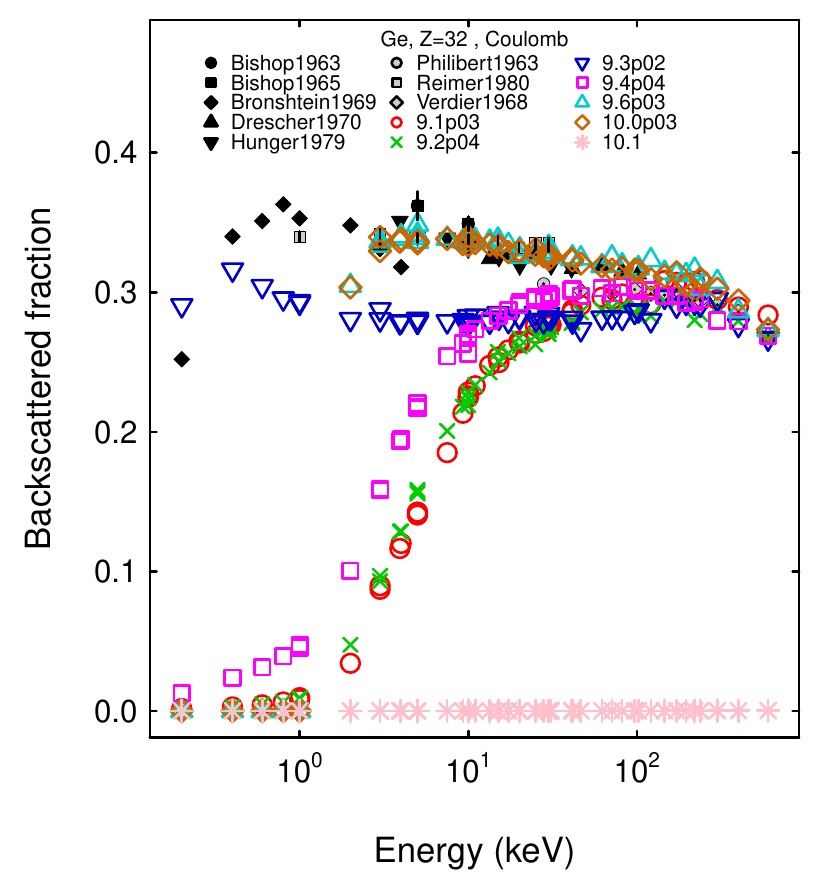}}
\caption{Fraction of electrons backscattered from a germanium target as a function
of the electron beam energy: experimental data (black and grey filled symbols)
and simulation results (empty symbols) with the Coulomb scattering model in
Geant4 version 9.1 (red circles), 9.2 (green crosses), 9.3 (blue upside down triangles), 
9.4 (magenta squares), 9.6 (turquoise triangles), 10.0 (brown diamonds) and 10.1 (pink asterisks). }
\label{fig_Coulomb_32}
\end{figure}

\begin{figure} 
\centerline{\includegraphics[angle=0,width=8.5cm]{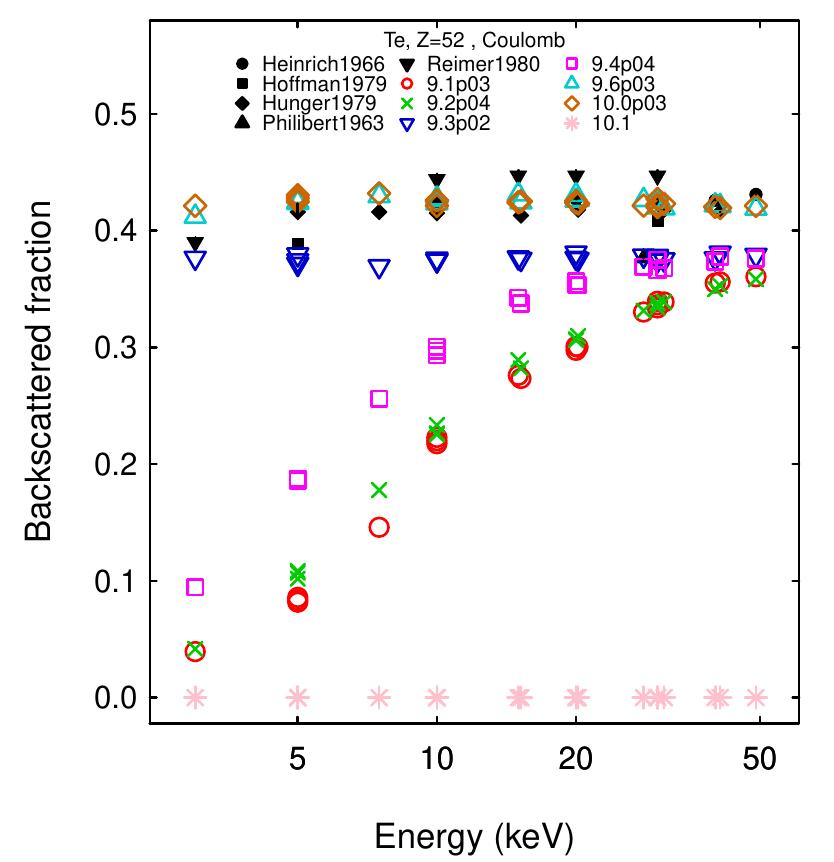}}
\caption{Fraction of electrons backscattered from a tellurium target as a function
of the electron beam energy: experimental data (black and grey filled symbols)
and simulation results (empty symbols) with the Coulomb scattering model in
Geant4 version 9.1 (red circles), 9.2 (green crosses), 9.3 (blue upside down triangles), 
9.4 (magenta squares), 9.6 (turquoise triangles), 10.0 (brown diamonds) and 10.1 (pink asterisks). }
\label{fig_Coulomb_52}
\end{figure}

A sample of the behaviour of this model with respect to experimental data is
illustrated in Figs.~\ref{fig_Coulomb_14}-\ref{fig_Coulomb_52}.
The single Coulomb scattering model appears to have evolved from Geant4 version 
9.1 to 10.0 towards better reproducing experimental measurements
especially at energies below a few tens of keV.
A degradation of the capability of the model in its default configuration 
is observed in Geant4 10.1.
Physical performance equivalent to that observed in Geant4 10.0
is achieved only by setting the ``$\theta$~limit'' parameter to zero
in the singleton G4EmParameters class: this simulation configuration 
is identified as CoulombP in this paper.
Despite our best efforts, we could not retrieve any documentation of the
semantic change of the construction of Coulomb scattering objects in Geant4
10.1, which is responsible for the modified behaviour in their default
configuration.

These qualitative considerations are confirmed by the results of
McNemar's test summarized in Table \ref{tab_mcnemar_coulomb}, which compares the
compatibility with experiment achieved by the Coulomb scattering confiburation
in Geant4 10.0 with the outcome of goodness of fit tests related to earlier
versions.

\begin{table}[htbp]
  \centering
  \caption{P-values of McNemar exact test comparing the compatibility with
experiment of the Coulomb configuration in Geant4 10.0 and in other Geant4 versions}
    \begin{tabular}{cccc}
    \hline
   Version & 1-20 keV & 20-100 keV & $\geq$100 keV \\
    \hline
     9.1 & $<0.001$ & $<0.001$ & $<0.001$ \\
 9.2 & $<0.001$ & $<0.001$ & $<0.001$ \\
     9.3 & $<0.001$ & $<0.001$ & $<0.001$ \\
    9.4 & $<0.001$ & 0.008 & $<0.001$ \\
     9.6 & 0.625 & 0.754 & 0.625 \\
    10.1 & $<0.001$ & $<0.001$ & $<0.001$ \\
    \hline
    \end{tabular}%
  \label{tab_mcnemar_coulomb}%
\end{table}%

The default Coulomb scattering configuration achieves the highest efficiency over all
energy ranges in Geant4 versions 9.6 and 10.0.
For convenience, the Coulomb configuration in Geant4 10.0 is defined
as the reference Coulomb configuration for further categorical data analysis.
Its compatibility with experimental data is compared with the achievements of other
multiple scattering configurations in Geant4 versions 9.6, 10.0 and 10.1
in Table \ref{tab_coulomb10.0}.
The hypothesis of equivalent compatibility with experiment is rejected for
all other models implemented in Geant4 10.1 at all energies, as
well as for the default Coulomb configuration of that version.
Equivalent behaviour is achieved with non-default settings in Geant4 10.1:
to the best of our efforts, we could not retrieve mention of the major semantic change 
of the default instantiation of single Coulomb scattering in Geant4 10.1, nor of 
the settings required to restore equivalent functionality to previous versions.
Regarding earlier Geant4 versions,
the hypothesis of equivalent compatibility with experiment is not rejected at
all energies with respect to the WentzelBRF configuration, which de facto
corresponds to enabling single scattering only in the default setting of
G4WentzelVIModel.
At higher energies the Urban configuration achieves equivalent compatibility
with experiment in Geant4 version 9.6, but not in Geant4 10.0.
Weak evidence of equivalent compatibility with experiment is reported in Table
\ref{tab_coulomb10.0} regarding the Goudsmit-Saunderson multiple scattering model
in versions 9.6 and 10.0; further
details are discussed in section \ref{sec_res_gs}.

The CoulombP configuration in Geant4 10.1 exhibits similar behaviour to the
reference Coulomb configuration of Geant4 10.0.
Its results are not shown in Figs. ~\ref{fig_Coulomb_14}-\ref{fig_Coulomb_52}
for better clarity of the plots, but they are included in Figs.
\ref{fig_wentzelcoulomb_v1010_6}-\ref{fig_wentzelcoulomb_v1010_79}.
Its associated efficiency is listed in Table \ref{tab_eff}.
McNemar's test confirms its compatibility with the reference Coulomb configuration 
of Geant4 10.0 with 0.01 significance.

The predefined G4EmStandardPhysics\_SS PhysicsConstructor, which instantiates
single scattering for various particle types, produces results statistically consistent
to the CoulombP configuration.
Since its functionality regarding electron scattering is equivalent to that of the 
CoulombP configuration, it will not be further discussed in the following sections.

\tabcolsep=2pt
\begin{table}[htbp]
  \centering
  \caption{P-values resulting from the comparison of compatibility with experiment
   between the single Coulomb scattering configuration in Geant4 10.0 and other unrelated configurations in
   versions 9.6, 10.0 and 10.1}
    \begin{tabular}{|c|clcccc|}
    \hline
    Energy 		& Geant4	&  Model 	& Fischer 	& Pearson $\chi^2$ 	& Barnard 	& Boschloo \\
    (kev) 		& Version		& 	& Test	& Test	& Test	& Test \\
    \hline
    {\multirow{9}[0]{*}{1-20}} & {\multirow{3}[0]{*}{9.6}} & Urban 			& $<0.001$ 	& 		& $<0.001$ 	& $<0.001$ \\
     										& & GSBRF  		& $<0.001$ 	& 		& $<0.001$ 	& $<0.001$ \\
   										& & WentzelBRF   	& 0.716 		& 0.627 	& 0.683 		& 0.666 \\
\cline{2-7}
 & {\multirow{3}[0]{*}{10.0}} 					& Urban  			& $<0.001$ 	&       	& $<0.001$ 	& $<0.001$ \\
    										&  & GSBRF  		& $<0.001$ 	&       	& $<0.001$ 	& $<0.001$ \\
  										& & WentzelBRF   	& 1 			& 0.903 	& 0.951 		& 1 \\
\cline{2-7}
     & {\multirow{3}[0]{*}{10.1}} & Urban 			& $<0.001$ &       	& $<0.001$ 	& $<0.001$ \\
     &  & GSBRF   								& $<0.001$ &       	& $<0.001$ 	& $<0.001$ \\
     & & WentzelBRF   							& $<0.001$ &   		& $<0.001$ 	& $<0.001$ \\
\hline    
{\multirow{9}[0]{*}{20-100}} & {\multirow{3}[0]{*}{9.6}} & Urban  & $<0.001$ & $<0.001$ & $<0.001$ & $<0.001$ \\
   & & GSBRF  & $<0.001$ &       & $<0.001$ & $<0.001$ \\
   &&  WentzelBRF   & 0.685 & 0.588 & 0.682 & 0.624 \\
\cline{2-7}
  & {\multirow{3}[0]{*}{10.0}} & Urban  & $<0.001$ &       & $<0.001$ & $<0.001$ \\
 & & GSBRF  & $<0.001$ &       & $<0.001$ & $<0.001$ \\
  & &  WentzelBRF   & 0.685 & 0.588 & 0.682 & 0.624 \\
\cline{2-7}
     & {\multirow{3}[0]{*}{10.1}} & Urban 			& $<0.001$ &       	& $<0.001$ 	& $<0.001$ \\
     &  & GSBRF   								& $<0.001$ &       	& $<0.001$ 	& $<0.001$ \\
     & & WentzelBRF   							& $<0.001$ &   		& $<0.001$ 	& $<0.001$ \\
\hline
{\multirow{30}[0]{*}{$>$100}} & {\multirow{10}[0]{*}{9.6}} & Urban  & 0.196 & 0.132 & 0.144 & 0.160 \\
     & &  GSBRF  & 0.014 & 0.008 & 0.009 & 0.009 \\
     & &  WentzelBRF   & 1 & 0.815 & 0.889 & 1 \\
     & &  WentzelBRFP   & $<0.001$ & $<0.001$ & $<0.001$ & $<0.001$ \\
     & &  EmLivermore   & $<0.001$ &       & $<0.001$ & $<0.001$ \\
     & &  EmStd   & $<0.001$ & $<0.001$ & $<0.001$ & $<0.001$ \\
     & &  EmOpt1   & $<0.001$ & $<0.001$ & $<0.001$ & $<0.001$ \\
     & &  EmOpt2   & $<0.001$ & $<0.001$ & $<0.001$ & $<0.001$ \\
     & &  EmOpt3   & $<0.001$ &       & $<0.001$ & $<0.001$ \\
     & &  EmOpt4   & $<0.001$ &       & $<0.001$ & $<0.001$ \\
\cline{2-7}
    & {\multirow{10}[0]{*}{10.0}} & Urban  & $<0.001$ & $<0.001$ & $<0.001$ & $<0.001$ \\
     & &  GSBRF  & 0.014 & 0.008 & 0.009 & 0.009 \\
     & &  WentzelBRF   & 1 & 1 & 1 & 1 \\
     & & WentzelBRFP   & $<0.001$ & $<0.001$ & $<0.001$ & $<0.001$ \\
     & & EmLivermore   & $<0.001$ &       & $<0.001$ & $<0.001$ \\
     & & EmStd   & $<0.001$ &       & $<0.001$ & $<0.001$ \\
     & & EmOpt1   & $<0.001$ & $<0.001$ & $<0.001$ & $<0.001$ \\
     & & EmOpt2   & $<0.001$ & $<0.001$ & $<0.001$ & $<0.001$ \\
     & & EmOpt3   & $<0.001$ &       & $<0.001$ & $<0.001$ \\
     & & EmOpt4   & $<0.001$ & $<0.001$ & $<0.001$ & $<0.001$ \\
  \cline{2-7}
     & {\multirow{10}[0]{*}{10.1}} & Urban  		& $<0.001$ 	&  		& $<0.001$ & $<0.001$ \\
     & & GSBRF   							& $<0.001$  	& $<0.001$ 	& $<0.001$  & $<0.001$  \\
     & & WentzelBRF   						& $<0.001$ 	& $<0.001$  	& $<0.001$  & $<0.001$  \\
     & & WentzelBRFP   					& $<0.001$ 	& $<0.001$ 	& $<0.001$ & $<0.001$ \\
     & & EmLivermore  						& $<0.001$ 	&       		& $<0.001$ & $<0.001$ \\
     & & EmStd    							& $<0.001$ 	&       		& $<0.001$ & $<0.001$ \\
     & & EmOpt1   						& $<0.001$ 	& $<0.001$ 	& $<0.001$ & $<0.001$ \\
     & & EmOpt2   						& $<0.001$ 	& $<0.001$ 	& $<0.001$ & $<0.001$ \\
     & & EmOpt3   						& $<0.001$ 	&       		& $<0.001$ & $<0.001$ \\
     & & EmOpt4   						& $<0.001$ 	&  			& $<0.001$ & $<0.001$ \\
     & & EmWVI	   						& $<0.001$ 	& $<0.001$ 	& $<0.001$ & $<0.001$ \\
 \hline
    \end{tabular}%
  \label{tab_coulomb10.0}%
\end{table}%
\tabcolsep=6pt

\subsection{Geant4 Goudsmit-Saunderson model}
\label{sec_res_gs}

\begin{figure} 
\centerline{\includegraphics[angle=0,width=8.5cm]{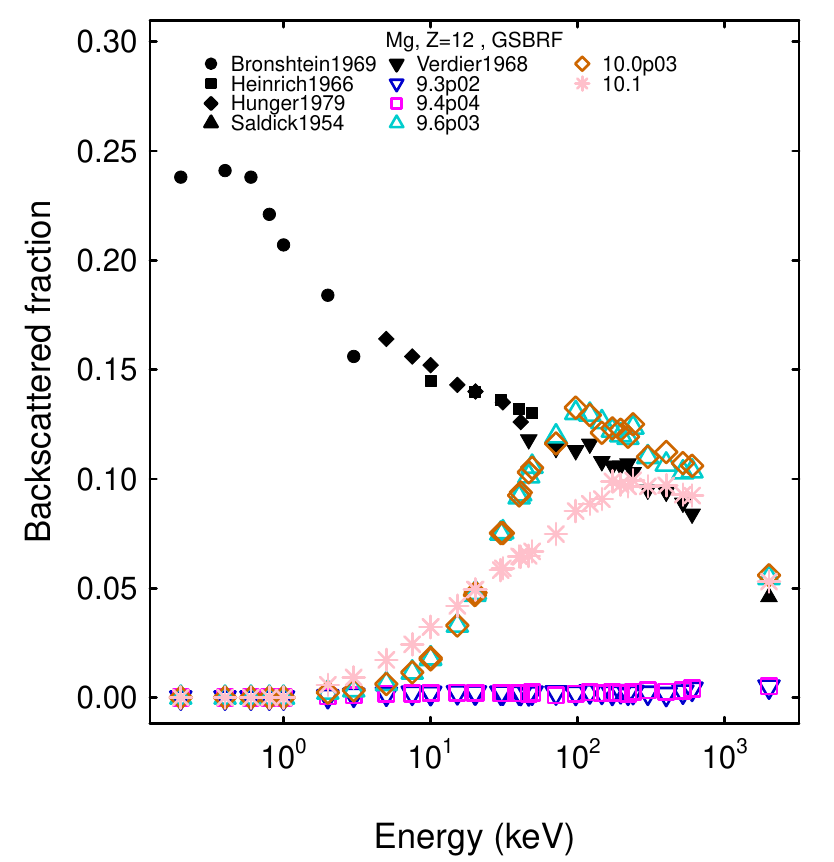}}
\caption{Fraction of electrons backscattered from a magnesium target as a function
of the electron beam energy: experimental data (black and grey filled symbols)
and simulation results (empty symbols) with the Goudsmit-Saunderson model in
Geant4 version 9.3 (blue upside down triangles), 9.4 (magenta squares), 
9.6 (turquoise triangles), 10.0 (brown diamonds) and 10.1 (pink asterisks). }
\label{fig_GSBRF_12}
\end{figure}

\begin{figure} 
\centerline{\includegraphics[angle=0,width=8.5cm]{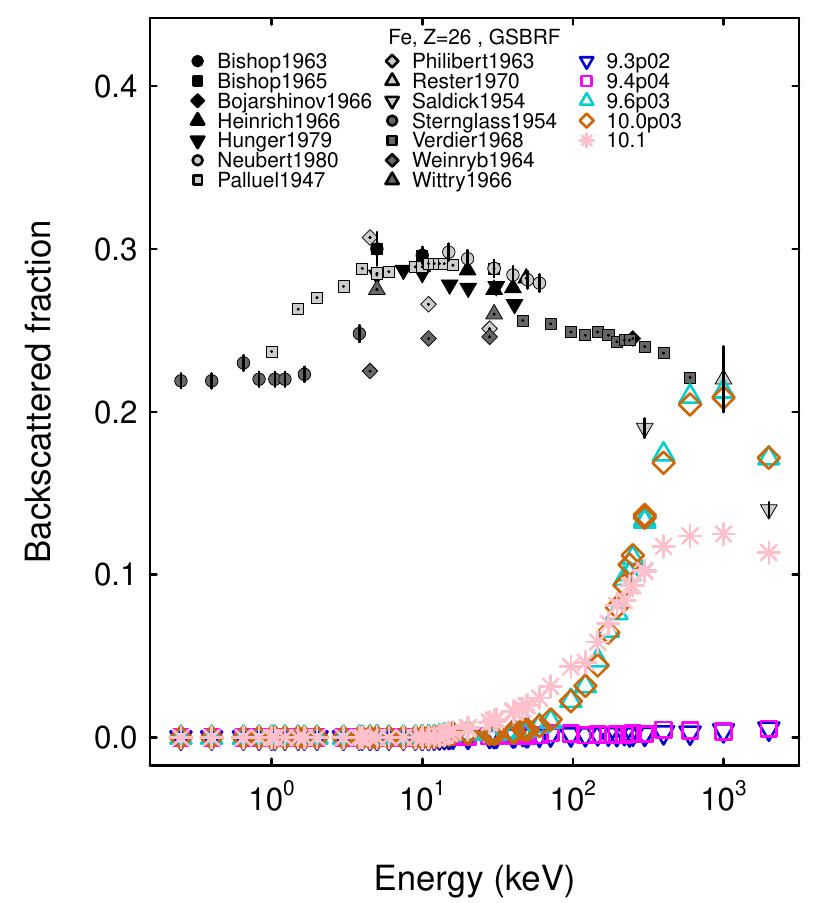}}
\caption{Fraction of electrons backscattered from an iron target as a function
of the electron beam energy: experimental data (black and grey filled symbols)
and simulation results (empty symbols) with the Goudsmit-Saunderson model in
Geant4 version 9.3 (blue upside down triangles), 9.4 (magenta squares), 
9.6 (turquoise triangles), 10.0 (brown diamonds) and 10.1 (pink asterisks). }
\label{fig_GSBRF_26}
\end{figure}

\begin{figure} 
\centerline{\includegraphics[angle=0,width=8.5cm]{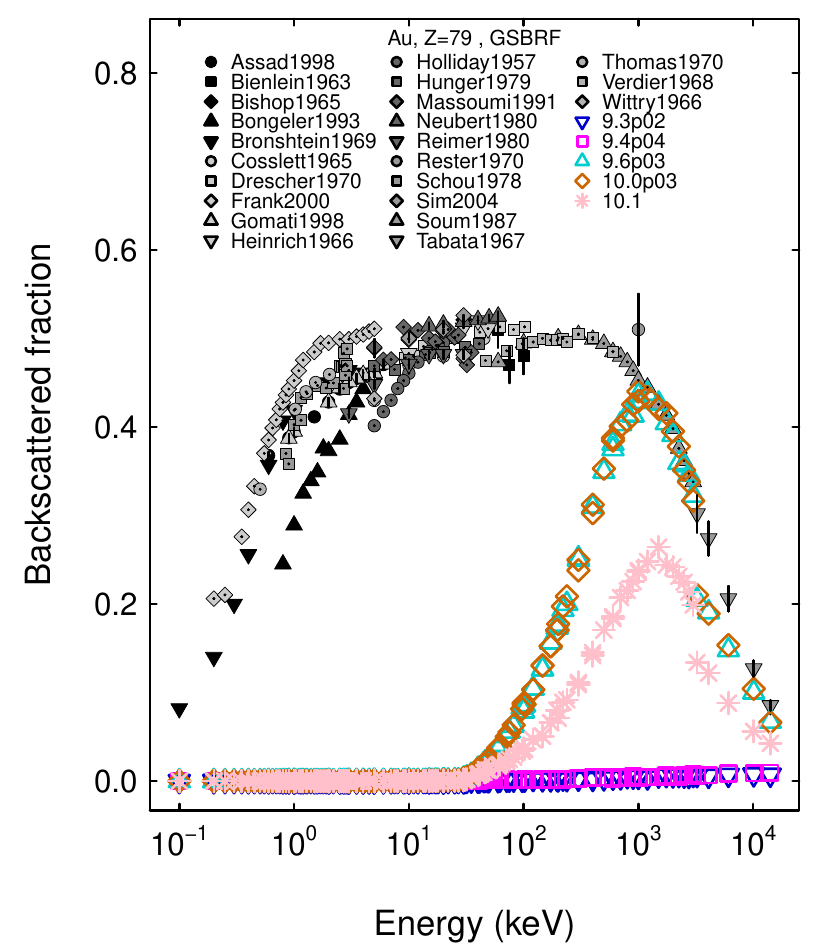}}
\caption{Fraction of electrons backscattered from a gold target as a
function of the electron beam energy: experimental data (black and grey filled
symbols) and simulation results (empty symbols) with the Goudsmit-Saunderson
model in Geant4 version 9.3 (blue upside down triangles), 9.4 (magenta squares),
9.6 (turquoise triangles), 10.0 (brown diamonds) and 10.1 (pink asterisks).. }
\label{fig_GSBRF_79}
\end{figure}

\begin{figure} 
\centerline{\includegraphics[angle=0,width=8.5cm]{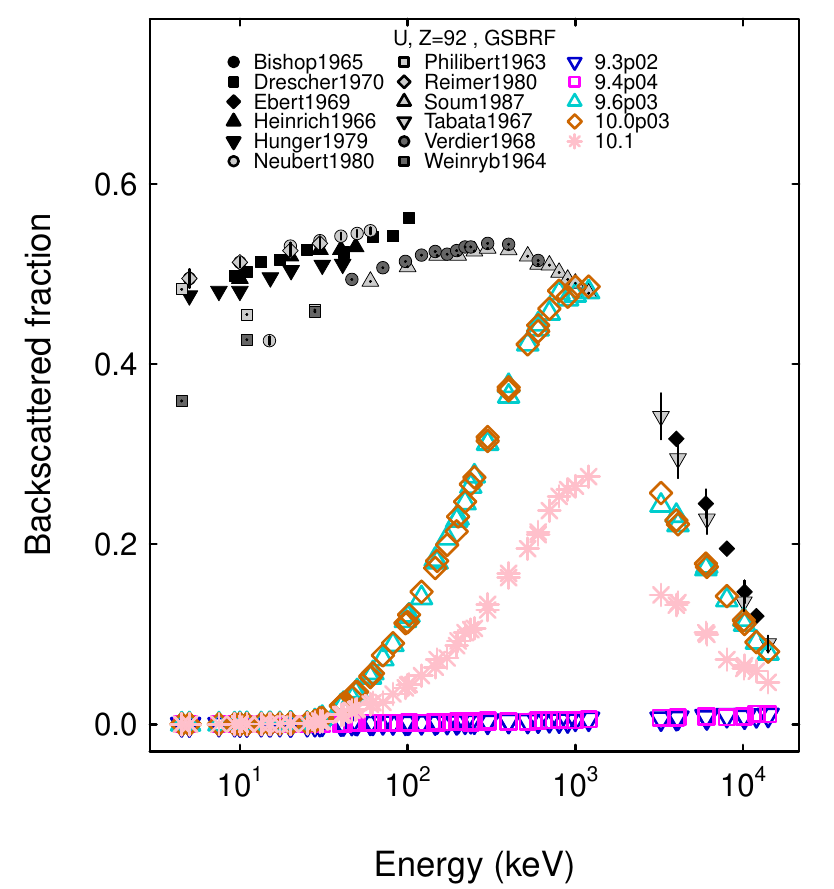}}
\caption{Fraction of electrons backscattered from an uranium target as a
function of the electron beam energy: experimental data (black and grey filled
symbols) and simulation results (empty symbols) with the Goudsmit-Saunderson
model in Geant4 version 9.3 (blue upside down triangles), 9.4 (magenta squares),
9.6 (turquoise triangles), 10.0 (brown diamonds) and 10.1 (pink asterisks).. }
\label{fig_GSBRF_92}
\end{figure}

The implementation of the Goudsmit-Saunderson multiple scattering model
appears to have evolved from its first release in Geant4 9.3 version to
the 10.1 version:
qualitatively, one can observe in Figs. \ref{fig_GSBRF_12}-\ref{fig_GSBRF_92} 
that at higher energies the evolution of this algorithm
has contributed to approach experimental backscattering values in
simulations based on Geant4 9.6 and 10.0, followed by an apparent 
deterioration of compatibility in Geant4 10.1.
These observations are confirmed by the statistical analysis of
compatibility with experiment summarized in Table \ref{tab_eff}.

The results of the tests reported in Table \ref{tab_urban91_range8} show
that the compatibility with experiment achieved with the Goudsmit-Saunderson
algorithm in Geant4 versions 9.6 and 10.0 is statistically equivalent to that
achieved with the Urban configuration in Geant4 version. 9.1 in the energy range
above 100 keV; nevertheless, equivalent compatibility with experiment with
respect to the Coulomb configuration of Geant4 10.0 in the same energy range is
assessed only by Fisher test, which is notoriously conservative, while more
powerful Z-pooled and Boschloo tests reject the hypothesis of equivalent
compatibility with experiment with 0.01 significance.
Comparisons at lower energies fail to establish equivalent compatibility
with experiment with respect to the 10.0 Coulomb scattering configuration, while they
are not physically relevant with respect to the 9.1 Urban configuration due to the
very low efficiency exhibited by both multiple scattering models in Table
\ref{tab_eff}.
The above mentioned comparisons with the Coulomb scattering configuration
of Geant4 10.0 are documented in Table \ref{tab_coulomb10.0}.

The evolution of the Goudsmit-Saunderson algorithm in Geant4 10.1 leads 
to incompatibility with both experimental data and the reference Urban and 
Coulomb scattering configurations.


\subsection{Geant4 WentzelVI Model}

\begin{figure} 
\centerline{\includegraphics[angle=0,width=8.5cm]{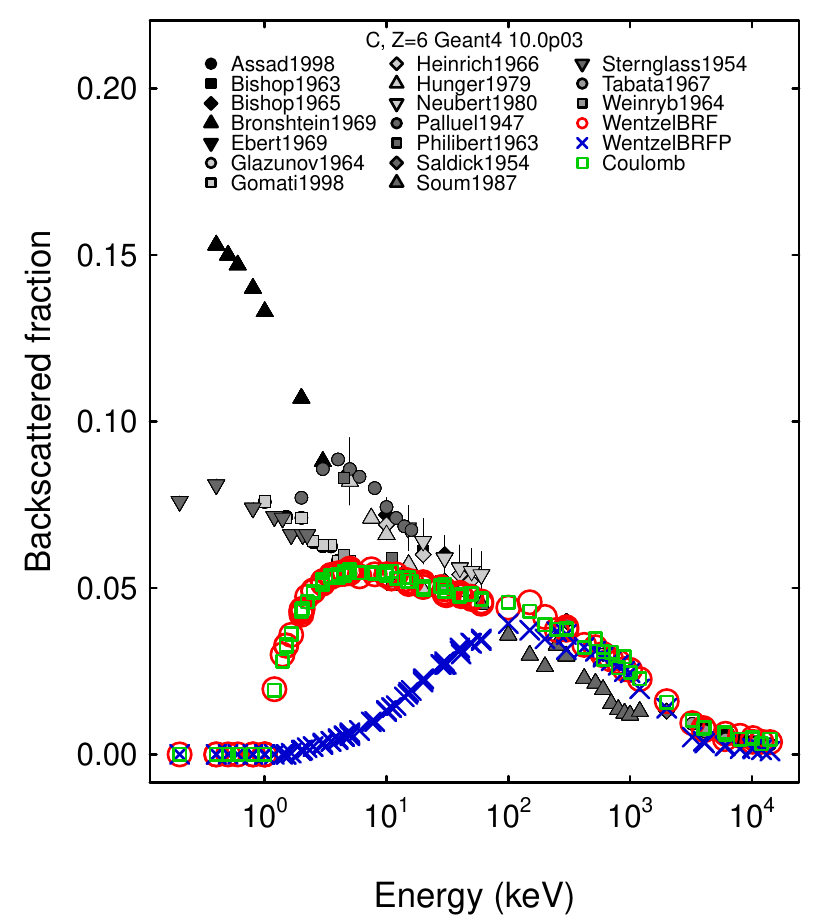}}
\caption{Fraction of electrons backscattered from a carbon target as a function
of the electron beam energy: experimental data (black and grey filled symbols)
and Geant4 10.0 simulation results with WentzelBRF (red empty circles) and WentzelBRFP (blue crosses)
multiple scattering settings and Coulomb single scattering model (green empty squares).}
\label{fig_wentzelcoulomb_v1003_6}
\end{figure}

\begin{figure} 
\centerline{\includegraphics[angle=0,width=8.5cm]{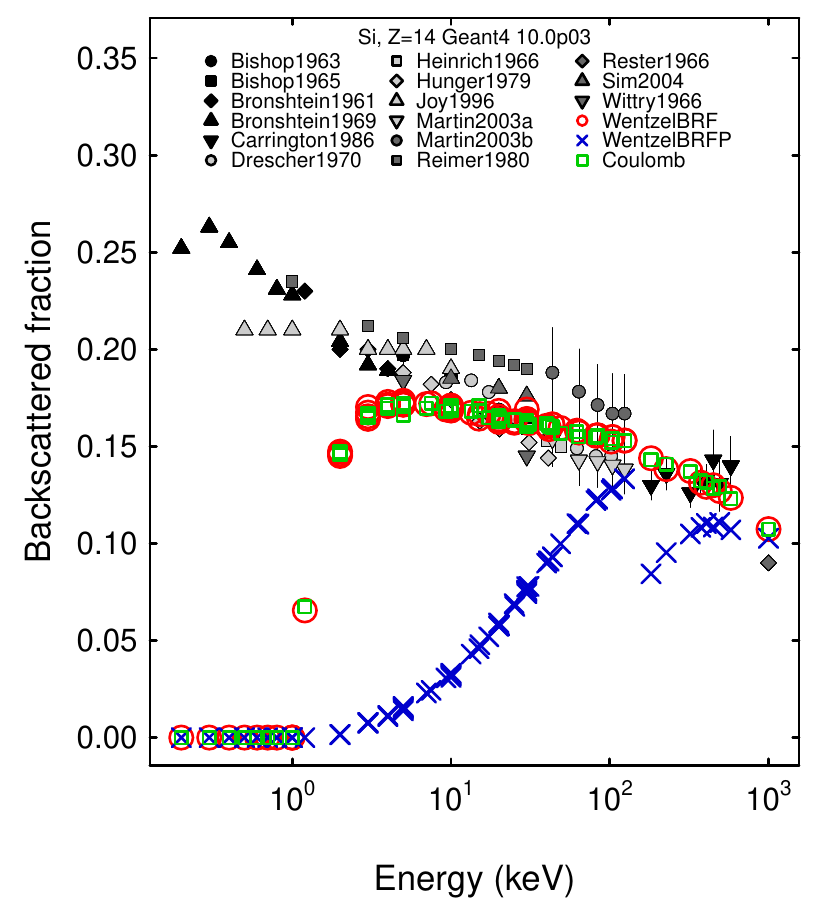}}
\caption{Fraction of electrons backscattered from a silicon target as a function
of the electron beam energy: experimental data (black and grey filled symbols)
and Geant4 10.0 simulation results with WentzelBRF (red empty circles) and WentzelBRFP (blue crosses)
multiple scattering settings and Coulomb single scattering model (green empty squares).}
\label{fig_wentzelcoulomb_v1003_14}
\end{figure}

\begin{figure} 
\centerline{\includegraphics[angle=0,width=8.5cm]{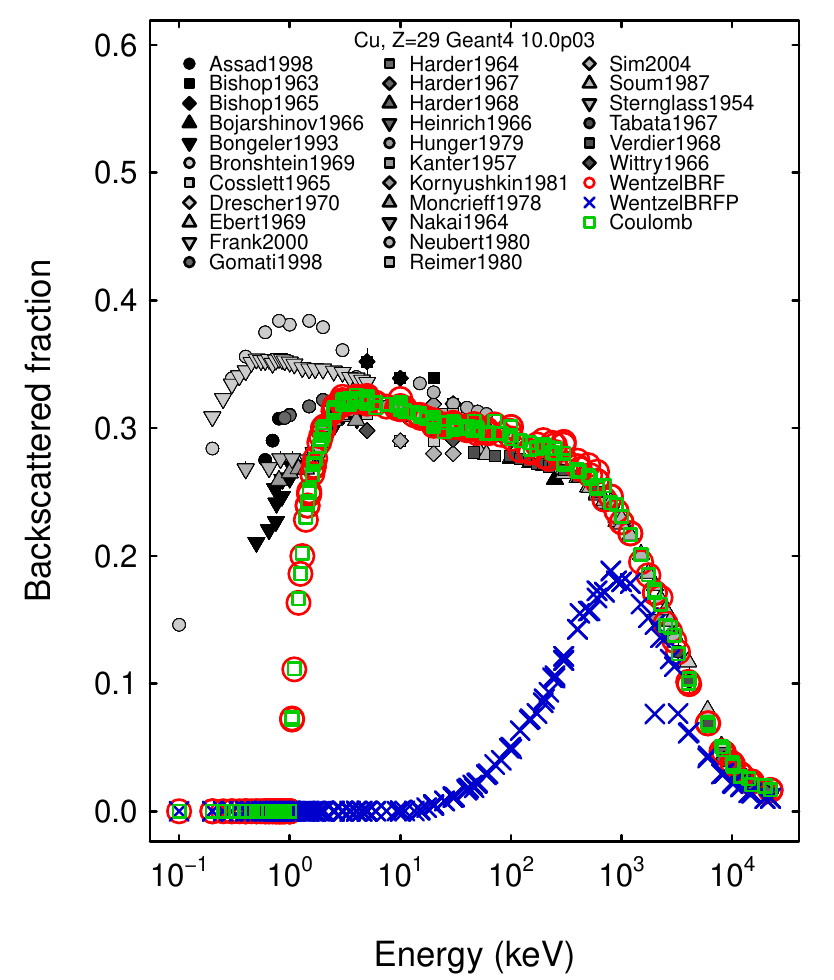}}
\caption{Fraction of electrons backscattered from a copper target as a function
of the electron beam energy: experimental data (black and grey filled symbols)
and Geant4 10.0 simulation results with WentzelBRF (red empty circles) and WentzelBRFP (blue crosses)
multiple scattering settings and Coulomb single scattering model (green empty squares).}
\label{fig_wentzelcoulomb_v1003_29}
\end{figure}

\begin{figure} 
\centerline{\includegraphics[angle=0,width=8.5cm]{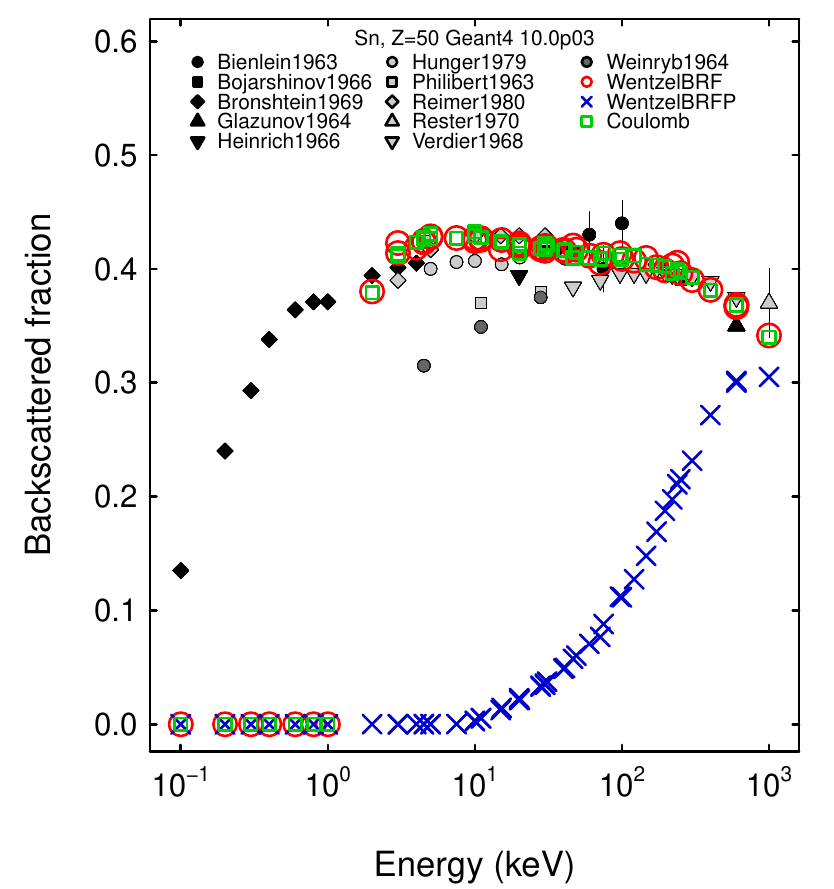}}
\caption{Fraction of electrons backscattered from a tin target as a function
of the electron beam energy: experimental data (black and grey filled symbols)
and Geant4 10.0 simulation results withWentzelBRF (red empty circles) and WentzelBRFP (blue crosses)
multiple scattering settings and Coulomb single scattering model (green empty squares).}
\label{fig_wentzelcoulomb_v1003_50}
\end{figure}

\begin{figure} 
\centerline{\includegraphics[angle=0,width=8.5cm]{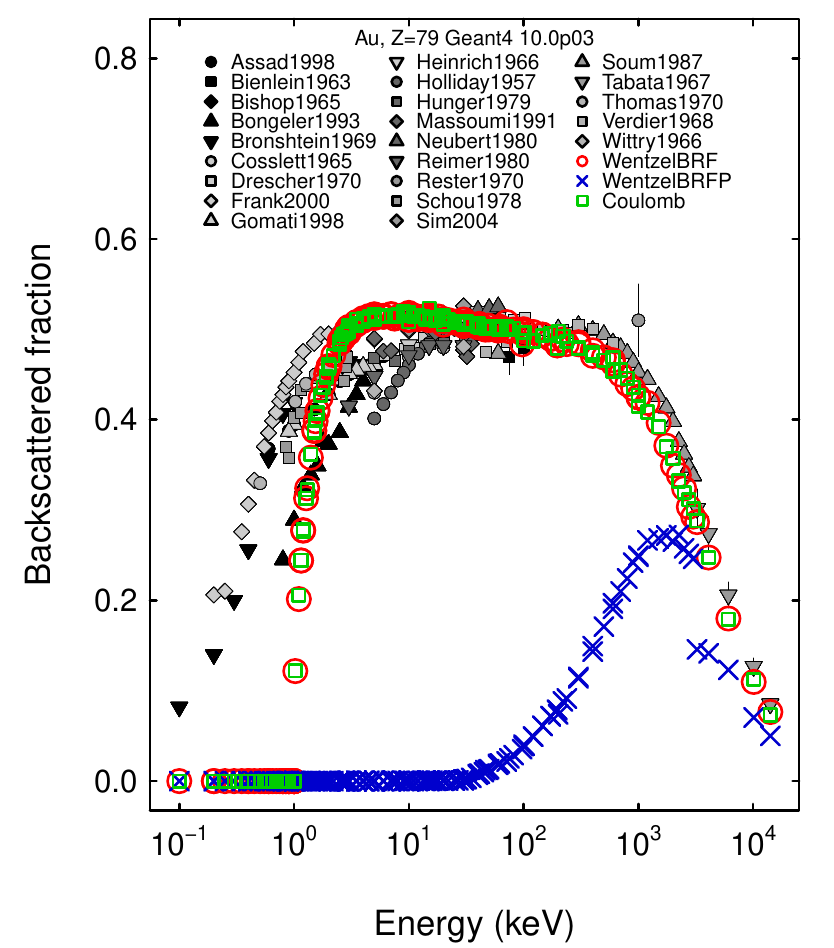}}
\caption{Fraction of electrons backscattered from a gold target as a function
of the electron beam energy: experimental data (black and grey filled symbols)
and Geant4 10.0 simulation results with WentzelBRF (red empty circles) and WentzelBRFP (blue crosses)
multiple scattering settings and Coulomb single scattering model (green empty squares).}
\label{fig_wentzelcoulomb_v1003_79}
\end{figure}

\begin{figure} 
\centerline{\includegraphics[angle=0,width=8.5cm]{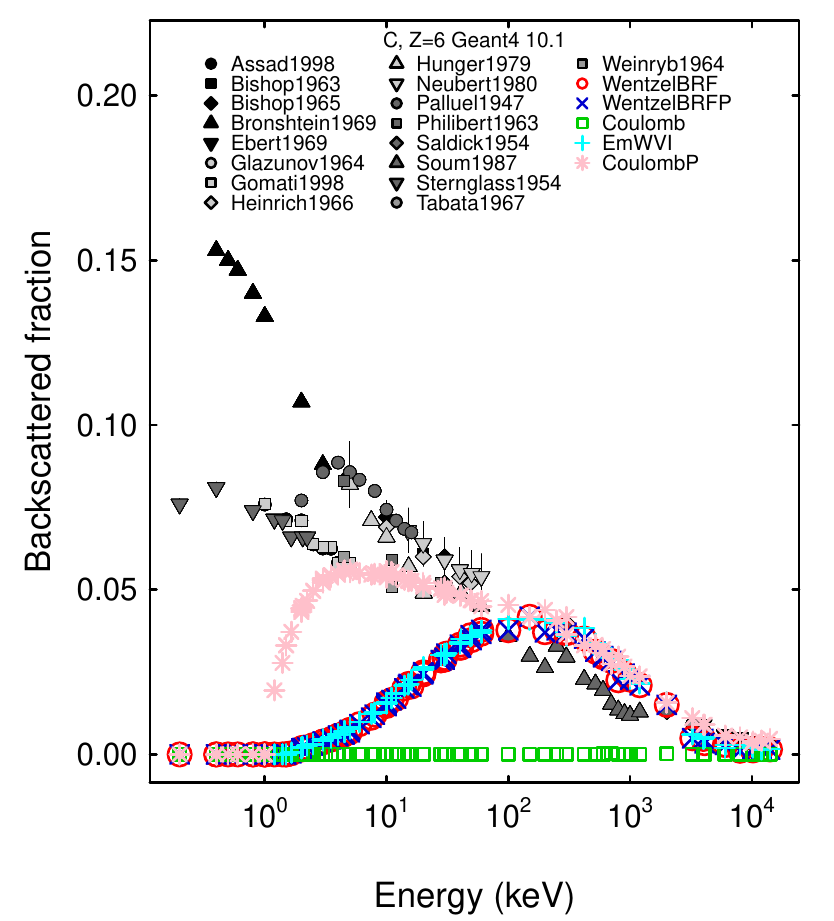}}
\caption{Fraction of electrons backscattered from a carbon target as a function
of the electron beam energy: experimental data (black and grey filled symbols)
and Geant4 10.1 simulation results with WentzelBRF (red empty circles), WentzelBRFP (blue crosses)
multiple scattering settings, Coulomb single scattering model in default configuration (green empty squares)
and with modified parameter settings (pink asterisks).}
\label{fig_wentzelcoulomb_v1010_6}
\end{figure}

\begin{figure} 
\centerline{\includegraphics[angle=0,width=8.5cm]{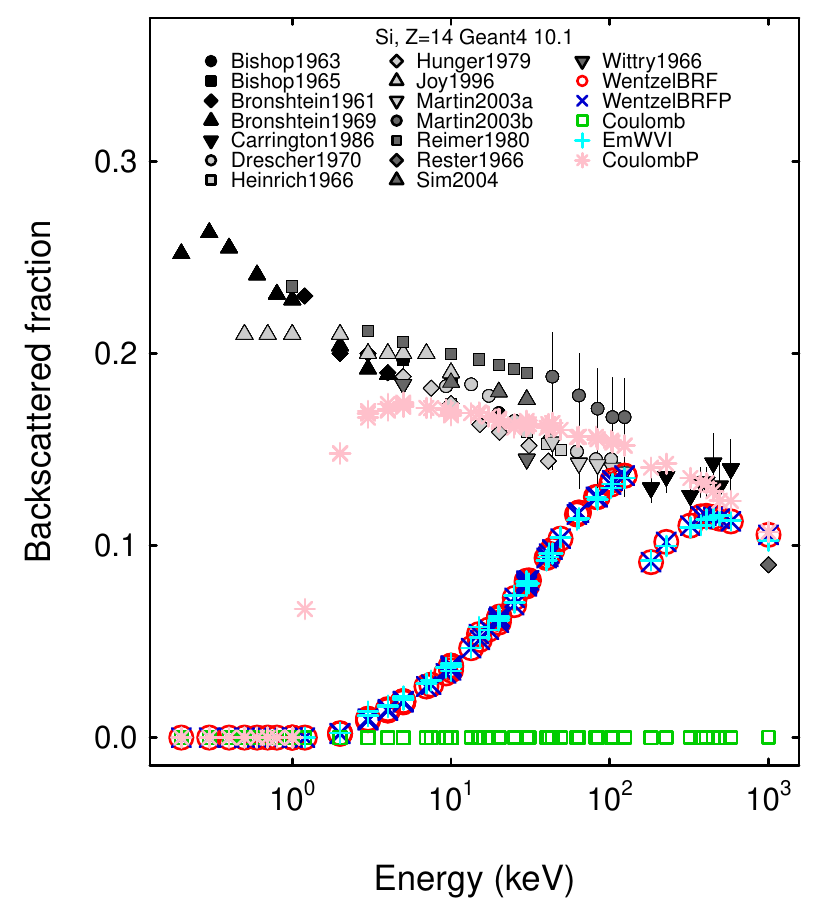}}
\caption{Fraction of electrons backscattered from a silicon target as a function
of the electron beam energy: experimental data (black and grey filled symbols)
and Geant4 10.1 simulation results with WentzelBRF (red empty circles), WentzelBRFP (blue crosses)
multiple scattering settings, Coulomb single scattering model in default configuration (green empty squares)
and with modified parameter settings (pink asterisks).}
\label{fig_wentzelcoulomb_v1010_14}
\end{figure}

\begin{figure} 
\centerline{\includegraphics[angle=0,width=8.5cm]{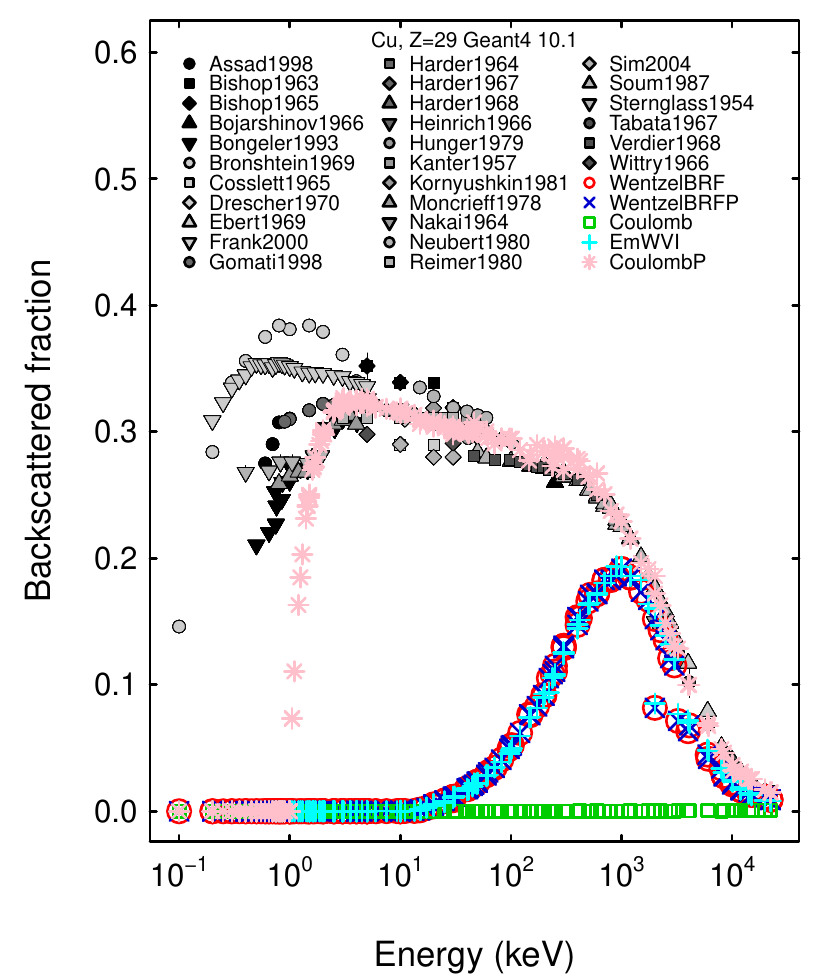}}
\caption{Fraction of electrons backscattered from a copper target as a function
of the electron beam energy: experimental data (black and grey filled symbols)
and Geant4 10.1 simulation results with WentzelBRF (red empty circles), WentzelBRFP (blue crosses)
multiple scattering settings, Coulomb single scattering model in default configuration (green empty squares)
and with modified parameter settings (pink asterisks).}
\label{fig_wentzelcoulomb_v1010_29}
\end{figure}

\begin{figure} 
\centerline{\includegraphics[angle=0,width=8.5cm]{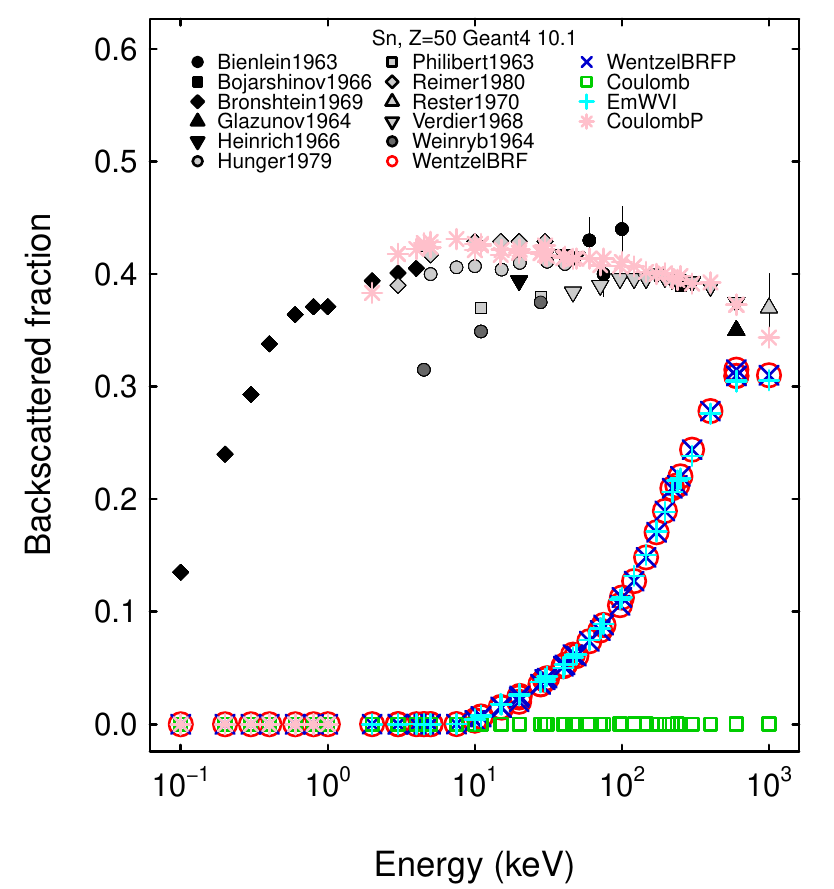}}
\caption{Fraction of electrons backscattered from a tin target as a function
of the electron beam energy: experimental data (black and grey filled symbols)
and Geant4 10.1 simulation results WentzelBRF (red empty circles), WentzelBRFP (blue crosses)
multiple scattering settings, Coulomb single scattering model in default configuration (green empty squares)
and with modified parameter settings (pink asterisks).}
\label{fig_wentzelcoulomb_v1010_50}
\end{figure}

\begin{figure} 
\centerline{\includegraphics[angle=0,width=8.5cm]{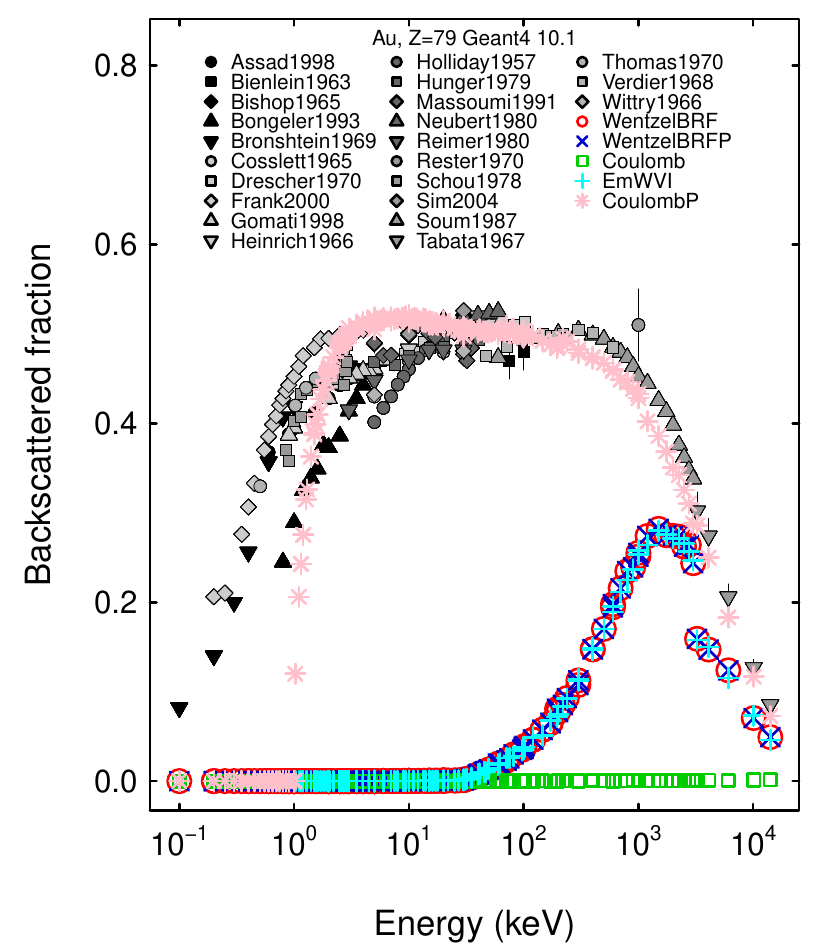}}
\caption{Fraction of electrons backscattered from a gold target as a function
of the electron beam energy: experimental data (black and grey filled symbols)
and Geant4 10.1 simulation results WentzelBRF (red empty circles), WentzelBRFP (blue crosses)
multiple scattering settings, Coulomb single scattering model in default configuration (green empty squares)
and with modified parameter settings (pink asterisks).}
\label{fig_wentzelcoulomb_v1010_79}
\end{figure}

The Geant4 WentzelVI model is used in two configurations evaluated in
this paper: WentzelBRF and WentzelBRFP, which use it in its default setting and with
a polar angle threshold, respectively.
The latter reflects the setting recommended in the electromagnetic
PhysicsConstructors that instantiate the WentzelVI model, such as
G4EmStandardPhysics and G4EmLivermorePhysics.
Their performance in the context of Geant4 10.0 and 10.1 is illustrated in Figs. 
\ref{fig_wentzelcoulomb_v1003_6}-\ref{fig_wentzelcoulomb_v1003_79}
and \ref{fig_wentzelcoulomb_v1010_6}-\ref{fig_wentzelcoulomb_v1010_79}, respectively.
In addition, the predefined G4EmStandardPhysics\_WVI PhysicsConstructor first released
in Geant4 10.1 configures multiple scattering with the WentzelVI model 
for various particle types, including electrons.

The WentzelBRF configuration appears to behave similarly to Geant4 single
Coulomb scattering configuration according to the statistical results collected
in Tables \ref{tab_eff} and \ref{tab_coulomb10.0} up to version 10; this similarity also concerns
its evolution over the Geant4 versions examined in this paper.
It achieves its highest efficiency in Geant4 10.0.

Large differences, especially visible at lower energies, are observed between
the results produced by the WentzelBRF and WentzelBRFP configurations
in Geant4 versions up to 10.0.
The two configurations behave similarly in Geant4 10.1.

The statistical analysis over the earlier Geant4 versions documented in Tables \ref{tab_eff},
\ref{tab_urban91_range8} and \ref{tab_coulomb10.0} confirms that, while the
WentzelBRF configuration produces statistically equivalent compatibility with
experiment with respect to the most efficient configurations (Urban in Geant4
9.1 and Coulomb in Geant4 10.0), the hypothesis of equivalent compatibility with
experiment is rejected for the WentzelBRFP one.

From these results one can infer that in the experimental scenarios
evaluated in this paper the recommended polar angle setting contributes
to worsening the reproduction of measured backscattering with respect
to the default settings.

The WentzelBRF and WentzelBRFP configurations produce statistically 
equivalent results in Geant4 10.1: McNemar's test fails to reject the hypothesis
of compatibility between the two categories of goodness-of-fit testing results
with 0.01 significance.

The behaviour of the predefined G4Em\-Standard\-Physics\_WVI Physics\-Constructor
is equivalent to that of the WentzelBRF and WentzelBRFP configurations
in Geant4 10.1: this conclusion is assessed by McNemar's test with 0.01
significance.



\subsection{Geant4 recommended PhysicsConstructors}

As stated in section \ref{sec_prod}, the evaluation of 
Geant4 recommended PhysicsConstructors regarding their 
simulation of electron backscattering concerns Geant4 versions
9.6, 10.0 and 10.1.

\begin{figure} 
\centerline{\includegraphics[angle=0,width=8.5cm]{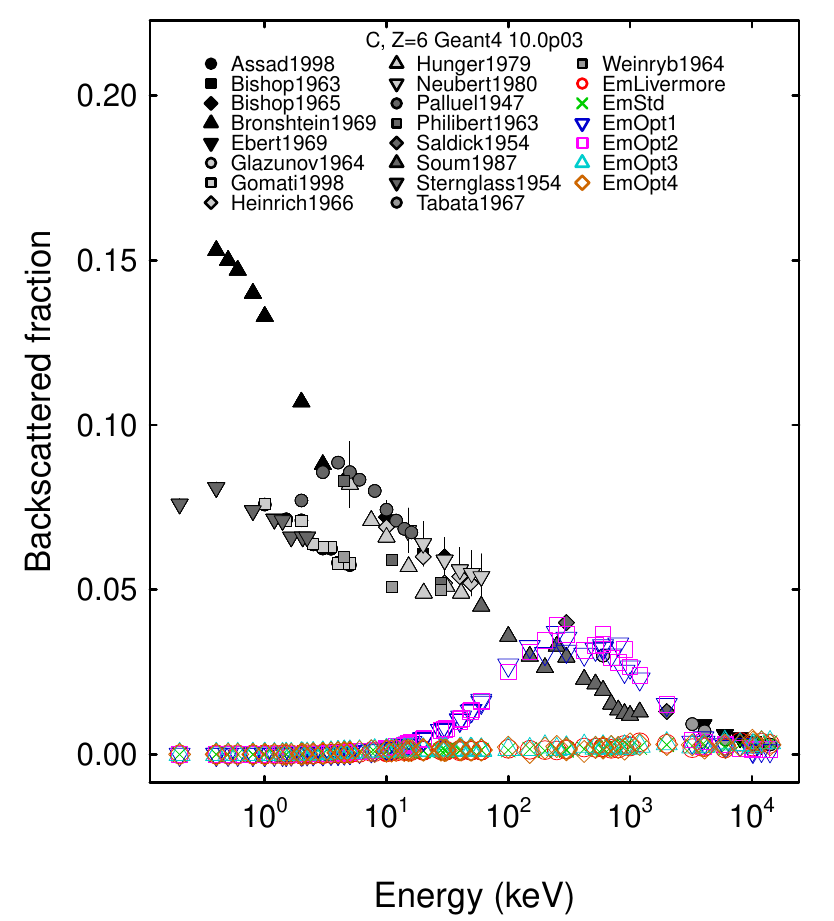}}
\caption{Fraction of electrons backscattered from a carbon target as a function
of the electron beam energy: experimental data (black and grey filled symbols)
and Geant4 10.0 simulation results with G4EmLivermorePhysics (red empty circles), 
G4EmStandardPhysics (green crosses),
G4EmStandardPhysics\_option1 (blue empty upside-down triangles),
G4EmStandardPhysics\_option2 (magenta empty squares),
G4EmStandardPhysics\_option3 (turquoise empty triangles)
and G4EmStandardPhysics\_option4 (brown empty diamonds)
PhysicsConstructors.}
\label{fig_physList_v1003_6}
\end{figure}

\begin{figure} 
\centerline{\includegraphics[angle=0,width=8.5cm]{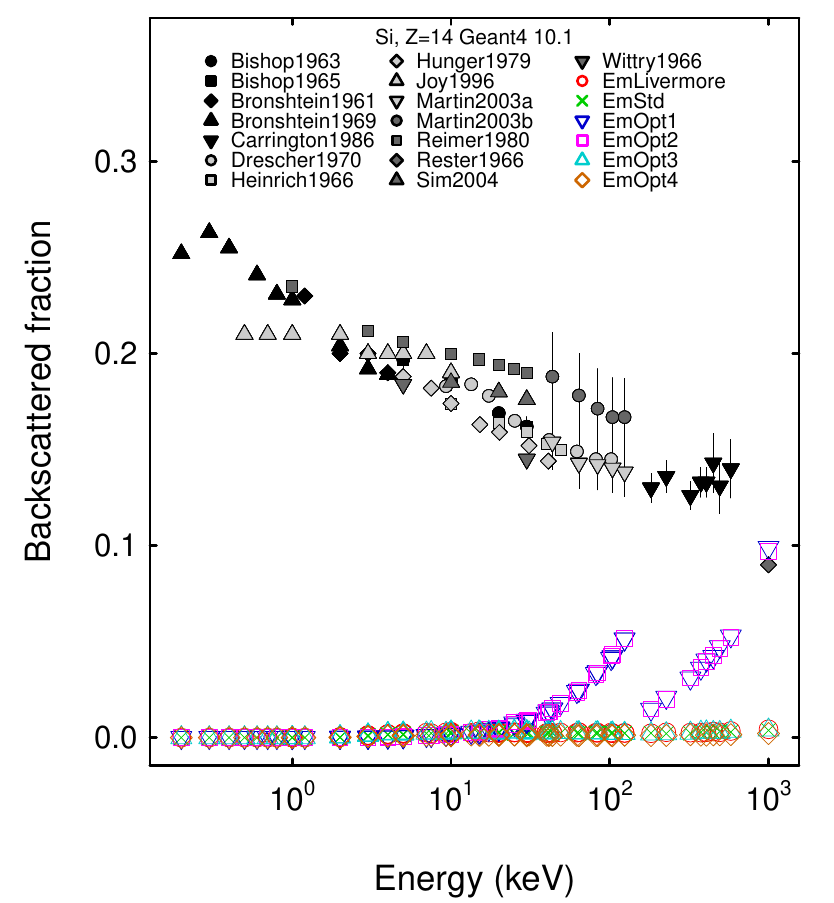}}
\caption{Fraction of electrons backscattered from a silicon target as a function
of the electron beam energy: experimental data (black and grey filled symbols)
and Geant4 10.1 simulation results with G4EmLivermorePhysics (red empty circles), 
G4EmStandardPhysics (green crosses),
G4EmStandardPhysics\_option1 (blue empty upside-down triangles),
G4EmStandardPhysics\_option2 (magenta empty squares),
G4EmStandardPhysics\_option3 (turquoise empty triangles)
and G4EmStandardPhysics\_option4 (brown empty diamonds)
PhysicsConstructors.}
\label{fig_physList_14}
\end{figure}

\begin{figure} 
\centerline{\includegraphics[angle=0,width=8.5cm]{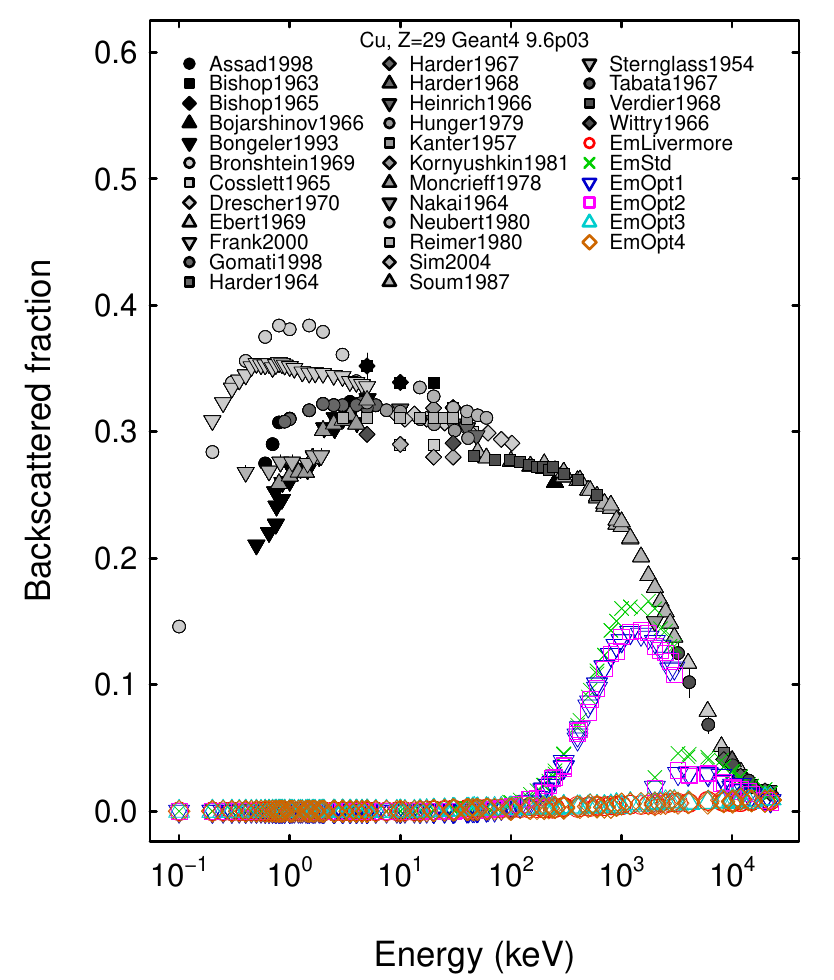}}
\caption{Fraction of electrons backscattered from a copper target as a function
of the electron beam energy: experimental data (black and grey filled symbols)
and Geant4 9.6 simulation results with G4EmLivermorePhysics (red empty circles), 
G4EmStandardPhysics (green crosses),
G4EmStandardPhysics\_option1 (blue empty upside-down triangles),
G4EmStandardPhysics\_option2 (magenta empty squares),
G4EmStandardPhysics\_option3 (turquoise empty triangles)
and G4EmStandardPhysics\_option4 (brown empty diamonds)
PhysicsConstructors.}
\label{fig_physList_29}
\end{figure}

\begin{figure} 
\centerline{\includegraphics[angle=0,width=8.5cm]{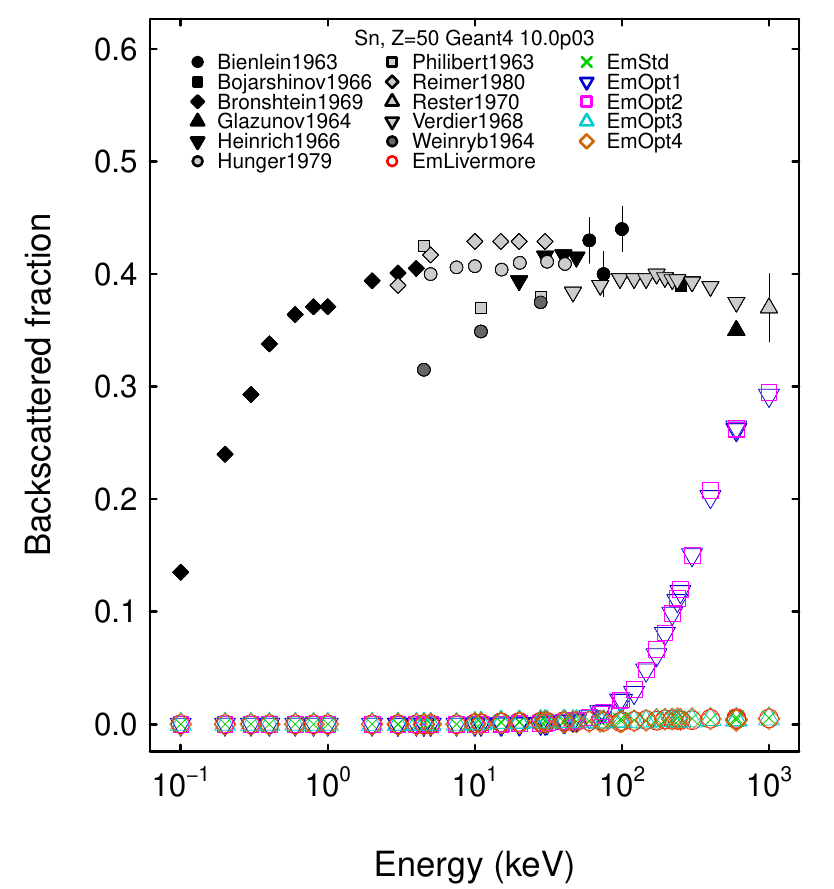}}
\caption{Fraction of electrons backscattered from a tin target as a function
of the electron beam energy: experimental data (black and grey filled symbols)
and Geant4 10.0 simulation results with G4EmLivermorePhysics (red empty circles), 
G4EmStandardPhysics (green crosses),
G4EmStandardPhysics\_option1 (blue empty upside-down triangles),
G4EmStandardPhysics\_option2 (magenta empty squares),
G4EmStandardPhysics\_option3 (turquoise empty triangles)
and G4EmStandardPhysics\_option4 (brown empty diamonds)
PhysicsConstructors.}
\label{fig_physList_50}
\end{figure}

\begin{figure} 
\centerline{\includegraphics[angle=0,width=8.5cm]{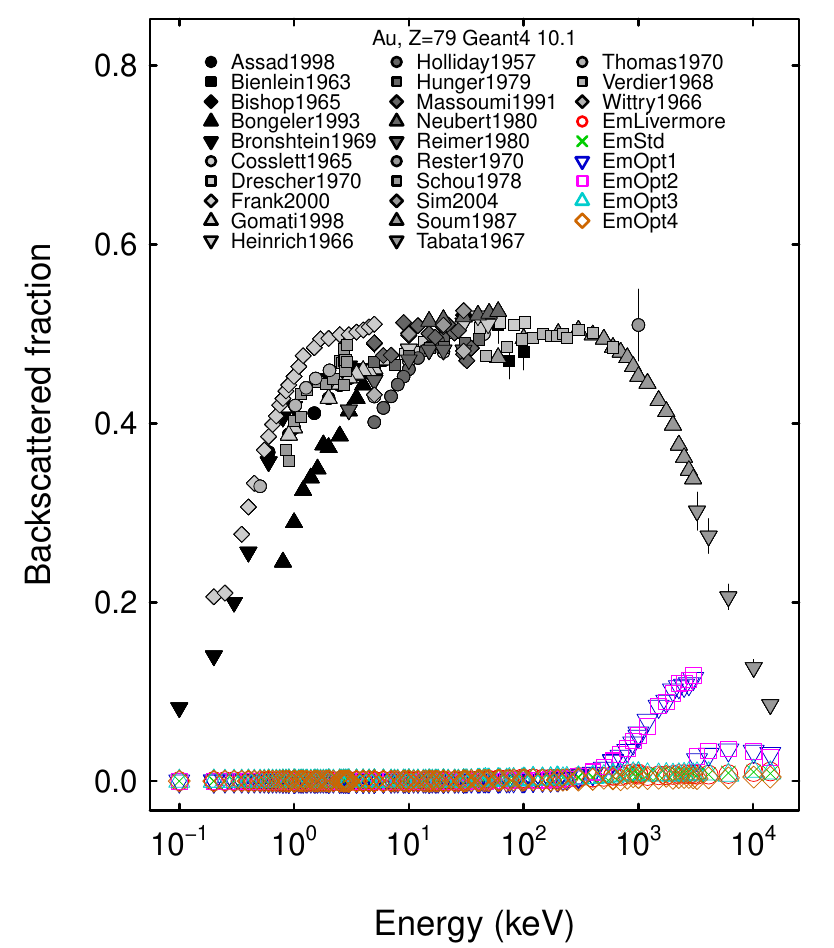}}
\caption{Fraction of electrons backscattered from a gold target as a function
of the electron beam energy: experimental data (black and grey filled symbols)
and Geant4 10.1 simulation results with G4EmLivermorePhysics (red empty circles), 
G4EmStandardPhysics (green crosses),
G4EmStandardPhysics\_option1 (blue empty upside-down triangles),
G4EmStandardPhysics\_option2 (magenta empty squares),
G4EmStandardPhysics\_option3 (turquoise empty triangles)
and G4EmStandardPhysics\_option4 (brown empty diamonds)
PhysicsConstructors.}
\label{fig_physList_79}
\end{figure}

The fraction of backscattered electrons produced by
G4Em\-Standard\-Physics, 
G4\-Em\-Standard\-Physics\_option1,
G4Em\-Standard\-Physics\_option2, 
G4\-Em\-Standard\-Physics\_option3
G4\-Em\-Standard\-Physics\_option4 and 
G4\-Em\-Livermore\-Physics
 with Geant4 10.0 version is illustrated in Figs.
\ref{fig_physList_v1003_6}-\ref{fig_physList_79}.
One can qualitatively observe that simulation results produced by 
G4EmStandardPhysics, G4EmStandardPhysics\_option3,
G4EmStandardPhysics\_option4 and G4EmLivermorePhysics
fail to reproduce the characteristics of experimental distributions
over the whole energy range.
Simulations using G4EmStandardPhysics\_option1 and
G4EmStandardPhysics\_option2 appear to approach experimental data
in the higher energy end above 1~MeV.

These qualitative considerations are reflected in the outcome of the statistical
data analysis.
At lower energies the recommended PhysicsConstructors appear incapable to 
reproduce experimental data similarly to other Geant4 multiple scattering configurations.
G4EmStandardPhysics\-\_option3 and  G4Em\-Standard\-Physics\-\_option4,
which according to Geant4 documentation are intended to produce high accuracy simulations,
exhibit negligible efficiency at all energies, as well as  G4EmLivermorePhysics.
At higher energies the efficiency achieved using
G4Em\-Standard\-Physics\-\_option1 and G4EmStandardPhysics\_option2
is higher than that associated with the other recommended
PhysicsConstructors in Geant4 9.6 and 10.0 is approximately a factor
two lower than that achieved in the most efficient scenarios (Urban multiple
scattering configuration in Geant4 9.1 and Coulomb single scattering
configuration in Geant4 10.0).
The performance of G4EmStandardPhysics\-\_option1 is consistent with the
statement in Geant4 user documentation that this PhysicsConstructor is
intended for fast, but less accurate simulation.
G4EmStandardPhysics exhibits a similar performance in Geant4 9.6, while
its efficiency drops in Geant4 10.0.

For all the recommended PhysicsConstructors the hypothesis of
equivalent compatibility with the most efficient configurations 
(Urban in Geant4 9.1 and Coulomb in Geant4 10.0)
is rejected with 0.0001 significance above 100 keV.

As a result of this analysis, one can conclude that in the scenario examined in
this test significantly better accuracy can be achieved with physics
configurations other than those recommended in Geant4 user documentation.

\subsection{Computational Performance}

\begin{figure}
\centerline{\includegraphics[angle=0,width=8.5cm]{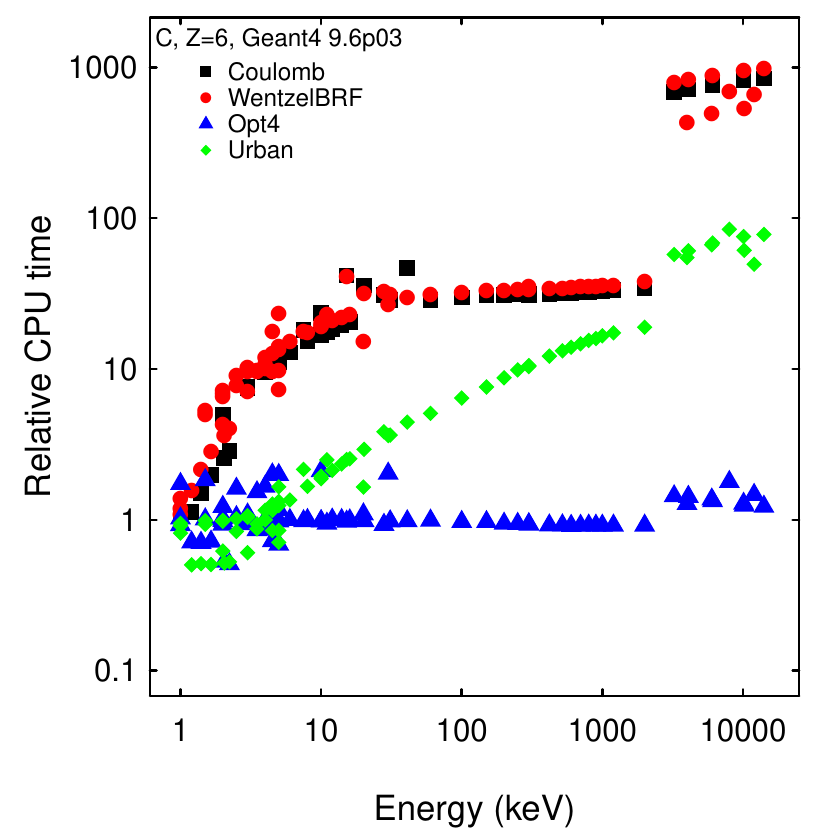}}
\caption{Relative execution time as a function of beam energy for a set of
physics configurations simulating electrons backscattered from a carbon target
with Geant4 9.6: single Coulomb scattering (black squares), WentzelBRF (red
circles), G4EmStandardPhysics\_option4 (blue triangles) and Urban
(green diamonds). The results in the plot are scaled with respect to the
execution time of a simulation with G4EmStandardPhysics. }
\label{fig_time_6}
\end{figure}

\begin{figure} 
\centerline{\includegraphics[angle=0,width=8.5cm]{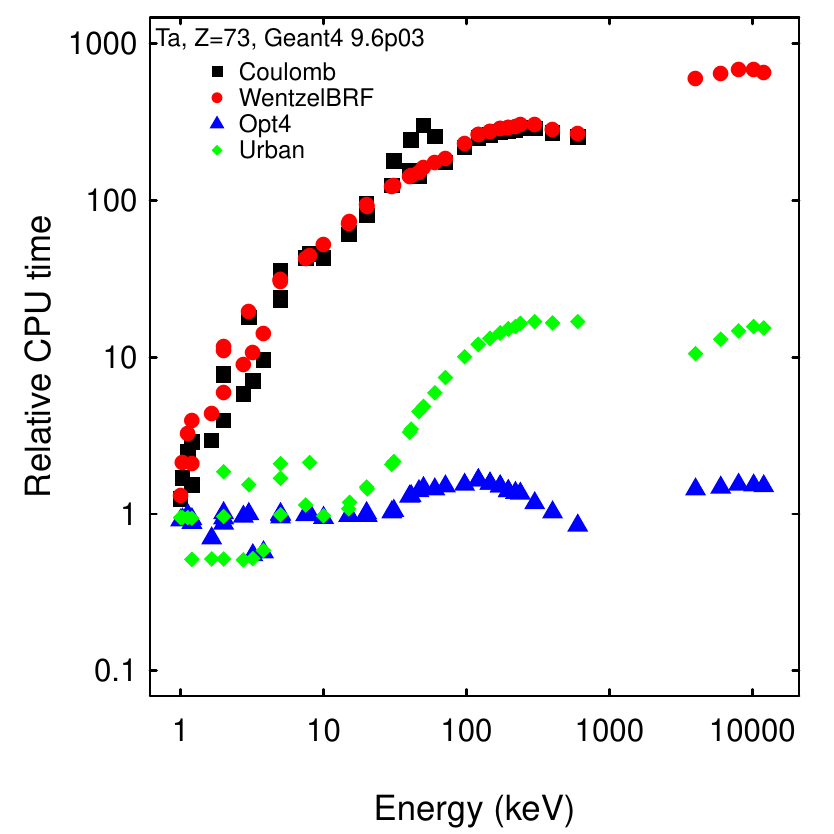}}
\caption{Relative execution time as a function of beam energy for a set of 
physics configurations simulating electrons backscattered from a tungsten target
with Geant4 9.6: single Coulomb scattering (black squares), WentzelBRF (red
circles), G4EmStandardPhysics\_option4 (blue triangles) and Urban
(green diamonds). The results in the plot are scaled with respect to the
execution time of a simulation with G4EmStandardPhysics. }
\label{fig_time_73}
\end{figure}

The heterogeneous production environment of the simulations documented in this
paper, consisting of different platforms, prevents the absolute comparison of
the computational performance of different physics modeling options.
Nevertheless, in most test cases the simulations corresponding to a group of
multiple scattering configurations (Urban model and its variants, WentzelVI model variants,
Coulomb, GSBRF) were executed on the same machine, thus enabling a
relative comparison of their computational performance at least at a qualitative level.

A sample of results in Figs. \ref{fig_time_6} and \ref{fig_time_73} illustrate the main
features regarding the computational performance of backscattering simulations,
namely the overhead associated with simulating single scattering
rather than multiple scattering, in Geant4 version 9.6.
One can observe that the Coulomb and WentzelBRF configurations exhibit similar
computational performance, while the Urban configuration is an order of
magnitude faster.
All recommended PhysicsConstructors are substantially faster than 
the other configurations shown in the plots, although at the price of
significantly degraded compatibility with experiment, as documented in the 
previous sections.

\section{Correlation between backscattering and energy deposition}

\begin{figure} 
\centerline{\includegraphics[angle=0,width=8.5cm]{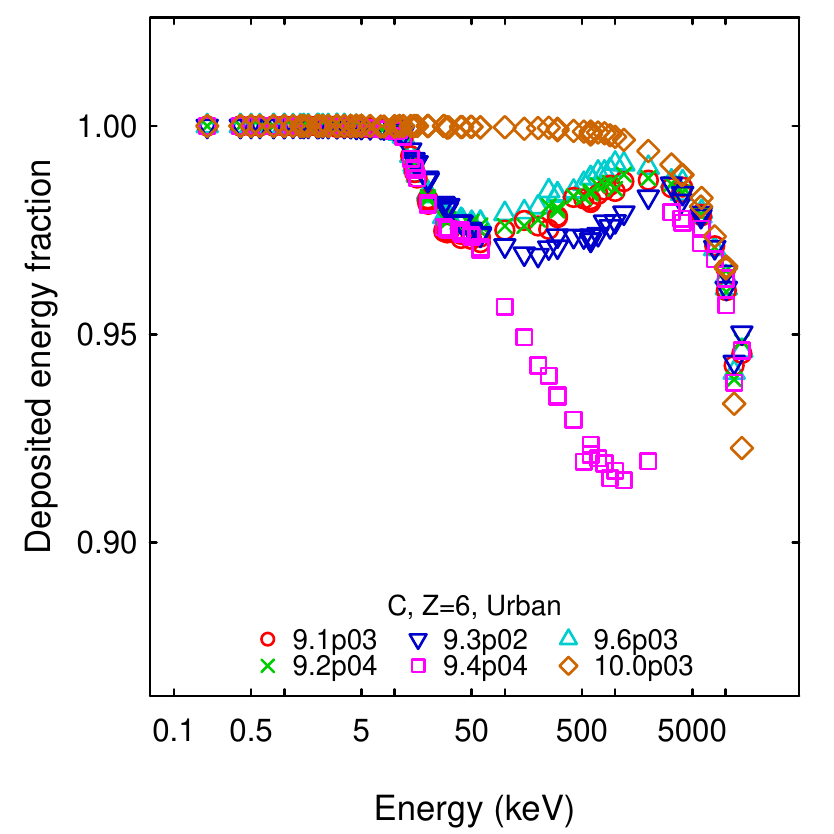}}
\caption{Energy deposited in a carbon target as a function of the electron beam
energy, resulting from simulations with the Urban multiple scattering
configuration in different Geant4 versions: 9.1 (red circles), 9.2 (green
crosses), 9.3 (blue upside down triangles), 9.4 (magenta squares), 9.6
(turquoise triangles) and 10.0 (brown diamonds).}
\label{fig_edep_Urban_6}
\end{figure}

\begin{figure} 
\centerline{\includegraphics[angle=0,width=8.5cm]{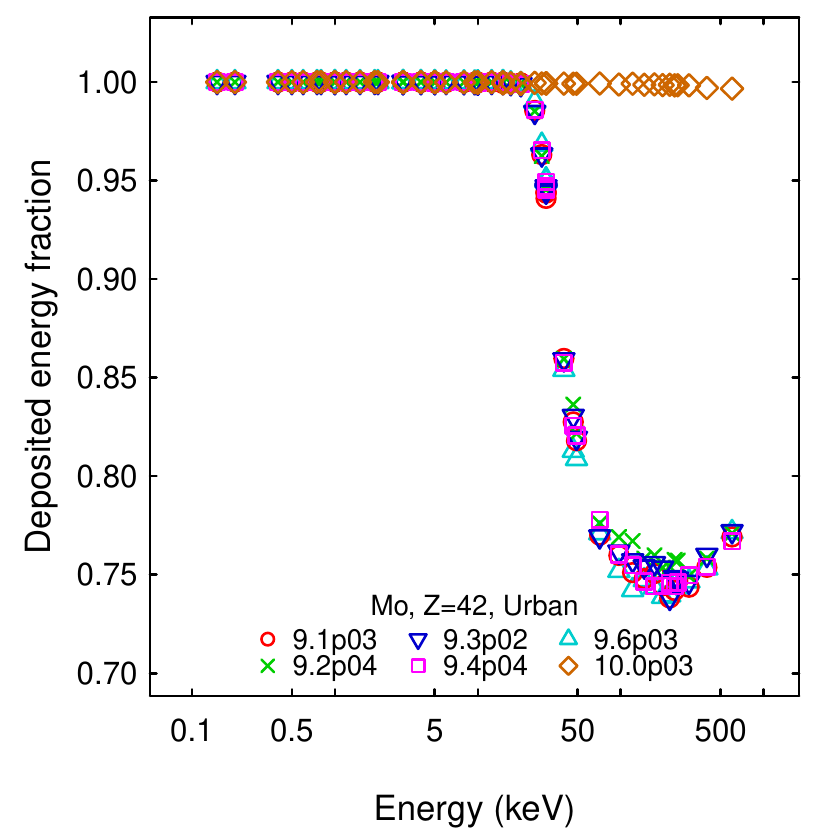}}
\caption{Energy deposited in a molybdenum target as a function of the electron beam
energy, resulting from simulations with the Urban multiple scattering
configuration in different Geant4 versions: 9.1 (red circles), 9.2 (green
crosses), 9.3 (blue upside down triangles), 9.4 (magenta squares), 9.6
(turquoise triangles) and 10.0 (brown diamonds).}
\label{fig_edep_Urban_42}
\end{figure}

\begin{figure} 
\centerline{\includegraphics[angle=0,width=8.5cm]{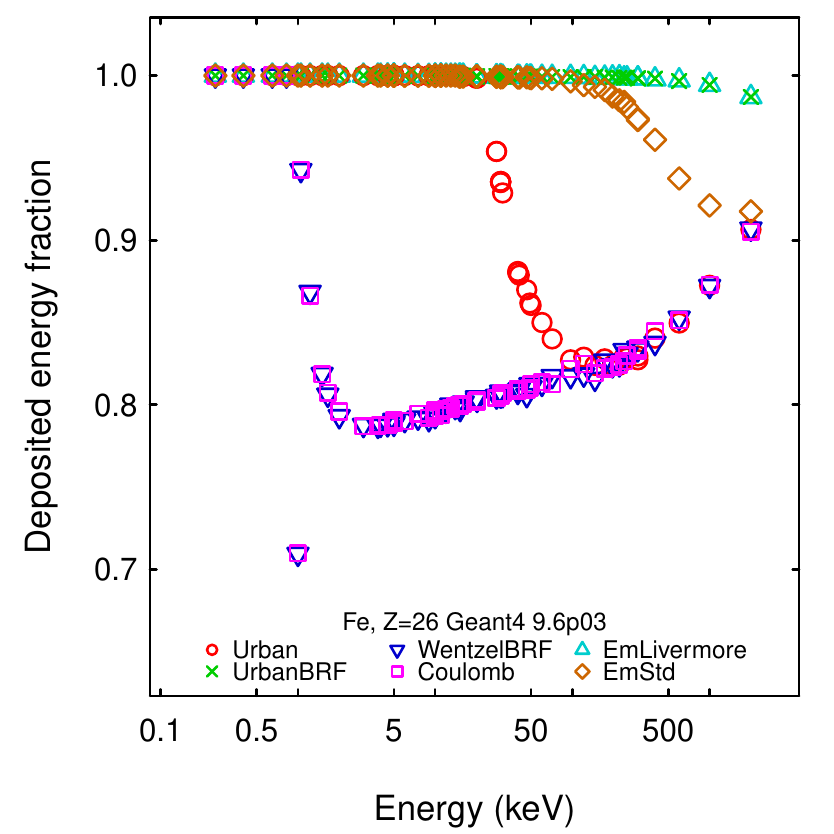}}
\caption{Energy deposited in an iron target as a function of the electron beam
energy, resulting from simulations with different multiple scattering
configurations in Geant4 version 9.6: Urban (red circles),
UrbanBRF (green crosses) WentzelBRF (blue upside down
triangles) and Coulomb (magenta squares) electron scattering
configurations, EmLivermore (turquoise triangles) and EmStandard
(brown diamonds) pre-packaged PhysicsConstructors).}
\label{fig_edep_v963_26}
\end{figure}

\begin{figure} 
\centerline{\includegraphics[angle=0,width=8.5cm]{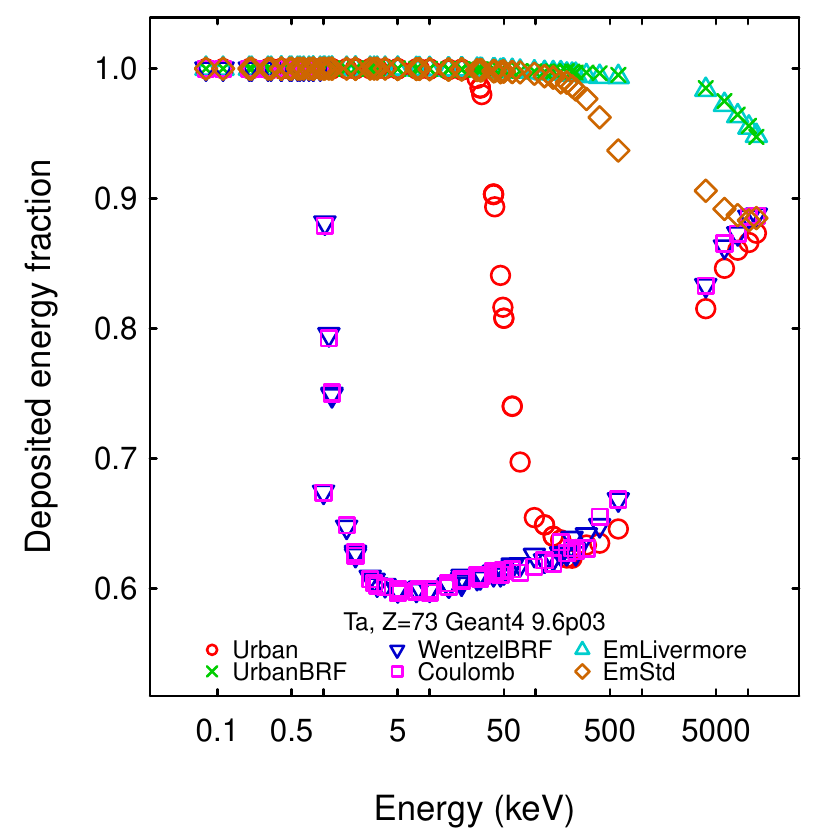}}
\caption{Energy deposited in a tantalum target as a function of the electron
beam energy, resulting from simulations with different multiple scattering
configurations in Geant4 version 9.6: Urban (red circles),
UrbanBRF (green crosses) WentzelBRF (blue upside down
triangles) and Coulomb (magenta squares) electron scattering
configurations, EmLivermore (turquoise triangles) and EmStandard
(brown diamonds) pre-packaged PhysicsConstructors.}
\label{fig_edep_v963_73}
\end{figure}

It is physically intuitive that a relation exists between the fraction of
electrons that are backscattered from a target volume and the energy deposited in it.
A few examples of the energy deposited in the targets used in the simulation of
electron backscattering are shown in Figs.
\ref{fig_edep_Urban_6}-\ref{fig_edep_v963_73}: they illustrate the dependence of
the energy deposition on the evolution of the Geant4 Urban model from
version 9.1 to 10.0 in Figs. \ref{fig_edep_Urban_6}-\ref{fig_edep_Urban_42} and
the effects associated with different physics configurations in the simulation
in Figs. \ref{fig_edep_v963_26}-\ref{fig_edep_v963_73}.
Visible differences are observed both among different Geant4 versions for the
same physics configuration, and among different physics configurations in the
same Geant4 version.

\begin{figure} 
\centerline{\includegraphics[angle=0,width=8.5cm]{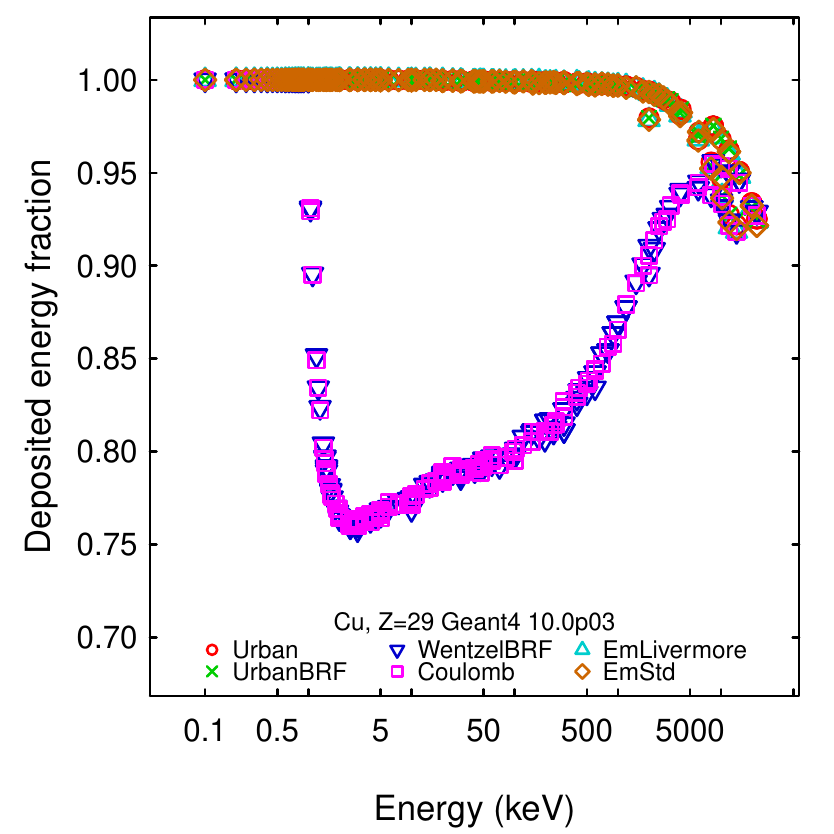}}
\caption{Energy deposited in a copper target as a function of the electron
beam energy, resulting from simulations with different multiple scattering
configurations in Geant4 version 10.0: Urban (red circles),
UrbanBRF (green crosses) WentzelBRF (blue upside down
triangles) and Coulomb (magenta squares) electron scattering
configurations, EmLivermore (turquoise triangles) and EmStandard
(brown diamonds) pre-packaged PhysicsConstructors.}
\label{fig_edep_v1003_29}
\end{figure}

\begin{figure} 
\centerline{\includegraphics[angle=0,width=8.5cm]{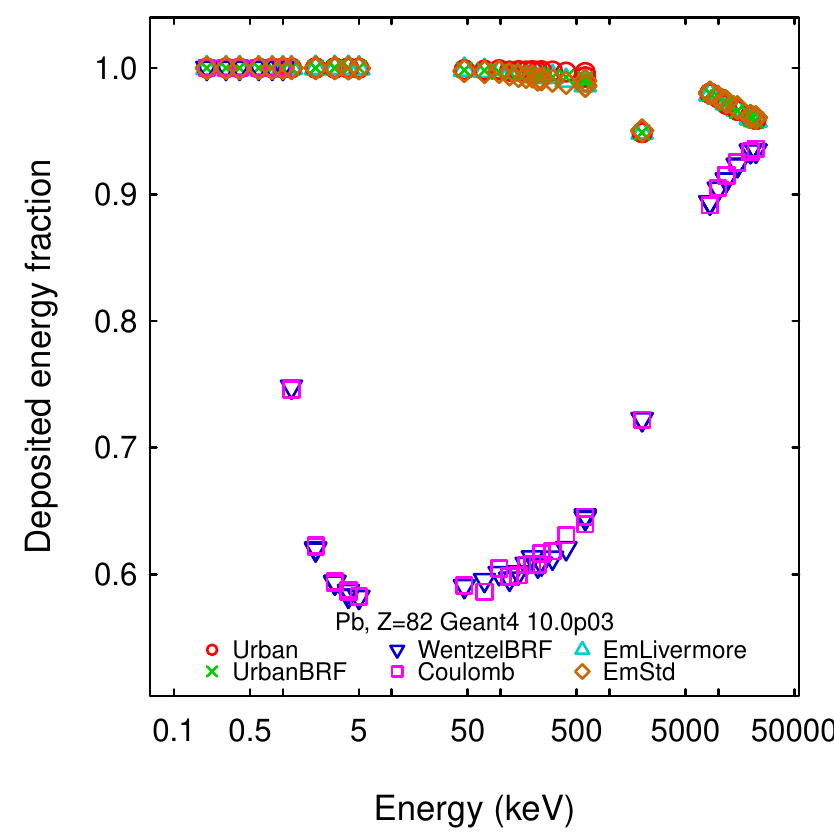}}
\caption{Energy deposited in a lead target as a function of the electron
beam energy, resulting from simulations with different multiple scattering
configurations in Geant4 version 10.0: Urban (red circles),
UrbanBRF (green crosses) WentzelBRF (blue upside down
triangles) and Coulomb (magenta squares) electron scattering
configurations, EmLivermore (turquoise triangles) and EmStandard
(brown diamonds) pre-packaged PhysicsConstructors.}
\label{fig_edep_v1003_82}
\end{figure}

The capability of Geant4 to reproduce experimental measurements of the energy
deposited by low energy electrons is quantitatively analyzed in
\cite{tns_sandia2013} with respect to two experimental configurations: a
longitudinally segmented detector and a detector consisting of a bulk volume.
The latter closely resembles the simulation configuration of the backscattering
test cases evaluated in this paper.
Reference \cite{tns_sandia2013} reports the efficiency, representing the
compatibility with experimental data of energy deposition simulations based on
Geant4 versions 9.1 to 9.6: these values were obtained using the default
Urban multiple scattering model corresponding to each Geant4 version
and imposing a user-defined step limitation, similar to the configuration
identified as Urban in this paper.
The hypothesis was formulated in \cite{tns_sandia2013} that the discrepancies in
compatibility with experiments obtained with different Geant4 versions could be
attributed to the evolution of the implementation of the Urban Geant4
multiple scattering algorithm.

The test of electron backscattering simulation, which is a direct effect of
multiple scattering algorithms implemented in Geant4, allows a quantitative test
of this hypothesis: the evaluation whether a correlation exists between the two
sets of efficiencies, derived from the validation of energy deposition and
backscattering simulations, respectively.
It is worthwhile to note that measures of correlation are not inferential
statistical tests, but are, instead, descriptive statistical quantities, which
represent the degree of relationship between two observables.
Statistical inference concerning the underlying population is enabled by the
analysis of the significance of the measured value through appropriate hypothesis
testing methods.

For the purpose of measuring the correlation between the two sets of
efficiencies, the data sample over which the backscattering efficiency is
calculated has been restricted to the energy range of the energy deposition test
described in \cite{tns_sandia2013}, and the correlation analysis is limited to
Geant4 versions common to both validation tests.

Measures of correlation related to these observables are calculated and
complemented by inferential statistical tests \cite{as89, hollander1973}.
The null hypothesis is formulated as the absence of any correlation between the
two sets of efficiencies, expressed by a measure of zero.
The alternative hypothesis concerns the existence of a positive correlation 
between the two sets of efficiencies associated with Geant4 versions;
it corresponds to the execution of one-tailed tests.

Two nonparametric correlation measures, Spearman $\rho$ correlation
\cite{spearman} and Kendall $\tau$ correlation \cite{kendall}, are reported in
Table \ref{tab_sandiacorr} along with the associated p-values.
Table \ref{tab_sandiacorr} also lists Pearson product-moment correlation
coefficient \cite{pearsoncorr} and the associated p-value: although more widely
known in the experimental physics environment, Pearson correlation coefficient
has a more limited scope, as it describes linear correlation only.

The null hypothesis of no correlation is rejected with 0.01 significance.
Consistent with the alternative hypothesis, one can infer that a positive
correlation exists.
From this analysis one can infer that the accuracy of simulation of electron
backscattering, which is a direct effect of multiple scattering modeling, and of
simulation of the energy deposited by low energy electrons are correlated.


\begin{table}[htbp]
  \centering
  \caption{Correlation between efficiency at simulating backscattering coefficient 
                and energy deposition compatible with experiment over Geant4 versions 9.1 to 9.6}
    \begin{tabular}{lcc}
    \hline
    Correlation  & Measure & p-value \\
    \hline
    Kendall $\tau$ & 1 & 0.008 \\
    Spearman $\rho$ & 1 & 0.008 \\
    Pearson correlation coefficient & 0.992 & 0.0005 \\
    \hline
    \end{tabular}%
  \label{tab_sandiacorr}%
\end{table}%

\section{Conclusion}

This paper has analyzed the capabilities of several Geant4 multiple scattering
models with respect to a large experimental data sample, and its evolution over Geant4
versions from 9.1 to 10.1.
In addition, a single scattering algorithm has been evaluated.
The fraction of electrons backscattered from a semi-infinite target has proven to
be a sensitive probe of multiple scattering algorithms available in Geant4.
A significant correlation between the accuracy of this observable and
the accuracy of the simulation of the energy deposition originating from
low energy electrons has been established on rigorous statistical grounds.

Large variability is observed in the performance of all models over the range
of Geant4 versions considered in this study.
The evolution does not always go in the direction of improving the compatibility
with experiment: statistically significant regressions are observed for some
Geant4 models with respect to their previous behaviour.
Although an analysis of the software development process in the electromagnetic
physics domain of Geant4 is outside the scope of this paper, the results of this
validation test highlight the opportunity to strengthen the discipline of change
management and the traceability of changes, including their side effects.
Established software process frameworks, such as the Unified Process \cite{UP},
CMMI (Capability Maturity Model Integration) \cite{cmmi} and the ISO/IEC 15504
Standard \cite{iso15504}, provide support for these and other related disciplines.


Due to the variability of physics performance affecting all models, it is not
possible to identify a single Geant4 version providing an optimal simulation 
environment over the whole energy range considered in the test.
All the evaluated multiple scattering models encounter difficulties at
reproducing low energy backscattering measurements: up to a few tens of keV only
the single Coulomb scattering model demonstrates some capability to describe
experimental data realistically, limited to Geant4 9.6 and 10.0 versions in its
default configuration,
although at the price of substantially slower computational performance.
Above a few tens of keV, the Urban model in Geant4 9.1, complemented by user
defined step limitation, demonstrates the best capability to reproduce
experimental data in a condensed transport scheme.
Recommended settings advertised for high accuracy, such as
\textit{UseDistanceToBoundary}, worsen the compatibility with experiment of
this configuration.


In the investigated scenario the predefined electromagnetic PhysicsConstructors
do not fulfill the expectations of accuracy deriving from their advertisement in
Geant4 documentation and the statement of their intensive validation in conference
papers.
The recommendations of their use should be revised and based on objective
grounds, documenting the scope to which they are pertinent and quantifying the
capabilities of the embedded physics configurations with respect to experimental
data.


The quantitative assessment documented in this paper, along with the results of
the validation test documented in \cite{tns_sandia2013}, allow Geant4 users to
optimize the physics configuration of simulation applications involving
electrons of energy up to a few tens of MeV, either as primary particles or as
secondary interaction products.


\section*{Acknowledgment}

The support of the CERN Library has been essential to this study. 
The authors thank Mihaly Novak for valuable discussions regarding Geant4
multiple scattering models, Sergio Bertolucci for support, Anita Hollier for
proofreading the manuscript, Fabio Pratolongo and the Computing Service at INFN
Genova for help with the computing hardware used for simulation productions.
S. H. Kim thanks CERN PH/SFT Group for the hospitality at CERN during summer
2014.


\end{document}